\documentclass{tran-l}

\usepackage{tabularx}
\usepackage{array}

\newcolumntype{L}{>{\raggedright\arraybackslash}X}

\makeatletter
\newcommand{\veryHuge}{\@setfontsize\veryHuge{40}{46}}
\makeatother

\usepackage{tikz}
\usetikzlibrary{positioning, shapes.geometric, shapes.misc, shapes.callouts}
\usetikzlibrary{arrows.meta}

\tikzset{
    mainbox/.style = {rectangle, draw, rounded corners=3pt, align=center, minimum width=14cm, minimum height=3.5cm, font=\veryHuge},
    cloudbox/.style = {cloud, cloud puffs=10, cloud ignores aspect, draw, align=center, font=\LARGE, minimum width=4.8cm, minimum height=2.0cm},
    ovalbox/.style = {ellipse, draw, align=center, font=\veryHuge, minimum width=8cm, minimum height=4.6cm},
    arrow/.style   = {->,
    >=Latex,
    line width=3.5pt,
    draw=black,
    scale=2.8}, 
    sidecomment/.style = {font=\Large, align=left}
}

\usepackage{bm}
\usepackage{mathrsfs}
\usepackage{amssymb}
\usepackage{enumitem}
\usepackage{times}
\usepackage{pdfpages}
\usepackage{float}
\usepackage{subfig,graphicx}
\usepackage{csquotes}
\usepackage{hyperref}
\usepackage{amsmath}
\usepackage{ulem}
\usepackage{lipsum} 
\usepackage{nccmath}
\usepackage{bigints}
\usepackage{natbib}
\usepackage{multirow}
\usepackage{adjustbox}
\usepackage{threeparttable}
\usepackage{bbm}

\usepackage{pdfpages}
\usepackage{wrapfig}
\usepackage{color}

\usepackage{float}

\usepackage{caption}
\usepackage{booktabs}
\usepackage{cite}

\usepackage{graphicx}

\newtheorem{theorem}{Theorem}
\newtheorem{lemma}{Lemma}

\newtheorem{proposition}{Proposition}
\newtheorem{definition}{Definition}

\newtheorem{remark}{Remark}

\numberwithin{property}{section}
\numberwithin{proposition}{section}
\numberwithin{equation}{section}
\numberwithin{theorem}{section}
\numberwithin{corollary}{section}
\numberwithin{lemma}{section}
\numberwithin{definition}{section}
\numberwithin{corollary}{section}
\numberwithin{example}{section}
\numberwithin{remark}{section}

\begin{document}

\title[Asymptotic Theory of $K$-fold CV in Lasso and the validity of Bootstrap]{Asymptotic Theory of $K$-fold Cross-validation in Lasso and the validity of Bootstrap}

\author{Mayukh Choudhury}
\address{Department of Mathematics, Indian Institute of Technology Bombay, Mumbai 400076, India}
\email{214090002@iitb.ac.in}

\author{Debraj Das}
\address{Department of Mathematics, Indian Institute of Technology Bombay, Mumbai 400076, India}
\email{debrajdas@math.iitb.ac.in}




\keywords{$K$-fold CV, Lasso, $n^{1/2}$-consistency, PB, Stochastic Equicontinuity, VSC.}

\begin{abstract}
Least absolute shrinkage and selection operator or Lasso is one of the widely used regularization methods in regression. Statisticians usually implement Lasso in practice by choosing the penalty parameter in a data-dependent way, the most popular being the $K-$fold cross-validation (or $K-$fold CV). However, inferential properties, such as the variable selection consistency and $n^{1/2}-$consistency, of the $K-$fold CV based Lasso estimator and validity of the Bootstrap approximation are still unknown. In this paper, we consider the heteroscedastic linear regression model and show only under some moment type conditions that the Lasso estimator with $K$-fold CV based penalty is $n^{1/2}-$consistent, but not variable selection consistent. Additionally, we establish the validity of Bootstrap in approximating the distribution of the $K-$fold CV based Lasso estimator. Therefore, our results theoretically justify the use of $K-$fold CV based Lasso estimator to perform statistical inference in linear regression. We validate our Bootstrap method for the $K-$fold CV based Lasso estimator in finite samples based on simulations. We also implement our Bootstrap based inference on a real data set.
\end{abstract}

\maketitle
\section{Introduction}\label{sec:intro}
Consider the heteroscedastic linear regression model:
\begin{align}\label{eqn:linearreg}
 y_i=\bm{x}_i^\top \bm{\beta}+\varepsilon_i,\;\;\;\; i\in \{1,..,n\},  \end{align}
where, $\{y_{1},..,y_{n}\}$ are responses, $\{\bm{x}_1,\dots, \bm{x}_n\}$ are non-random covariates with design vector $\bm{x}_i=(x_{i1},...,x_{ip})^\top$ and $\bm{\beta}=(\beta_1,..,\beta_p)^\top$ is the regression parameter vector. Throughout the paper, we assume that the dimension $p$ is fixed and the design matrix $\bm{X}= (\bm{x}_1,\dots, \bm{x}_n)^\top$ has full column rank. Moreover, $\{\varepsilon_1,...,\varepsilon_n\}$ are independent (but not necessarily identically distributed) errors with $\mathbf{E}(\varepsilon_i)=0$ for all $i$. When we try to draw an inference about the parameter in a regression set-up, the first thing we generally check is whether all the covariates are relevant, i.e. whether the parameter $\bm{\beta}$ actually sits in a lower dimensional space. One of the widely used methods to identify the sparsity and to draw inference is to employ Lasso, introduced by \citet{tibshirani1996regression}. The Lasso estimator $\hat{\bm{\beta}}_n $ with the penalty parameter $\lambda (> 0)$ is defined as
\begin{align}\label{eqn:def}
\hat{\bm{\beta}}_{n} \equiv \hat{\bm{\beta}}_{n}(\lambda)= \mbox{Argmin}_{\bm{\beta}} \Big\{\frac{1}{2}\sum_{i=1}^{n}\Big(y_{i} - \bm{x}_{i}^\top\bm{\beta}\Big)^2 +\lambda\sum_{j=1}^{p}|\beta_{j}|\Big\}.
\end{align}
Here $\lambda $ is inducing sparsity in the model (\ref{eqn:linearreg}) and hence $\hat{\bm{\beta}}_{n}$ can be thought of as a function of $\lambda$. An important part of the practical implementation of Lasso is to identify a suitable choice of the penalty parameter $\lambda$. Practitioners usually choose the penalty parameter in a data-dependent way. Among different data-dependent methods, the most popular is cross-validation (hereafter referred to as CV), specifically the $K$-fold method. We define the $K-$fold CV based penalty $\hat{\lambda}_{n, K}$ in the next subsection.

\subsection{\bf The setup}
Suppose that $K$ is some positive integer. Without loss of generality, we can assume that $n=mK$ and consider $[I_k:k\in \{1,..,K\}]$ to be a partition of the set $\{1,..,n\}$ with $m=|I_k|$ for all $k\in \{1,\dots,K\}$. Otherwise, we can simply consider $(K-1)$ many partitions of the same size and put the remaining elements of $\{1,\dots, n\}$ in another partition. Throughout the paper, we keep $K$ fixed. For each $k\in \{1,\ldots,K\}$ and a fixed penalty $\lambda (> 0)$, the Lasso estimator corresponding to all observations except those in $I_k$ is defined as
\begin{align}\label{eqn:cvlassoestlm}
  \hat{\bm{\beta}}_{n,-k}(\lambda) = \operatorname*{Argmin}_{\bm{\beta}} \Big\{\frac{1}{2}\sum_{i \notin I_k}\Big(y_{i} - \bm{x}_{i}^\top\bm{\beta}\Big)^2 +\lambda\sum_{j=1}^{p}|\beta_{j}|\Big\}.
  \end{align}
Then we can define the cross-validated prediction error for the Lasso estimator at penalty $\lambda (> 0)$ as 
\begin{align}\label{eqn:CVprediction}
H_{n,K}(\lambda) =\frac{1}{2}\sum_{k=1}^{K}\sum_{i \in I_k}\Big[y_i - \bm{x_i}^\top \hat{\bm{\beta}}_{n,-k}(\lambda)\Big]^2.
\end{align}
Subsequently, the $K$-fold CV based penalty $\hat{\lambda}_{n,K}$ is defined as 
\begin{align}\label{eqn:deflambdacv}
    \hat{\lambda}_{n,K}\in \operatorname*{Argmin}_{\lambda \ge  0} H_{n,K}(\lambda).
\end{align}
Practitioners usually draw inferences about $\bm{\beta}$ based on the $K-$fold CV based Lasso estimator $\hat{\bm{\beta}}_n(\hat{\lambda}_{n, K})$. However, to our knowledge, the asymptotic distributional properties of $\hat{\bm{\beta}}_n(\hat{\lambda}_{n, K})$ are not yet known, which we investigate in this paper. We also develop a Bootstrap based approximation of the distribution of $\hat{\bm{\beta}}_n(\hat{\lambda}_{n, K})$. We present the details of our contributions in the next subsection.

\subsection{\bf Our Contributions}

Among the different asymptotic distributional properties, variable selection consistency (or VSC) and $n^{1/2}-$consistency of a regression estimator are the most important from practical point of view. This is because that these two properties generally dictate how the underlying estimator can be used to draw inferences based on an observed dataset. We call $\hat{\bm{\beta}}_n$  to be VSC if $\mathbf{P}(\{j: \beta_j \neq 0\} = \{j: \hat{\beta}_{n,j} \neq 0\}) \rightarrow 1$, as $n\rightarrow \infty$, and to be $n^{1/2}-$ consistent if $n^{1/2}(\hat{\bm{\beta}}_n - \bm{\beta})$ is stochastically bounded.  However, results on VSC and $n^{1/2}-$consistency of Lasso are available only when the penalty is nonrandom. In a seminal paper, \citet{knight2000asymptotics} established $n^{1/2}-$consistency and found the asymptotic distribution of the Lasso estimator $\hat{\bm{\beta}}_n(\lambda_n)$ when $\{n^{-1/2}\lambda_n\}_{n \geq 1}$ is convergent. Later \citet{zhao2006model} assumed some irrepresentable type conditions on the design vectors and established VSC of $\hat{\bm{\beta}}_n(\lambda_n)$, when $\{n^{-1/2}\lambda_n \}_{n\geq 1}$ is divergent. Recently, \citet{lahiri2021necessary}, among other results, showed that divergence of $\{n^{-1/2}\lambda_n\}_{n\geq 1}$ is also necessary for $\hat{\bm{\beta}}_n(\lambda_n)$ to be VSC. Therefore, the Lasso estimator cannot simultaneously be VSC and be $n^{1/2}-$consistent. 
In view of this, a natural question that we would like to answer here is\\
\begin{center}
  \textit{ Does $K-$fold CV based Lasso perform variable selection or is it $n^{1/2}-$consistent?}
\end{center}
\vspace*{3mm}

The answer crucially depends on how $n^{-1/2}\hat{\lambda}_{n, K}$ behaves as a sequence of $n$. First, we show that $$\mathbf{P}\big(n^{-1}\hat{\lambda}_{n, K}\rightarrow 0\;\text{as}\;n\to\infty\big) = 1.$$ This result essentially implies that $\hat{\bm{\beta}}_n(\hat{\lambda}_{n, K})$ is consistent for $\bm{\beta}$, based on Theorem 1 of \citet{knight2000asymptotics}. Subsequently, using the consistency of $\hat{\bm{\beta}}_n(\hat{\lambda}_{n, K})$, we explore the asymptotic nature of the sequence $\{n^{-1/2}\hat{\lambda}_{n, K}\}_{n\geq 1}$.
We essentially establish that 
\begin{align}\label{eqn:lambdahatmainresult}
\mathbf{P}\Big(\limsup_{n\rightarrow \infty}n^{-1/2}\hat{\lambda}_{n, K}< \infty\Big) = 1,
\end{align}
only under some mild moment conditions on the design vectors and regression errors. This result, jointly with Theorem 4.1 of \citet{lahiri2021necessary}, imply that $\hat{\bm{\beta}}_n(\hat{\lambda}_{n, K})$ cannot be VSC. 

In the next step, we explore whether $\hat{\bm{\beta}}_n(\hat{\lambda}_{n,K})$ is $n^{1/2}$-consistent.  
Towards that first we investigate the convergence of the sequence $\{n^{-1/2}\hat{\lambda}_{n,K}\}_{n\geq 1}$, as an improvement of (\ref{eqn:lambdahatmainresult}). However, it is not possible to study the convergence of the sequence $\{n^{-1/2}\hat{\lambda}_{n,K}\}_{n\geq 1}$ without some smoothness condition on the Lasso solution path and some form of uniqueness of the minimizer of an asymptotic objectiove function, obtained based on $H_{n, K}(\lambda)$, defined in (\ref{eqn:CVprediction}).
Indeed, we establish the stochastic equicontinuity of a centered and scaled version of $\{ \hat{\bm{\beta}}_n(\lambda)\}_{n\geq 1}$ as a function of $\lambda$ (cf. Proposition \ref{prop:eqicon}), by utilizing strong convexity of the Lasso objective function. This proposition is basically an improvement of the results on continuity of the Lasso solution path, established in \citet{efron2004least} and \citet{tibshirani2011solution}, and may be of independent interest in other related problems on Lasso.  Next in Proposition \ref{prop:C.4}, we establish the well-separated uniqueness of the asymptotic objective function, as desired, by analyzing the piecewise linearity of the Lasso estimator $\hat{\bm{\beta}}_n(\lambda)$ as a function of $\lambda$.
We utilize this stochastic equicontinuity to establish the following. 
\begin{align}\label{eqn:dc}
n^{-1/2}\hat{\lambda}_{n,K}\xrightarrow{d}\hat{\Lambda}_{\infty,K},
\end{align}
where $\xrightarrow{d}$ denotes the convergence in distribution. The limit $\hat{\Lambda}_{\infty,K}$ is defined in Table \ref{tab:unified}. Based on the convergence in distribution of $n^{-1/2}\hat{\lambda}_{n, K}$ in (\ref{eqn:dc}), we then explore the $n^{1/2}-$consistency of the Lasso estimator $\hat{\bm{\beta}}_{n}(\hat{\lambda}_{n,K})$. We basically establish that 
\begin{align}\label{eqn:distconv}
n^{1/2}(\hat{\bm{\beta}}_{n}(\hat{\lambda}_{n,K})-\bm{\beta})\xrightarrow{d}\operatorname*{argmin}_{\bm{u}}V_{\infty}(\bm{u},\hat{\Lambda}_{\infty,K}),
\end{align}
where the limiting objective function $V_{\infty}(\cdot, \lambda)$ is defined in \eqref{eqn:nn}. 
Thus, we are able to answer the aforementioned question on VSC and $n^{1/2}-$consistency of $\hat{\bm{\beta}}_n(\hat{\lambda}_{n, K})$. We can conclude that the Lasso estimator with $K-$fold CV based penalty is $n^{1/2}-$consistent, but not VSC.
Clearly, the form of $V_{\infty}(\cdot, \lambda)$ is complicated and, hence, the asymptotic distribution of $n^{1/2}(\hat{\bm{\beta}}_{n}(\hat{\lambda}_{n,K})-\bm{\beta})$ cannot generally be used to draw inferences about $\bm{\beta}$. As an alternative, we develop a Bootstrap method based on the Perturbation bootstrap method of \citet{das2019distributional} and establish its validity in approximating the distribution of $n^{1/2}\big(\hat{\bm{\beta}}_{n}(\hat{\lambda}_{n, K}) - \bm{\beta}\big)$.
An integral step in our Bootstrap approximation is the recalculation of the $K-$fold CV based Lasso penalty based on the pseudo responses created at the Bootstrap stage. See Section \ref{sec:bootstrap} for  the details on our proposed Bootstrap approximation. To the best of our knowledge, this is the first theoretical result on the validity of a Bootstrap approximation of the distribution of the Lasso estimator when the penalty parameter is data dependent, essentially bridging the gap between theory and practice. Our simulation results also validate the Bootstrap approximation in finite samples.

\subsection{\bf Our techniques} In this section, we give an overview of the techniques that we use to establish our results. Note that both the Lasso estimator and the $K-$fold CV based penalty are $Argmin$ functionals and do not have explicit
forms. In order to establish (\ref{eqn:lambdahatmainresult}),  
we explore these $Argmin$ functionals on different favorable sets. We show that $\hat{\bm{\beta}}_n(\lambda_n)$ belongs to different sets based on how the sequence $\{n^{-1/2}\lambda_n\}_{n\geq 1}$ behaves and
the set where $\hat{\bm{\beta}}_n(\lambda_{n})$ lies under boundedness of $\{n^{-1/2}\lambda_n\}_{n\geq 1}$ is more favorable for the $K-$fold CV prediction error ($H_{n, K}(\lambda_n)$), presented in (\ref{eqn:cvlassoestlm}), to
achieve smaller values. We establish several concentration bounds on the Lasso estimator $\hat{\bm{\beta}}_n(\lambda_n)$ based on the asymptotic nature of the penalty parameter $\lambda_n$ to analyze those favorable sets. An important aspect of these concentration bounds is that they also capture the dependence on the dimension $p$. We present these concentration bounds in the supplementary file to save space.

One of the key ingredients in establishing the distribution convergence results (\ref{eqn:dc}) and (\ref{eqn:distconv}), is the Dudley's almost sure representation theorem based on the notion of perfect maps (cf. \citet{dudley1985extended,kim1990cube,kato2009asymptotics}). To establish (\ref{eqn:dc}), first we obtain a version of $\{H_{n,K}(\lambda)\}_{n\geq 1}$ by Dudley's almost sure representation theorem and then we obtain the stochastic equicontinuity of that version over compact sets by utilizing the  stochastic equicontinuity of the centered and scaled Lasso solution path established in Proposition \ref{prop:eqicon}. The stochastic equicontinuity over compact sets translates the pointwise convergence, in probability, to the uniform convergence on compact sets, in probability, due to Theorem 2.1 of \citet{newey1991uniform}. This along-with the unique well separated minimizer of an asymptotic objective function, establied in Proposition \ref{prop:C.4}, essentially establish (\ref{eqn:dc}).
\begin{figure}[H]

\noindent\hspace*{-3.2cm}%
\resizebox{1.15\linewidth}{!}{%
\begin{tikzpicture}[node distance=5.5cm]


\node[mainbox] (A) {\veryHuge{Is $\hat{\beta}_{n}(\hat{\lambda}_{n,K})$  VSC or $n^{1/2}$-consistent?}};

\node[mainbox, below=of A] (B) {\veryHuge{$\mathbf{P}\!\left(n^{-1}\hat{\lambda}_{n,K}\to 0 \;\;\text{as}\;\; n\to\infty\right)=1$}};
\node[mainbox, right=5.5cm of B] (B1) {\veryHuge{$\hat{\beta}_{n}(\hat{\lambda}_{n,k})$ is consistent for $\bm{\beta}$}};
\node[ovalbox, left=2.5cm of B] (B2) {\veryHuge{Theorem  \ref{prop:cv}}};

\node[mainbox, below=of B] (C) {$\mathbf{P}\!\left(\limsup_{n\to\infty} n^{-1/2}\hat{\lambda}_{n,K}<\infty\right)=1$};
\node[mainbox, right=6.5cm of C] (C1) {\veryHuge{$\hat{\beta}_{n}(\hat{\lambda}_{n,K})$ is not VSC}};
\node[ovalbox, left=2.5cm of C] (C2) {\veryHuge{Theorem \ref{thm:cvthm}}};

\node[mainbox, below left=2cm and 5cm of C] (D) { Stochastic equicontinuity of Lasso solution path\\ and\\
Almost sure well-separated unique minimizer of $\hat{H}^\prime_{\infty,K}(\cdot)$};

\node[ovalbox, below left=0.25cm and 0.26cm of D] (D1) {\veryHuge{Proposition \ref{prop:eqicon}}\\\veryHuge{and}\\ \veryHuge{Proposition \ref{prop:C.4}}};

\node[mainbox, right=of D, xshift=0.15cm, below=4cm of C] (E) {\veryHuge{$n^{-1/2}\hat{\lambda}_{n,K} \;\;\overset{d}{\longrightarrow}\;\; \hat{\Lambda}_{\infty,K}$}};
\node[ovalbox, right=2.5cm of E] (E1) {\veryHuge{Theorem \ref{thm: lambdaCVconv}}};

\node[mainbox, below=of E] (F) {$n^{1/2}(\hat{\beta}_n(\hat{\lambda}_{n,K})-\beta)\;\;\overset{d}{\longrightarrow}\;\;\arg\min_u V_\infty(u,\hat{\Lambda}_{\infty,K})$};
\node[mainbox, right=6.5cm of F] (F1) {\veryHuge{$\hat{\beta}_{n}(\hat{\lambda}_{n,K})$ is $n^{1/2}$-consistent}};
\node[ovalbox, left=1.85cm of F] (F2) {\veryHuge{Theorem \ref{thm:distconvcv}}};

\node[mainbox, below=of F] (G) {\veryHuge{$n^{1/2}\Big(\hat{\bm{\beta}}_n^*(\hat{\lambda}_{n,K}^*)-\tilde{\bm{\beta}}_n(\hat{\lambda}_{n,K})\Big)\;\;\overset{d_*}{\longrightarrow}\;\;\arg\min_u V_\infty(u,\hat{\Lambda}_{\infty,K})$}}; 
\node[ovalbox, left=2.5cm of G] (G2){\veryHuge{{\color{black}Theorem \ref{thm:bootconsistency}}}};
\node[mainbox, right=6.5cm of G] (G1) {\veryHuge{Perturbation Bootstrap is valid}};


\draw[arrow] (A) -- (B) node[midway, right, sidecomment] {};
\draw[arrow] (B) -- (B1) node[midway, above, sidecomment] {{\color{black}\Huge{Conclusion}}};
\draw[arrow] (B) -- (C) node[midway, right, sidecomment] {};
\draw[arrow] (C) -- (C1) node[midway, above, sidecomment] {{\color{black}\Huge{Conclusion}}};
\draw[arrow] (A.west) .. controls +(-12,0) and +(-12,-2) .. (D.west)
    node[midway, left=3.2mm, sidecomment]{};
\draw[arrow] (C) -- (E) node[midway, above, sidecomment] {};
\draw[arrow] (D) -- (E) node[midway, above, sidecomment] {};
\draw[arrow] (E) -- (F) node[midway, right, sidecomment] {};
\draw[arrow] (F) -- (F1) node[midway, above, sidecomment] {{\color{black}\Huge{Conclusion}}};
\draw[arrow] (F) -- (G)node[midway, right, sidecomment] {};
\draw[arrow] (G) -- (G1) node[midway, above, sidecomment] {{\color{black}\Huge{Conclusion}}};

\end{tikzpicture}
}
\caption{Schematic representation of our findings.  $\hat{\bm{\beta}}_n^*(\hat{\lambda}_{n,K}^*)$ and $\tilde{\bm{\beta}}_n(\hat{\lambda}_{n,K})$ are defined in section \ref{sec:bootstrap} and other notations are defined in Table \ref{tab:unified})}
\label{fig:cv-lasso-flowchart}

\end{figure}
To prove (\ref{eqn:distconv}), we utilize the $Argmin$ theorem of convex stochastic processes, established in \citet{choudhury2024bootstrapping}. Unlike traditional epiconvergence tools (cf. \citet{geyer1994asymptotics,geyer1996asymptotics}), this $Argmin$ theorem is based on finite-dimensional convergence and stochastic equicontinuity on compact sets and hence blends well with the techniques used to establish (\ref{eqn:dc}). 
Finite dimensional convergence of the objective function of $\hat{\bm{\beta}}_n(\hat{\lambda}_{n, K})$ follows from the joint distribution convergence of the random linear term present in the objective function and $\{n^{-1/2}\hat{\lambda}_{n, K}\}_{n\geq 1}$. We again utilize the Dudley's almost sure representation theorem to prove this joint convergence. 
On the other hand, the stochastic equicontinuity of the objective function of $\hat{\bm{\beta}}_n(\hat{\lambda}_{n, K})$, over compact sets, is due to  (\ref{eqn:lambdahatmainresult}) and the tightness of the random linear term of the objective function of $\hat{\bm{\beta}}_n(\hat{\lambda}_{n, K})$. 
In case of the Bootstrap approximation, we essentially establish the distribution convergence of the Bootstrap pivotal quantity to the limit same as in (\ref{eqn:distconv}), essentially proving the validity of Bootstrap. Based on these discussions, we summarize our findings of this paper schematically in Figure \ref{fig:cv-lasso-flowchart}.

\subsection{\bf Related literature}
The rationale for using CV for selecting the optimal penalty parameter based on empirical evidences in the case of Lasso and other penalized regression methods have been studied by many authors, including \citet{zou2007degrees,friedman2010regularization,buhlmann2011statistics,fan2012variance,van2013lasso,hastie2015statistical,giraud2021introduction}. However, on the theoretical side of the CV, the literature is not at all substantial and also the focus was mostly on establishing risk consistency and oracle-type bounds, rather than on distributional properties. In particular, most of the available results establish upper bounds on prediction or estimation errors without addressing the limiting behavior of the estimator itself. Early contributions by \citet{lecue2012oracle} and  \citet{homrighausen2013lasso, homrighausen2014leave, homrighausen2017risk} explored the risk function for $\hat{\bm{\beta}}_n(\hat{\lambda}_{n, K})$ and established risk consistency under certain regularity conditions. \citet{chatterjee2015prediction} established a non-asymptotic upper bound on the mean squared prediction error for a variant of $2-$fold CV based Lasso estimator. Recently, \citet{chetverikov2021cross} and \citet{chaudhuri2022cross} derived the non-asymptotic oracle inequalities in terms of prediction error and $L^2$ \& $L^1$ estimation errors for $\hat{\bm{\beta}}_n(\hat{\lambda}_{n, K})$ when $\hat{\lambda}_{n, K}$ has the candidate set which is, respectively, a polynomial and exponentially growing grid.

Despite these advances, distributional approximations for CV-based penalized estimators remain largely unexplored. The existing oracle inequalities and risk bounds of $\hat{\bm{\beta}}_n(\hat{\lambda}_{n, K})$ are not useful for constructing confidence intervals or performing hypothesis tests on $\bm{\beta}$. In contrast, here, we explore the distributional properties of the CV-selected Lasso estimator. Thus, our results can be used to draw valid statistical inferences on $\bm{\beta}$. Moreover, we do not place any restriction on the candidate set of $\hat{\lambda}_{n,K}$, unlike some of the existing works on CV based Lasso. Our analysis is also non-standard in nature, since the data-dependent choice of penalty introduces an additional layer of randomness that cannot be handled using existing asymptotic distribution theory in Lasso. 

\subsection{\bf Organization of the paper}
The remainder of this paper is organized as follows. The regularity conditions are presented in Section \ref{sec:assum}. The main results on the failure of VSC of the Lasso estimator $\hat{\bm{\beta}}_n(\lambda)$ when $\lambda=\hat{\lambda}_{n,K}$, are provided in Section \ref{sec:crossv}. The main results on the $n^{1/2}-$consistency of $\hat{\lambda}_{n,K}$ and of $\hat{\bm{\beta}}_n(\hat{\lambda}_{n,K})$ are presented in Section \ref{sec:crossvconv}. The result on validity of Bootstrap approximation of the distribution of $\hat{\bm{\beta}}_n(\hat{\lambda}_{n,K})$ is addressed in Section \ref{sec:bootstrap}. A moderate simulation study on the Bootstrap approximation is provided in Section \ref{sec:simstudy}. Analysis of a real data set based on our proposed Bootstrap method is performed in Section \ref{sec:realdat}. Proofs of all the requisite lemmas and that of main results viz. Theorem \ref{prop:cv}, Theorem \ref{thm:cvthm}, Proposition \ref{prop:eqicon}, Proposition \ref{prop:C.4}, Theorem \ref{thm: lambdaCVconv}, Theorem \ref{thm:distconvcv} and Theorem \ref{thm:bootconsistency} have been addressed in section \ref{sec:proof2}. 

\section{Preliminaries and regularity conditions}\label{sec:assum}

In this section for completeness, we define all the major quantities of this paper in Table \ref{tab:unified}. Additional quantities required for Bootstrap approximation are defined in section \ref{sec:bootstrap}. 

\begin{table}[H]
\centering
\caption{Notations}
\label{tab:unified}

\begin{tabularx}{\linewidth}{l L}
\hline
Notation  & Description \\ \hline

\ensuremath{H_{n,K}(\lambda)}
& $:=\frac{1}{2}\sum_{k=1}^{K}\sum_{i \in I_k}\Big[y_i - \bm{x_i}^\top \hat{\bm{\beta}}_{n,-k}(\lambda)\Big]^2$ \\

\ensuremath{\hat{\lambda}_{n,K}
}
& $\in \operatorname*{Argmin}_{\lambda\ge0} H_{n,K}(\lambda)$ \\

\ensuremath{\hat{H}_{n, K}^\prime(\lambda)} & $: = H_{n, K}(n^{1/2}\lambda)$\\

\ensuremath{n^{-1/2}\hat{\lambda}_{n, K}} & $\in \operatorname*{Argmin}_{\lambda}\hat{H}_{n, K}^\prime(\lambda)$\\

\ensuremath{\bm{W}_n = n^{-1/2}\sum_{i=1}^n \varepsilon_i \bm{x}_i}
& Random linear term \\

\ensuremath{\bm{W}_{n,k}=m^{-1/2}\sum_{i\in I_k}\varepsilon_i\bm{x}_i} & Group level of $\bm{W}_n$ for $I_k$\\

\ensuremath{\bm{W}_{n,-k}=(n-m)^{-1/2}\sum_{i\notin I_k}\varepsilon_i\bm{x}_i} & Group level of $\bm{W}_n$ except $I_k$\\

\ensuremath{\bm{W}_{\infty},\bm{W}_{\infty,k},\bm{W}_{\infty,-k}}
& Dependent copies of $N(\bm{0},\bm{S})$\\

\ensuremath{\bm{S}_n=n^{-1}\sum_{i=1}^n\bm{x}_i\bm{x}_i^\top\mathbf{E}(\varepsilon_i^2)} & Variance of $\bm{W}_n$\\

\ensuremath{\bm{S}_{n, k}=m^{-1}\sum_{i\in I_k}\bm{x}_i\bm{x}^\top_i\mathbf{E}(\varepsilon_i^2)} & Group level of $\bm{S}_n$ for $I_k$\\

\ensuremath{\bm{S}_{n, -k}=(n-m)^{-1}\sum_{i\notin I_k}\bm{x}_i\bm{x}^\top_i\mathbf{E}(\varepsilon_i^2)} & Group level of $\bm{S}_n$ except $I_k$\\

\ensuremath{\bm{L}_n=n^{-1}\sum_{i=1}^n\bm{x}_i\bm{x}_i^\top} & Gram matrix\\

\ensuremath{\bm{L}_{n, k}=m^{-1}\sum_{i\in I_k}\bm{x}_i\bm{x}^\top_i} & Group level of $\bm{L}_n$ for $I_k$\\ 

\ensuremath{\bm{L}_{n, -k}=(n-m)^{-1}\sum_{i\notin I_k}\bm{x}_i\bm{x}^\top_i} & Group level of $\bm{L}_n$ except $I_k$\\

\ensuremath{\hat{V}_{\infty,-k}(\bm{u},\lambda)} & $:=(1/2)\bm{u}^\top\bm{L}\bm{u} -\bm{u}^\top \bm{W}_{\infty,-k} + \lambda\Big\{\sum_{j=1}^{p_0}u_jsgn({{\beta}_{j}})+\sum_{j=p_0+1}^{p}|u_j|\Big\}$\\

\ensuremath{\hat{\bm{u}}_{\infty, - k}(\lambda)} & $ := \operatorname*{Argmin}_{\bm{u}} \hat{V}_{\infty,-k}(\bm{u},\lambda)$\\

\ensuremath{\hat{H}_{\infty, K}^\prime(\lambda)} & $:=\sum_{k=1}^{K}\Big[2^{-1}\big(\hat{\bm{u}}_{\infty, -k}(\lambda)\big)^{\top} \bm{L}\big(\hat{\bm{u}}_{\infty, -k}(\lambda)\big)-\big(\hat{\bm{u}}_{\infty, -k}(\lambda)\big)^{\top} \bm{W}_{\infty, k}\Big]$\\

\ensuremath{\hat{\Lambda}_{\infty,K} } & $\in  \operatorname*{Argmin}_{\lambda}\hat{H}_{\infty, K}^\prime(\lambda)$\\
\hline
\end{tabularx}
\end{table}
Suppose that $(\Omega,\mathcal{F},\mathbf{P})$ is the underlying probability space. Without loss of generality, assume that the set of relevant covariates is $\mathcal{A}=\{j: \beta_{j}\neq 0\}=\{1,\ldots,p_0\}$, where $p_0\leq p$. Now we list the regularity conditions on the design vectors and the regression errors:
\begin{enumerate}[label = (C.\arabic*)]
    \item $\bm{S}_{n, -k},\bm{S}_{n, k} \rightarrow \bm{S}$ and $\bm{L}_{n, -k},\bm{L}_{n, k} \rightarrow \bm{L}$ as $n\to \infty$, for all $k\in \{1,..,K\}$, where $\bm{S,L}$ are positive definite matrices.
    \item $n^{-1}\sum_{i=1}^{n}\mathbf{E}|\varepsilon_i|^{6}=O(1).$
    \item $n^{-1}\sum_{i=1}^n\|\bm{x}_i\|^6=O(1).$ 

\end{enumerate}

Condition (C.1) is natural to assume and in the homoscedastic case simply means that the group levels of the design vectors do not matter asymptotically. For example, if the design vectors are independent observations from some distribution, condition (C.1) holds. Condition (C.1) is primarily needed to obtain upper and lower bounds on the leave $k$ out Lasso objective functions uniformly over $k$. Conditions (C.2) and (C.3) are moment conditions, respectively, on the design vectors and the regression errors. These conditions are required to establish distribution convergence of the sequences $\{\bm{W}_{n,-k}\}_{n\geq 1}$ and $\{\bm{W}_{n,k}\}_{n\geq 1}$ and their Bootstrap counterparts, for all $k\in\{1,...,K\}$. 

\section{Failure of VSC of $\hat{\bm{\beta}}_n(\hat{\lambda}_{n,K}$)}\label{sec:crossv}
One of the important properties of the Lasso estimator is the variable selection consistency or VSC. However, whether Lasso achieves VSC or not crucially depends on the asymptotic nature of the underlying penalty sequence $\{\lambda_n\}_{n\geq 1}$.
\citet{zhao2006model} were the first to establish that $\hat{\bm{\beta}}_n(\lambda_n)$ attains VSC whenever $\lambda_n$ grows faster than $n^{1/2}$ and the design matrix has some irrepresentable type conditions. 
Recently, \citet{lahiri2021necessary} performed a more nuanced asymptotic analysis of Lasso and showed that the divergence of the sequence $\{n^{-1/2}\lambda_n\}_{n\geq 1}$ is also necessary for $\hat{\bm{\beta}}_n(\lambda_n)$ to achieve VSC. Therefore, to find out whether $\hat{\bm{\beta}}_n(\hat{\lambda}_{n, K})$ achieves VSC, it is essential to study whether the sequence $\{n^{-1/2}\hat{\lambda}_{n, K}\}_{n\geq 1}$ is bounded or not, which we do in this section. However, the analysis of the sequence $\{n^{-1/2}\hat{\lambda}_{n,K}\}_{n\geq 1}$ intrinsically depends on the consistency of $\hat{\bm{\beta}}_{n}(\hat{\lambda}_{n, K})$, for which convergence of the sequence $\{n^{-1}\hat{\lambda}_{n, K}\}_{n\geq 1}$ is important, due to the results of \citet{knight2000asymptotics}. This is what we explore as the first result of this section.

\begin{theorem}\label{prop:cv}
Suppose that the regularity conditions (C.1),(C.2) and (C.3) are true. Then we have $$\mathbf{P}\Big(n^{-1}\hat{\lambda}_{n,K} \rightarrow 0\ \text{as}\ n\ \to\ \infty \Big)= 1.$$     
\end{theorem}
The proof of Theorem \ref{prop:cv} is presented in section \ref{sec:prop3.1}. To prove this result, we basically show that the $K-$fold CV objective function $H_{n, K}(\lambda_n)$, defined in Table \ref{tab:unified}, is smaller when $\lambda_n = o(n)$ compared to other possibilities. Now note that \citet{knight2000asymptotics} established consistency of $\hat{\bm{\beta}}_n(\lambda_n)$ whenever $\lambda_n = o(n)$. Therefore, Theorem \ref{prop:cv} essentially implies that the Lasso estimator with the $K-$fold CV based penalty is consistent for $\bm{\beta}$. Therefore, the Lasso estimator satisfies one of the desired properties of an estimator even with a data based choice of penalty. We now explore the asymptotic nature of the sequence $\{n^{-1/2}\hat{\lambda}_{n, K}\}_{n\geq 1}$, essential to study VSC as well as $n^{1/2}-$consistency of $\hat{\bm{\beta}}_{n}(\hat{\lambda}_{n, K})$, in the next theorem. 
\begin{theorem}\label{thm:cvthm}
Suppose that assumptions (C.1),(C.2) and (C.3) are true. Then we have $$\mathbf{P}\Big(\limsup_{n\rightarrow \infty}n^{-1/2}\hat{\lambda}_{n,K}< \infty \Big)=1.$$
\end{theorem}
The proof of Theorem \ref{thm:cvthm} is provided in section \ref{sec:thm3.1}. The proof critically utilizes the fact that the $K-$fold CV objective function $H_{n, K}(\lambda_n)$, defined in Table \ref{tab:unified}, behaves differently depending on whether the sequence $\{n^{-1/2}\lambda_n\}_{n\geq 1}$ is bounded or not. In particular, we show that $\hat{\bm{\beta}}_n(\lambda_n)$ belongs to different sets based on properties of $\{n^{-1/2}\lambda_n\}_{n\geq 1}$. The set where $\hat{\bm{\beta}}_n(\lambda_n)$ lies under boundedness of $\{n^{-1/2}\lambda_n\}_{n\geq 1}$ is more favorable for $H_{n, K}(\lambda_n)$ to achieve smaller values. 
Theorem \ref{thm:cvthm} establishes that $\{n^{-1/2}\hat{\lambda}_{n,K}\}_{n\geq 1}$ is almost surely bounded. On the other hand, \citet{lahiri2021necessary} showed that the sequence $\{n^{-1/2}\lambda_n\}_{n\geq 1}$ must diverge for $\hat{\bm{\beta}}_n(\lambda_n)$ to achieve VSC. Therefore, Theorem \ref{thm:cvthm} implies that the Lasso estimator with the $K-$fold CV based penalty is not variable selection consistent, although it is consistent. In the next section, we explore whether $\hat{\bm{\beta}}_n(\hat{\lambda}_{n, K})$ is $n^{1/2}-$consistent. 

\section{$n^{1/2}$-Consistency of $\hat{\bm{\beta}}_n(\hat{\lambda}_{n,K})$}\label{sec:crossvconv}
The most important property of an estimator from the point of view of statistical inference is probably the $n^{1/2}-$consistency, since then the confidence region can be constructed and tests can be performed on the unknown parameter based on the asymptotic distribution. In this section, we study the $n^{1/2}-$consistency of the $K-$fold CV based Lasso estimator $\hat{\bm{\beta}}_n(\hat{\lambda}_{n,K})$. Towards that, let us again recall the asymptotic analysis of \citet{knight2000asymptotics}, who showed that $\hat{\bm{\beta}}_{n}(\lambda_n)$ is $n^{1/2}-$consistent provided $\{n^{-1/2}\lambda_n\}_{n\geq 1}$ converges. In the same spirit, here we need to first explore the convergence in distribution of the random sequence $\{n^{-1/2}\hat{\lambda}_{n, K}\}_{n\geq 1}$. Note that 
\begin{align*}
    n^{-1/2}\hat{\lambda}_{n, K} \in \operatorname*{Argmin}_{\lambda}\hat{H}_{n, K}^\prime(\lambda),
\end{align*}
where $\hat{H}_{n, K}^\prime(\lambda) = H_{n, K}(n^{1/2}\lambda)$ with $H_{n, K}(\lambda)$ being defined in Table \ref{tab:unified}. Note that a natural candidate for the weak limit of $H^\prime_{n, K}(\lambda)$ is $\hat{H}_{\infty, K}^\prime(\lambda)$, defined in Table \ref{tab:unified}. Thus, we expect that $n^{-1/2}\hat{\lambda}_{n, K}$ converges in distribution to $\hat{\Lambda}_{\infty,K}$. To prove this convergence, we use Dudley's almost sure representation theorem and hence we essentially establish the probability convergence of a version of $n^{-1/2}\hat{\lambda}_{n, K}$ to a version of $\hat{\Lambda}_{\infty, K}$. To keep the discussion simple, assume that we would like to establish the probability convergence of $n^{-1/2}\hat{\lambda}_{n, K}$ to $\hat{\Lambda}_{\infty, K}$. To that end, let us consider the following inequality (cf. \citet{van1996weak}) which represents the closeness of the $Argmin$s in terms of the closeness of the respective objective functions: For any $\varepsilon > 0$,
\begin{align}\label{eqn:10002}
  &\mathbf{P}\Big(|n^{-1/2}\hat{\lambda}_{n, K} - \hat{\Lambda}_{\infty, K}| > \varepsilon\Big)\nonumber\\ \leq &\;\mathbf{P}\Big(2\sup_{\lambda}\Big|\hat{H}_{n, K}^\prime(\lambda) - \hat{H}_{\infty, K}^\prime(\lambda)\Big|  > \inf_{|\lambda - \hat{\Lambda}_{\infty,K}| > \varepsilon} \hat{H}_{\infty, K}^\prime(\lambda) - \hat{H}_{\infty, K}^\prime(\hat{\Lambda}_{\infty, K}) \Big),
\end{align}
where all the quantities can be found in Table \ref{tab:unified}. Moreover, we have the stochastic boundedness of the sequence $\{n^{-1/2}\hat{\lambda}_{n, K}\}_{n\geq 1}$  due to Theorem \ref{thm:cvthm}. 
Therefore, to establish the probability convergence of $n^{-1/2}\hat{\lambda}_{n, K}$ to $\hat{\Lambda}_{\infty, K}$ from (\ref{eqn:10002}), we essentially require to have\\
\emph{(I)} the uniform convergence of $\{\hat{H}_{n, K}^\prime(\lambda)\}_{n\geq 1}$ to $\hat{H}_{\infty, K}^\prime(\lambda)$ over compact sets, in probability, and\\
\emph{(II)} $\big[\inf_{|\lambda - \hat{\Lambda}_{\infty,K}| > \varepsilon} \hat{H}_{\infty, K}^\prime(\lambda) - \hat{H}_{\infty, K}^\prime(\hat{\Lambda}_{\infty, K})\big] > 0$, i.e., the minimizer $\hat{\Lambda}_{\infty, K}$ of $\hat{H}_{\infty, K}^\prime(\cdot)$ is unique and well-separated, almost surely.\\

Note that \emph{(I)} is implied by the stochastic equicontinuity of $\{\hat{H}_{n, K}^\prime(\cdot)\}_{n\geq 1}$, over compact sets, due to Theorem 2.1 of \citet{newey1991uniform}. Again the stochastic equicontinuity of $\{\hat{H}_{n, K}^\prime(\cdot)\}_{n\geq 1}$ follows from the stochastic equicontinuity of the lasso solution path $\{\hat{\bm{u}}_{n,-k}(\cdot)\}_{n\geq 1} := \{n^{1/2}(\hat{\bm{\beta}}_{n, -k}(n^{1/2}\lambda) - \bm{\beta})\}_{n\geq 1}$, for all $k \in \{1,\dots, K\}$. Recall the definition of $\hat{H}_{n,K}^\prime(\lambda)$ from Table \ref{tab:unified} and note that for any fixed $\lambda>0$ with $|\lambda^\prime-\lambda|<\delta$, we have for large enough $n$,
\begin{align}\label{eqn:mkop}
\big{|}\hat{H}_{n,K}^\prime(\lambda^\prime)-\hat{H}_{n,K}^\prime(\lambda)\big{|}&\leq \tilde{\gamma_1}\sum_{k=1}^{K}\Big{\|}\hat{\bm{u}}_{n,-k}(\lambda^\prime)-\hat{\bm{u}}_{n,-k}(\lambda)\Big{\|}\Bigg[\Big{\|}\hat{\bm{u}}_{n,-k}(\lambda^\prime)-\hat{\bm{u}}_{n,-k}(\lambda)\Big{\|}\nonumber\\
&\;\;\;\;\;\;\;\;\;\;\;\;\;\;\;\;\;\;+2\big{\|}\hat{\bm{u}}_{n,-k}(\lambda)\big{\|}+\tilde{\gamma_1}^{-1}\big{\|}\bm{W}_{n,k}\big{\|}\Bigg].
\end{align}
Here $\tilde \gamma_1$ is the largest eigen value of $L$. Thus, the stochastic equicontinuity of $\{\hat{H}_{n,K}^\prime(\lambda)\}_{n\geq 1}$ follows from that of $\{\hat{\bm{u}}_{n,-k}(\lambda)\}_{n\geq 1}$, for all $k \in \{1,\dots, K\}$, along with tightness of $\{\bm{W}_{n,k}\}_{n\geq 1}$ and of $\{\hat{\bm{u}}_{n,-k}(\lambda)\}_{n\geq 1}$. Hence it essentially remains to establish the stochastic equicontinuity of $\{\hat{\bm{u}}_n(\cdot)\}_{n\geq 1}$ in order to claim \emph{(I)}. We establish it in the next proposition.
\begin{proposition}\label{prop:eqicon}
Assume that the regularity conditions (C.1), (C.2) and (C.3) hold. Then $ \{\hat{\bm{u}}_n(\lambda)\}_{n\geq 1}$ is stochastically equicontinuous in $\lambda$. 
\end{proposition}
The proof of Proposition \ref{prop:eqicon} is presented in section \ref{sec:prop4.1}. The stochastic equicontinuity of the Lasso solution path across all sample sizes, i.e., of $\{\hat{\bm{\beta}}_n(\lambda)\}_{n\geq 1}$ follows analogously to Proposition \ref{prop:eqicon}. See Remark \ref{remark:steqiconoriginalbeta} for details. Now we explore whether \emph{(II)} is also true. Note that \emph{(II)} asserts two properties of the minimizer of the limiting CV prediction error
$\hat H_{\infty,K}'(\cdot)$ in almost sure sense: uniqueness and well separatedness of the unique minimizer. One may easily analyze these two properties of any function if the analytical form is available. 
However, an analytical form of $\hat H_{\infty,K}'(\lambda)$ is generally not available unless the Lasso estimator has
a closed-form expression which is not there except when the design matrix has a simple structure, like orthogonality. We circumvent this issue by visualizing the Lasso optimization problem as a quadratic programming problem and then analyzing the KKT conditions related to it. Based on this analysis, we obtain an implicit form of $\hat H_{\infty,K}'(\lambda)$ in terms of $\lambda$, which we analyze to claim \emph{(II)}. 
We summarize the result as the following proposition.
\begin{proposition}\label{prop:C.4}
Suppose that the model (\ref{eqn:linearreg}) is not null, i.e., $\mathcal{A} = \{j: \beta_j \neq 0\}$ is non-empty. Also assume that the regularity conditions (C.1)-(C.3) are true. Then \emph{(II)} holds.  \end{proposition}
The detailed proof of Proposition~\ref{prop:C.4} is presented in
Section~\ref{sec:appBprop3.3}. The proof proceeds in two main steps.
First, using an equivalent constrained formulation of the Lasso problem
together with its KKT conditions, we show that, for each validation fold,
the limiting estimator
$\hat{\bm u}_{\infty,-k}(\lambda)$ is a piecewise affine function of
$\lambda$. Consequently,
$\hat H_{\infty,K}'(\lambda)$ is a piecewise quadratic function whose
break points are determined by changes in the active constraint set.
Moreover, the locations of these break points and the coefficients of the
quadratic pieces are measurable functions of the Gaussian vector
$\bigl(\bm W_{\infty,-1}^{\top},\ldots,\bm W_{\infty,-K}^{\top}\bigr)^{\top}$.
Each quadratic piece is shown to be strictly convex and therefore possesses
a unique local minimizer. In the second step, we prove that the equality of the minimum values
corresponding to any two distinct local minimizers is equivalent to the
vanishing of a non-trivial polynomial in
$\bigl(\bm W_{\infty,-1}^{\top},\ldots,\bm W_{\infty,-K}^{\top}\bigr)^{\top}$.
Since this Gaussian random vector has an absolutely continuous
distribution, such an event occurs with probability zero. Therefore,
$\hat H_{\infty,K}'(\cdot)$ admits a unique global minimizer
$\hat\Lambda_{\infty,K}$ almost surely. Finally, the strict convexity of
each quadratic piece implies that every local minimizer is well separated,
which yields the desired well-separatedness of the unique global minimizer.
Based on the notion of perfect maps, Proposition \ref{eqn:10002} and Proposition \ref{prop:C.4} can be established (and hence \emph{(I)} and \emph{(II)} can be concluded) for the versions in the new probability space and then we can utilize them to establish the probability convergence of a version of $n^{-1/2}\hat{\lambda}_{n, K}$ to a version of $\hat{\Lambda}_{\infty, K}$. This will imply the distribution convergence of $n^{-1/2}\hat{\lambda}_{n, K}$ to $\hat{\Lambda}_{\infty,K}$. We state the result as the following theorem. 
\begin{theorem}\label{thm: lambdaCVconv}
Suppose that the regularity conditions (C.1)-(C.3) hold. 
Then we have $$n^{-1/2}\hat{\lambda}_{n, K} \xrightarrow{d} \hat{\Lambda}_{\infty,K}.$$
\end{theorem}


The proof of Theorem \ref{thm: lambdaCVconv} is presented in section \ref{sec:thm4.1}. Given the $n^{1/2}-$consistency of the $K-$the fold CV based penalty $\hat{\lambda}_{n, K}$, we are in a position to explore the $n^{1/2}-$consistency of $\hat{\bm{\beta}}_n(\hat{\lambda}_{n, K})$ based on the $Argmin$ theorem of convex stochastic processes. Note that $n^{1/2}(\hat{\bm{\beta}}_{n}(\hat{\lambda}_{n,K})-\bm{\beta}) = \operatorname*{Argmin}_{\bm{u}}V_n(\bm{u}, \hat{\lambda}_{n, K})$ where 
\begin{align}\label{eqn:mmmmm}
V_n(\bm{u}, \hat{\lambda}_{n, K}) &= \bigg\{\Big[(1/2)\bm{u}^\top\bm{L}_n\bm{u} -\bm{u}^\top \bm{W}_{n}\Big] + \Big[\hat{\lambda}_{n,K}\sum_{j=1}^{p}\Big(|\beta_{j}+\dfrac{u_{j}}{n^{1/2}}|-|\beta_{j}|\Big)\Big]\bigg\}.
\end{align}
Clearly, if $\{j: \beta_j \neq 0\} = \{1,\dots, p_0\}$ is the set of indices of the relevant covariates, then it is natural to expect that the weak limit of $V_n(\bm{u}, \hat{\lambda}_{n, K})$ is $V_{\infty}(\bm{u},\hat{\Lambda}_{\infty,K})$, where 
\begin{align}\label{eqn:nn}
V_{\infty}(\bm{u},\hat{\Lambda}_{\infty,K})=(1/2)\bm{u}^\top\bm{L}\bm{u} -\bm{u}^\top \bm{W}_{\infty} + \hat{\Lambda}_{\infty,K}\bigg\{\sum_{j=1}^{p_0}u_jsgn({{\beta}_{j}})+\sum_{j=p_0+1}^{p}|u_j|\bigg\}.
\end{align}
The definitions of the quantities $\bm{L}_n$, $\bm{W}_n$, $\bm{L}$, $\bm{W}_{\infty}$ can be found in Table \ref{tab:unified}. To that end, we have the following theorem on the $n^{1/2}-$consistency of $n^{1/2}(\hat{\bm{\beta}}_{n}(\hat{\lambda}_{n,K})-\bm{\beta})$.
\begin{theorem}\label{thm:distconvcv}
Suppose that the regularity conditions (C.1)-(C.3) hold. Then we have   
$$n^{1/2}(\hat{\bm{\beta}}_{n}(\hat{\lambda}_{n,K})-\bm{\beta})\xrightarrow{d}\operatorname*{argmin}_{\bm{u}}V_{\infty}(\bm{u},\hat{\Lambda}_{\infty,K}).$$
\end{theorem}

The proof of Theorem \ref{thm:distconvcv} is presented in section \ref{sec:thm4.2} and has the following main steps. 
\begin{enumerate}[label=(\alph*)]
\item Establishing finite dimensional convergence of $V_n(\cdot, \hat{\lambda}_{n,K})$ to $V_{\infty}(\cdot,\hat{\Lambda}_{\infty,K})$.
\item Showing the stochastic equicontinuity of $\{V_n(\cdot, \hat{\lambda}_{n,K})\}_{n\geq 1}$ over compact sets.
\item Employing the $Argmin$ theorem of \citet{choudhury2024bootstrapping}.
\end{enumerate}
The first two steps essentially imply the convergence in distribution of the process $V_n(\cdot, \hat{\lambda}_{n,K})$ to the process $V_{\infty}(\cdot,\hat{\Lambda}_{\infty,K})$ and the third step concludes the proof by translating this distribution convergence to that of $n^{1/2}(\hat{\bm{\beta}}_{n}(\hat{\lambda}_{n,K})-\bm{\beta})$. Therefore, we can give a definitive answer to the following question which we raised in Section \ref{sec:intro}:
\begin{center}
  \textit{ Does $K-$fold CV based Lasso perform variable selection or is it $n^{1/2}-$consistent?}
\end{center}
The answer is that the Lasso estimator with $K-$fold CV based penalty is $n^{1/2}-$consistent, but does not achieve variable selection consistency.

\begin{remark}\label{remark:steqiconoriginalbeta}
The Stochastic equicontinuity of the solution path of the Lasso estimator $\{\hat{\bm{\beta}}_n(\lambda)\}_{n\geq 1}$ in terms of penalty $\lambda$ can be established analogously to Proposition \ref{prop:eqicon}. Note that the centered Lasso estimator at penalty $\lambda$ is given by $\hat{\bm{v}}_n(\lambda)=(\hat{\bm{\beta}}_{n}(\lambda)-\bm{\beta})=\mbox{Argmin}_{\bm{u}} \hat{U}_n(\bm{u},\lambda)$,
where,
$$\hat{U}_{n}(\bm{u},\lambda)=(1/2)\bm{u}^\top\big[\sum_{i=1}^{n}\bm{x}_i\bm{x}_i^\top\big]\bm{u} -\bm{u}^\top \big[\sum_{i=1}^{n}\varepsilon_i\bm{x}_i\big] +\lambda\Big\{\sum_{j=1}^{p}\big(|u_j+\beta_j|-|\beta_j|\big)\Big\}.$$
Now for any two $\lambda,\lambda^\prime$ we have;
\begin{align*}
 & \Big[\hat{U}_{n}(\hat{\bm{v}}_{n}(\lambda^\prime),\lambda)-\hat{U}_{n}(\hat{\bm{v}}_{n}(\lambda^\prime),\lambda^\prime)\Big]-\Big[\hat{U}_{n}(\hat{\bm{v}}_{n}(\lambda),\lambda)-\hat{U}_{n}(\hat{\bm{v}}_{n}(\lambda),\lambda^\prime)\Big]\nonumber\\
 &=(\lambda-\lambda^\prime)\Bigg[\Big\|\hat{\bm{v}}_{n}(\lambda^\prime)+\bm{\beta}\Big\|_1-\Big\|\hat{\bm{v}}_{n}(\lambda)+\bm{\beta}\Big\|_1\Bigg]\leq p^{1/2}|\lambda-\lambda^\prime|\; \|\hat{\bm{v}}_{n}(\lambda)-\hat{\bm{v}}_{n}(\lambda^\prime)\|.
\end{align*}
Under the existing assumption, $\hat{U}_{n}(\bm{u},\lambda)$ is strongly convex in $\bm{u}$ for fixed $\lambda$. Denote the minimum eigen value of $X^\top X$ by $\xi_0$. Then strong convexity will imply that,
\begin{align*}
\frac{\xi_0}{8}||\hat{\bm{v}}_{n}(\lambda)-\hat{\bm{v}}_{n}(\lambda^\prime)||^2&\le \Big[\hat{U}_{n}(\hat{\bm{v}}_{n}(\lambda^\prime),\lambda)-\hat{U}_{n}(\hat{\bm{v}}_{n}(\lambda^\prime),\lambda^\prime)\Big]-\Big[\hat{U}_{n}(\hat{\bm{v}}_{n}(\lambda),\lambda)-\hat{U}_{n}(\hat{\bm{v}}_{n}(\lambda),\lambda^\prime)\Big]\\
&\le p^{1/2}|\lambda-\lambda^\prime|\; \|\hat{\bm{v}}_{n}(\lambda)-\hat{\bm{v}}_{n}(\lambda^\prime)\|.
\end{align*}
Now, for any $\varepsilon>0$, choosing $\delta:=\frac{\varepsilon \xi_0}{16p^{1/2}}$, the above calculations
 imply the stochastic equicontinuity of $\{\hat{\bm{v}}_n(\lambda)\}_{n\ge 1}$
and hence that of $\{\hat{\bm{\beta}}_n(\lambda)\}_{n\geq 1}$.
\end{remark}

\section{Bootstrap approximation}\label{sec:bootstrap}
We have established $n^{1/2}$-consistency of the Lasso estimator when the penalty is chosen based on $K$-fold CV. However, $n^{1/2}-$consistency is only useful for drawing valid inferences if the asymptotic distribution is tractable, which is clearly not the case here. Therefore, it would be important to explore an alternative distributional approximation method, such as Bootstrap, to capture the distribution of $\hat{\bm{\beta}}_n(\hat{\lambda}_{n, K})$. We follow the construction of the Bootstrap method of \citet{das2019distributional}, who showed the validity of Perturbation Bootstrap in Lasso with non-random penalty. Perturbation Bootstrap (hereafter referred to as PB) is generally defined based on a collection of random weights $G_1^*,\ldots, G_n^*$ which are basically a collection of independent copies of a non-negative and non-degenerate random variable $G^*$. $G^*$ is independent of the data generation process and has the property that the mean of $G^*$ is $\mu_{G^*}$, $Var(G^*)=\mu_{G^*}^2$  and $\mathbf{E}(G_1^{*3})< \infty$. Subsequently, for each $i\in\{1,..,n\}$, define the $i$th Bootstrap observation as $(z_i, \bm{x}_i^\top)$ where $$z_i=\bm{x}_i^\top\tilde{\bm{\beta}}_n(\hat{\lambda}_{n, K})+(y_i-\bm{x}_i^\top\tilde{\bm{\beta}}_n(\hat{\lambda}_{n, K}))(G_i^*-\mu_{G^*})\mu_{G^*}^{-1}.$$ Here $\tilde{\bm{\beta}}_n(\hat{\lambda}_{n,K})=\Big(\tilde{\beta}_{n,1}(\hat{\lambda}_{n,K}),...,\tilde{\beta}_{n,p}(\hat{\lambda}_{n,K})\Big)^\top$ is the thresholded Lasso estimator with $j$th component $\tilde{\beta}_{n,j}(\hat{\lambda}_{n,K})=\hat{\beta}_{n,j}(\hat{\lambda}_{n,K})\mathbbm{1}\Big\{\Big{|}\hat{\beta}_{n,j}(\hat{\lambda}_{n,K})\Big{|}>a_n\Big\}$. The sequence $\{a_n\}_{n\geq 1}$ is a sequence of positive constants such that $a_n+(n^{-1/2}\log n)a_n^{-1} \rightarrow 0$ as $n\rightarrow \infty$. Then the PB-Lasso estimator at penalty term $\lambda_n^*>0$ is defined as
\begin{align}\label{eqn:defpblasso}
\hat{\bm{\beta}}_n^*(\lambda_n^*)=\mbox{argmin}_{\bm{\beta}}\Bigg[\frac{1}{2}\sum_{i=1}^{n}\Big(z_{i} - \bm{x}_{i}^\top\bm{\beta}\Big)^2 +\lambda_n^*\sum_{j=1}^{p}|\beta_{j}|\Bigg].
\end{align}
Subsequently, the Bootstrap pivotal quantity is $\bm{T}_{n}^*(\lambda_n^*) = n^{1/2}\big[\hat{\bm{\beta}}_{n}^*\big(\lambda_n^*\big)-\tilde{\bm{\beta}}_n(\hat{\lambda}_{n,K})\big].$ The reason for using the thresholded Lasso estimator $\tilde{\bm{\beta}}_n(\hat{\lambda}_{n,K})$ instead of the original Lasso estimator $\hat{\bm{\beta}}_n(\hat{\lambda}_{n,K})$ to define the Bootstrap observations and to center the pivotal quantity is that $\hat{\bm{\beta}}_n(\hat{\lambda}_{n,K})$ is not VSC, as observed in Section \ref{sec:crossv}. Without VSC, Lasso may not capture the true signs of the regression coefficients and hence limits of $l_1$ penalty part of the objective function in the original and Bootstrap regime may not be the same. Thus, Bootstrap approximation may fail in Lasso without such thresholding. This phenomenon was first observed by \citet{chatterjee2011bootstrapping}.

Now obviously an important step is to identify a suitable choice of the penalty term $\lambda_n^*$ based on the data. Essentially, we need to establish a version of Theorem \ref{thm:distconvcv} for $\bm{T}_n^*(\lambda_n^*)$ with the same limit. Thus, it is natural to define $\lambda_n^*$ as the $K-$fold CV based penalty based on Bootstrap observations $(z_i, \bm{x}_i^\top)_{i=1}^{n}$, hereafter denoted by $\hat{\lambda}_{n,K}^*$. 
To that end, define the PB-Lasso estimator for all observations except those are in $I_k$ (defined in Section \ref{sec:intro} for the original K-fold CV based penalty) at some penalty $\lambda^* > 0$ as
\begin{align}\label{eqn:900}
\hat{\bm{\beta}}_{n,-k}^*(\lambda^*)=\mbox{argmin}_{\bm{\beta}}\Bigg[\frac{1}{2}\sum_{i\notin I_k}\Big(z_{i} - \bm{x}_{i}^\top\bm{\beta}\Big)^2+\lambda^*\sum_{j=1}^{p}|\beta_{j}|\Bigg]   . 
\end{align}
Then the Bootstrapped cross-validation error at penalty $\lambda^*>0$ is defined as
\begin{align}\label{eqn:901}
 H_{n,K}^*(\lambda^*) =\frac{1}{2}\sum_{k=1}^{K}\sum_{i \in I_k}\Big[y_i - \bm{x_i}^\top \hat{\bm{\beta}}_{n,-k}^*(\lambda^*)\Big]^2  . 
\end{align}
Finally, the Bootstrapped $K$-fold CV estimator $\hat{\lambda}_{n,K}^*$ is defined as
\begin{align}\label{eqn:902}
 \hat{\lambda}_{n,K}^*\in \mbox{argmin}_{\lambda^*\ge 0} H_{n,K}^*(\lambda^*)  .
\end{align}
To state the result on Bootstrap, let $\mathcal{B}(\mathbb{R}^p)$ be the Borel sigma-field defined on $\mathbb{R}^p$ and $\rho(\cdot,\cdot)$ be the Prokhorov metric defined on the space of all probability measures on $\big(\mathbb{R}^p,\mathcal{B}(\mathbb{R}^p)\big)$.
 Suppose that $F_n$ is the distribution of $\bm{T}_{n}(\hat{\lambda}_{n, K})$ and $\hat{F}_n$ is the conditional distribution of $\bm{T}_n^*(\hat{\lambda}_{n,K}^*)$ given the regression errors $\{\varepsilon_1,\dots, \varepsilon_n\}$. Now we are ready to state the result on the validity of our proposed Bootstrap approximation.
\begin{theorem}\label{thm:bootconsistency}
 Suppose that the regularity conditions (C.1)-(C.3) are true. Then we have
$$\rho\big(\hat{F}_n, F_n\big) \rightarrow 0, \; \text{almost surely},\; \text{as}\; n \rightarrow \infty.$$
\end{theorem} 
The proof of Theorem \ref{thm:bootconsistency} is presented in section \ref{sec:thm5.1}. To prove the above theorem, we essentially show that the conditional distribution of $\bm{T}_n^*(\hat{\lambda}_{n,K}^*)$ given $\{\varepsilon_1,\dots, \varepsilon_n\}$ converges to the distribution of $\operatorname*{argmin}_{\bm{u}}V_{\infty}(\bm{u},\hat{\Lambda}_{\infty,K}),$ almost surely, which essentially follows based on the same line of arguments as in case of Theorem \ref{thm:distconvcv}. An important implication of Theorem \ref{thm:bootconsistency} is that unlike all the existing literature, the original $K-fold$ CV-based penalty obtained based on the observed data should not be used at the Bootstrap stage. A fresh calculation of the $K-$fold CV is required based on the Bootstrap observations. Moreover, Theorem~\ref{thm:bootconsistency} provides a principled framework for conducting valid inference on the true parameter $\bm{\beta}$ when the penalty is selected in a data-dependent fashion. This result therefore closes the longstanding theoretical gap underlying the practical deployment of the $K$-fold cross-validated Lasso estimator.

\section{Simulation Study}\label{sec:simstudy}

We investigate the finite sample performance of the Bootstrap method developed in the previous section. In particular, we find empirical coverages of 90\% one-sided and two-sided confidence intervals for individual coefficients as well as of the 90\% confidence regions of the entire parameter vector under both homoscedastic and heteroscedastic linear regression setup. We used the Bootstrap percentile method. We try to capture finite sample performance under the following set-up: 
$(p,p_0)=(25,12)$ over $n \in \{30, 50, 75, 100\}$ separately when (A.1) the regression model is homoscedastic with errors $\epsilon_i$'s being iid standard Gaussian and (A.2) the regression model is heteroscedastic with $\epsilon_i$ being independently chosen from $N(0,\sigma_i)$ with $\sigma_i=p^{-1}\|\bm{x}_i\|^{3/2}$. In each setup, we consider the regression parameter vector to be $\bm{\beta}=(\beta_1, \dots, \beta_{p})^\top$ with
$\beta_j=0.5(-1)^j j \mathbbm{1}(1 \leq j \leq p_0)$. The design vectors are generated independently from the zero mean $p$-variate normal distribution with the covariance structure:
$cov(x_{ij},x_{ik})= \mathbbm{1}(j=k)+0.3^{|j-k|}  \mathbbm{1}(j \neq k)$, for all $1\leq j,k\leq p$, and kept fixed throughout the entire simulation. Based on these considerations, we pull out $n$ independent copies of response variables $y_1,\ldots,y_n$ from the model (\ref{eqn:linearreg}) in each of the cases (A.1) and (A.2).  

\begin{table}[H]
\centering
\caption{Empirical coverage probabilities of $90\%$ confidence region for $\bm{\beta}$ under Set-ups (A.1) and (A.2).}
\label{tab:coverage}

\begin{minipage}{0.48\linewidth}
\centering
\textbf{Set-up (A.1)}\\[0.5ex]
\small
\begin{tabular}{cccc}
\toprule
$n=30$ & $n=50$ & $n=75$ & $n=100$\\
\midrule
0.610 & 0.784 & 0.846 & 0.886 \\
\bottomrule
\end{tabular}
\end{minipage}
\hfill
\begin{minipage}{0.48\linewidth}
\centering
\textbf{Set-up (A.2)}\\[0.5ex]
\small
\begin{tabular}{cccc}
\toprule
$n=30$ & $n=50$ & $n=75$ & $n=100$\\
\midrule
0.548 & 0.768 & 0.858 & 0.892 \\
\bottomrule
\end{tabular}
\end{minipage}

\end{table}

In either setup, we keep the design matrix fixed and generate the entire data set 500 times to compute empirical coverage probability of one-sided and both sided confidence intervals and average width of the both sided confidence intervals over those above mentioned settings of $(n,p,p_0)$. The optimal $K$-fold CV based penalty $\hat{\lambda}_{n,K}$ as in equation (\ref{eqn:deflambdacv}), is obtained through $10$-fold CV using 
\texttt{cv.glmnet} in \texttt{R}. We minimize the objective function in (\ref{eqn:def}) using this CV-based choice of penalty to obtain $\hat{\bm{\beta}}_n(\hat{\lambda}_{n,K})$. Then thresholded Lasso estimator $\tilde{\bm{\beta}}_n(\hat{\lambda}_{n,K})$ is obtained based on the thresholding parameter $a_n=n^{-1/3}$ in either case.

Next, to define the PB quantities, we generate $n$ iid copies of $\texttt{Exp}(1)$ random variable. We consider $B=n$ many Bootstrap iterations to obtain PB-based percentile intervals. At each Bootstrap iteration, we calculate $\hat{\lambda}^*_{n,K}$ which is nothing but the $K$-fold CV base Lasso penalty based on the Bootstrapped observations defined in Section \ref{sec:bootstrap}. Then we utilize this choice to obtain $\hat{\bm{\beta}}^*_n(\hat{\lambda}^*_{n,K})$. We first present the empirical coverages of 90\% confidence regions and then move to the same for individual coefficients. Let $\big(\|\bm{T}_{n}^*(\hat{\lambda}_{n,K}^*)\|\big)_{\alpha}$ be the $\alpha$-th quantile of the Bootstrap distribution of $\|\bm{T}_{n}^*(\hat{\lambda}_{n,K}^*)\|$. Then the nominal $100(1-\alpha) \%$ confidence region of $\bm{\beta}$ is given by the set $C_{1 - \alpha}\subset \mathbb{R}^p$ where
$C_{1-\alpha} = \bigg\{\bm{\beta}:\|\bm{T}_{n}(\hat{\lambda}_{n, K})\|\leq \big(\|\bm{T}_{n}^*(\hat{\lambda}_{n,K}^*)\|\big)_{1-\alpha}\bigg\}.$
\begin{table}[H]
\centering
\caption{Empirical coverage probabilities and average widths of $90\%$ confidence intervals for $\bm{\beta}$ under Set-ups (A.1) and (A.2).}
\label{tab:ciwidths}

\begin{minipage}{0.48\linewidth}
\centering
\textbf{Homoscedastic set-up (A.1)}\\[0.6ex]
\small
\setlength{\tabcolsep}{4pt}
\renewcommand{\arraystretch}{1.1}
\begin{tabular}{ccccc}
\toprule
 & \multicolumn{4}{c}{\textbf{Both-sided}} \\
\cmidrule(rl){2-5}
$\beta_j$ & $n=30$ & $n=50$ & $n=75$ & $n=100$ \\
\midrule
-0.5  & 0.570 & 0.778 & 0.842 & 0.882 \\
$(\beta_{1})$    & (1.048) & (0.512) & (0.473) & (0.361) \\
1.0 & 0.736 & 0.808 & 0.864 & 0.898 \\
$(\beta_{2})$     & (1.121) & (0.534) & (0.369) & (0.304) \\
-5.5 & 0.712 & 0.842 & 0.856 & 0.868 \\
$(\beta_{11})$    & (1.183) & (0.661) & (0.433) & (0.379) \\
6.0  & 0.712 & 0.826 & 0.862 & 0.880 \\
$(\beta_{12})$     & (1.248) & (0.619) & (0.438) & (0.390) \\
0    & 0.808 & 0.866 & 0.878 & 0.904 \\
$(\beta_{13})$    & (1.019) & (0.568) & (0.442) & (0.388) \\
0    & 0.806 & 0.824 & 0.862 & 0.884 \\
$(\beta_{25})$     & (0.846) & (0.524) & (0.366) & (0.284) \\
\bottomrule
\end{tabular}
\end{minipage}
\hfill
\begin{minipage}{0.48\linewidth}
\centering
\textbf{Heteroscedastic set-up (A.2)}\\[0.6ex]
\small
\setlength{\tabcolsep}{4pt}
\renewcommand{\arraystretch}{1.1}
\begin{tabular}{ccccc}
\toprule
 & \multicolumn{4}{c}{\textbf{Both-sided}} \\
\cmidrule(rl){2-5}
$\beta_j$ & $n=30$ & $n=50$ & $n=75$ & $n=100$ \\
\midrule
-0.5  & 0.672 & 0.744 & 0.824 & 0.882 \\
$(\beta_{1})$    & (1.228) & (0.662) & (0.561) & (0.381) \\
1.0 & 0.810 & 0.838 & 0.884 & 0.896 \\
$(\beta_{2})$     & (0.964) & (0.734) & (0.449) & (0.318) \\
-5.5 & 0.778 & 0.852 & 0.864 & 0.888 \\
$(\beta_{11})$    & (1.108) & (0.781) & (0.542) & (0.366) \\
6.0  & 0.802 & 0.846 & 0.876 & 0.894 \\
$(\beta_{12})$     & (1.204) & (0.606) & (0.568) & (0.482) \\
0    & 0.824 & 0.866 & 0.884 & 0.902 \\
$(\beta_{13})$    & (1.049) & (0.668) & (0.542) & (0.324) \\
0    & 0.812 & 0.856 & 0.882 & 0.892 \\
$(\beta_{25})$     & (0.906) & (0.624) & (0.446) & (0.246) \\
\bottomrule
\end{tabular}
\end{minipage}

\end{table}
The results on the empirical coverage probability of $90\%$ confidence region of $\bm{\beta}$ are displayed in Table \ref{tab:coverage}. The coverage performance of $\bm{\beta}$ in both homoscedastic and heteroscedastic set-up, continues to improve and moves toward the nominal level of $0.9$ as we increase $n$.

Next, we present the component wise both-sided and one-sided empirical coverage probabilities. We denote $\alpha-$th Bootstrap quantile of $j-$th component of $\bm{T}_{n}^*(\hat{\lambda}_{n,K}^*)$ as $\big(\bm{T}_{n,j}^*(\hat{\lambda}_{n,K}^*)\big)_{\alpha}$. Then $100(1-\alpha)\%$ both-sided Bootstrap percentile interval for $\beta_j$ is given by: $\Big[\hat{\beta}_{n,j}-\frac{\big(\bm{T}_{n,j}^*(\hat{\lambda}_{n,K}^*)\big)_{1-\frac{\alpha}{2}}}{n^{1/2}},\hat{\beta}_{n,j}-\frac{\big(\bm{T}_{n,j}^*(\hat{\lambda}_{n,K}^*)\big)_{\frac{\alpha}{2}}}{n^{1/2}}\Big].$
\begin{table}[H]
\centering
\caption{Empirical coverage probabilities of $90\%$ right-sided confidence intervals for $\bm{\beta}$ under Set-ups (A.1) and (A.2).}
\label{tab:rightsided}

\begin{minipage}{0.48\linewidth}
\centering
\textbf{Homoscedastic set-up (A.1)}\\[0.6ex]
\small
\setlength{\tabcolsep}{4pt}
\renewcommand{\arraystretch}{1.1}
\begin{tabular}{ccccc}
\toprule
 & \multicolumn{4}{c}{\textbf{Right-sided}}\\
\cmidrule(rl){2-5} 
$\beta_j$ & $n=30$ & $n=50$ & $n=75$ & $n=100$\\
\midrule
-0.5  & 0.802 & 0.832 & 0.870 & 0.886 \\ 
1.0 & 0.746 & 0.812 & 0.856 & 0.882 \\ 
-5.5 & 0.824 & 0.832 & 0.888 & 0.906 \\ 
6.0  & 0.804 & 0.850 & 0.880 & 0.882 \\ 
0    & 0.784 & 0.836 & 0.878 & 0.888 \\ 
0    & 0.824 & 0.846 & 0.878 & 0.904 \\
\bottomrule
\end{tabular}
\end{minipage}
\hfill
\begin{minipage}{0.48\linewidth}
\centering
\textbf{Heteroscedastic set-up (A.2)}\\[0.6ex]
\small
\setlength{\tabcolsep}{4pt}
\renewcommand{\arraystretch}{1.1}
\begin{tabular}{ccccc}
\toprule
 & \multicolumn{4}{c}{\textbf{Right-sided}}\\
\cmidrule(rl){2-5} 
$\beta_j$ & $n=30$ & $n=50$ & $n=75$ & $n=100$\\
\midrule
-0.5  & 0.722 & 0.812 & 0.874 & 0.896 \\ 
1.0 & 0.788 & 0.842 & 0.880 & 0.912 \\ 
-5.5 & 0.802 & 0.846 & 0.874 & 0.886 \\ 
6.0  & 0.798 & 0.856 & 0.886 & 0.904 \\ 
0    & 0.816 & 0.832 & 0.868 & 0.882 \\ 
0    & 0.780 & 0.826 & 0.858 & 0.896 \\
\bottomrule
\end{tabular}
\end{minipage}

\end{table}
To save space, we demonstrate the empirical coverage probabilities of the following regression components for both sided and right sided 90\% confidence intervals through tables: $\{\beta_{1}=-0.5, \beta_{2}=1.0, \beta_{11}=-5.5, \beta_{12}=6.0, \beta_{13}=\beta_{25}=0\}$ among all $25$ components. The average widths of the two-sided intervals for these components of $\bm{\beta}$ are mentioned in parentheses under the empirical coverage probabilities. We observe in the set-up (A.1) of Table \ref{tab:ciwidths} and \ref{tab:rightsided} that as $n$ increases, the empirical coverage probabilities get closer and closer to the nominal confidence level of 90\% for all the aforementioned regression coefficients. Table \ref{tab:ciwidths} also shows that the average width of all the two-sided confidence intervals decreases as $n$ increases. Similar traits pertain to empirical coverages of other individual components and also in heteroscedastic setup. These conclusions are validated by results presented in set-up (A.2) of Table \ref{tab:ciwidths} and \ref{tab:rightsided}. All the reproducible codes are available at \footnote{\url{https://github.com/mayukhc13/Asymptotic-Theory-of-K-fold-CV-in-Lasso.git}}.

\section{Real Data Analysis}\label{sec:realdat}
We have applied our Bootstrap method to well-studied prostate cancer data which is an original study by \citet{stamey1989prostate} and later familiarized by \citet{tibshirani1996regression}. This data is available at \url{https://web.stanford.edu/~hastie/ElemStatLearn/datasets/prostate.data}. The prostate cancer data set examined the relationship between prostate-specific antigen (PSA) levels and several clinical measurements in men prior to undergoing radical prostatectomy. PSA is a protein produced by the prostate gland, and elevated levels can indicate the presence or progression of prostate cancer. The primary objective of this analysis is to model the log-transformed PSA levels (\texttt{lpsa}) as a function of various clinical predictors, using Lasso regression. This helps to identify the most influential clinical factors related to PSA levels through our proposed Bootstrap method. The dataset comprises $97$ observations with the following variables: Log cancer volume (\texttt{lcavol}), Log prostate weight (\texttt{lweight}), Age of patient (\texttt{age}), Log of benign prostatic hyperplasia amount (\texttt{lbph}), Seminal vesicle invasion (\texttt{svi}), Log capsular penetration (\texttt{lcp}), Gleason score (\texttt{gleason}) and
Percentage of Gleason scores 4 or 5 (\texttt{pgg45}). The response variable is Log of prostate-specific antigen (\texttt{lpsa}). We fit Lasso regression using 10-fold CV to select the regularization parameter (see \url{https://github.com/mayukhc13/Asymptotic-Theory-of-K-fold-CV-in-Lasso.git} for details).
\begin{table}[htbp]
\centering
\caption{Estimated Lasso Coefficients \& 90\% Bootstrap Percentile Confidence Intervals for the Prostate Cancer Data}\par 
\vspace{0.65ex}
\label{tab:lasso-prostate}
\begin{tabular}{lcccc}
\toprule
\textbf{Covariates} & $\hat{\beta}_j$ & \textbf{Both Sided} & \textbf{Left Sided} & \textbf{Right Sided} \\
\midrule
lcavol   & 0.4851 & [0.3975, 0.7102]     & [0.4122, $\infty$)     & ($-\infty$, 0.6638] \\
lweight  & 0.4599 & [0.2123, 0.9198]     & [0.3583, $\infty$)     & ($-\infty$, 0.9198] \\
age      & 0.0000 & [$-$0.0083, 0.0088]  & [$-$0.0050, $\infty$)  & ($-\infty$, 0.0000] \\
lbph     & 0.0161 & [$-$0.0439, 0.0595]  & [$-$0.0108, $\infty$)  & ($-\infty$, 0.0269] \\
svi      & 0.5039 & [0.2299, 1.0077]     & [0.2996, $\infty$)     & ($-\infty$, 1.0077] \\
lcp      & 0.0000 & [$-$0.1157, 0.0000]  & [$-$0.0986, $\infty$)  & ($-\infty$, 0.0000] \\
gleason  & 0.0000 & [$-$0.0421, 0.0185]  & [$-$0.0148, $\infty$)  & ($-\infty$, 0.0000] \\
pgg45    & 0.0008 & [$-$0.0022, 0.0012]  & [$-$0.0015, $\infty$)     & ($-\infty$, 0.0008] \\

\bottomrule
\end{tabular}
\end{table}
The Lasso regression identifies three variables with non-zero coefficients: \texttt{lcavol} (log cancer volume), \texttt{lweight} (log prostate weight), and \texttt{svi} (seminal vesicle invasion). All three covariates have strictly positive coefficient estimates and 90\% Bootstrap confidence intervals that exclude zero in both the two-sided and one-sided settings, providing strong evidence of their influence in predicting PSA levels. In fact, \texttt{lcavol} and \texttt{lweight} consistently rank as the important predictors across studies. The right-sided and the left-sided CI further reinforce this conclusion, as also found in both \citet{stamey1989prostate} and \citet{tibshirani1996regression}. The Lasso assigns a coefficient of $0.5039$ to seminal vesicle invasion. Its both-sided and left-sided confidence intervals  again exclude zero, supporting the clinical understanding that invasion of seminal vesicles is associated with advanced cancer and elevated PSA levels. On the contrary, the covariates (\texttt{age}, \texttt{lcp}, \texttt{gleason} and \texttt{pgg45}) are all estimated to have coefficients of zero (or close to zero). For each, the two-sided confidence intervals include zero. Furthermore, the right-sided intervals indicate a lack of strong positive association with PSA. In some cases, the left-sided intervals do not convincingly exclude zero either, suggesting a weak or negligible negative effect. In particular, this aligns well with analyses of \citet{tibshirani1996regression} 
and \citet{stamey1989prostate}, where they were not identified as dominant predictors in the presence of \texttt{lcavol}, \texttt{lweight}, and \texttt{svi}. In summary, the Bootstrap confidence intervals validate Lasso's ability to select meaningful predictors in real biomedical data, while offering principled uncertainty quantification.

\section{Proofs}\label{sec:proof2}

In this section, precisely in section \ref{sec:appC} we provide the the proofs of main results viz. Theorem \ref{prop:cv}, Theorem \ref{thm:cvthm}, Proposition \ref{prop:eqicon}, Proposition \ref{prop:C.4},  Theorem \ref{thm: lambdaCVconv}, Theorem \ref{thm:distconvcv} and Theorem \ref{thm:bootconsistency}. All the requisite lemmas, needed for our main results, needed for our main results are addressed in Sections \ref{sec:appA}, \ref{sec:appB}, \ref{sec:appCBoot} and \ref{sec:app4.c} sequentially. In section \ref{subsec:measurability}, we consider the measurability aspects of the Argmin functionals which will help utilizing Dudley's almost sure representation.

\subsection{Measurability of Argmin Maps}\label{subsec:measurability}
Before moving to the requisite lemmas, we describe some notions which are crucial in understanding the measurability of different argmin maps and suprema (or infima) discussed in the paper. We mostly follow \citet{aliprantis2006infinite} and \citet{rockafellar2009variational} for defining these notions. In the following remarks, without loss of generality we assume that the underlying
probability space $(\Omega,\mathcal F,\mathbf P)$ is complete. This assumption is required specially to handle the measurability of infimum and supremum arising in the relation
\begin{align}\label{eqn:100025}
  &\mathbf{P}\Big(|n^{-1/2}\hat{\lambda}_{n, K} - \hat{\Lambda}_{\infty, K}| > \varepsilon\Big)\nonumber\\ \leq &\;\mathbf{P}\Big(2\sup_{\lambda}\Big|\hat{H}_{n, K}^\prime(\lambda) - \hat{H}_{\infty, K}^\prime(\lambda)\Big|  > \inf_{|\lambda - \hat{\Lambda}_{\infty,K}| > \varepsilon} \hat{H}_{\infty, K}^\prime(\lambda) - \hat{H}_{\infty, K}^\prime(\hat{\Lambda}_{\infty, K}) \Big),
\end{align}
for any $\varepsilon > 0$. All the quantities are defined in Table \ref{tab:unified}.

\begin{definition}\label{def:measurabilitycorrespondence}
Let $(T,\mathcal{A})$ be some non-empty measure space. For a set-valued map or correspondence $G$ from $T$ to $\mathbb{R}^p$ (usually denoted as $G:T\rightrightarrows\mathbb R^p$) is said to be measurable if for every open set $O\subset \mathbb{R}^p$, the set $G^{-1}(O)=\Big\{t\in T: G(t)\cap O\neq \phi\Big\}\in \mathcal{A}.$  When the image of $G$ is singleton, then this notion reduces to the notion of measurability in usual sense.
\end{definition}

\begin{definition}\label{def:normalintegrand}
The map $f:T\times\mathbb{R}^p\mapsto \mathbb{R}$ is said to be a \underline{proper normal integrand} if its epigraphical mapping $epi f(\cdot):T\mapsto \mathbb{R}^p\times \mathbb{R}$ is closed-valued and measurable in the sense of definition \ref{def:measurabilitycorrespondence} where we define 
 $$epi f(t):=\Big\{(\bm{x},\alpha)\in \mathbb{R}^p\times \mathbb{R}:f(t,\bm{x})\leq \alpha \Big\}.$$
\end{definition}
\begin{definition}\label{def:caratheodoryintegrand}
The map $f:T\times\mathbb{R}^p\mapsto \mathbb{R}$ is called a \underline{Carath{\'e}odory integrand} if $f(t,\bm{x})$ is measurable in $t$ for each $\bm{x}$ and continuous in $\bm{x}$ for each $t$. 
\end{definition}

\begin{remark}{\underline{$\big($$W_n$ and $W_{n,k}$ are measurable w.r.t. $\sigma(W_{n, -1}, \dots, W_{n, -K})$$\big)$}}\label{remark:measurabilitywrttraining} 

Recall that the each fold has size \(m=n/K\). Recall that \(\bm{W}_{n,k}
=
m^{-1/2}\sum_{i\in I_k}\varepsilon_i\bm{x}_i\), $\bm{W}_{n,-k}
=
(n-m)^{-1/2}\sum_{i\notin I_k}\varepsilon_i\bm{x}_i$ and
\(\bm{W}_n
=
n^{-1/2}\sum_{i=1}^{n}\varepsilon_i\bm{x}_i.\) Then
\[
\sum_{k=1}^{K}\bm W_{n,-k}
=
\sqrt{K(K-1)}\,\bm W_n,
\]
and hence
\(\bm W_n
=
\frac{1}{\sqrt{K(K-1)}}
\sum_{k=1}^{K}\bm W_{n,-k}.
\) Moreover, for each \(k=1,\ldots,K\),
\[
\bm W_{n,k}
=
\sqrt{K}\,\bm W_n
-
\sqrt{K-1}\,\bm W_{n,-k},
\]
which, upon substituting the above expression for \(\bm W_n\), yields
\[
\bm W_{n,k}
=
\frac{1}{\sqrt{K-1}}
\left(
\sum_{\ell\neq k}\bm W_{n,-\ell}
-
(K-2)\bm W_{n,-k}
\right).
\]
Consequently, both \(\bm W_n\) and \(\bm W_{n,k}\), \(k=1,\ldots,K\), are linear, and hence measurable with respect to 
$\sigma(\bm W_{n,-1},\ldots,\bm W_{n,-K}).$
\end{remark}

\begin{remark}{\underline{$\big($Measurability of $\hat{\bm{u}}_{n,-k}(\lambda)=(n-m)^{1/2}(\hat{\bm{\beta}}_{n,-k}(n^{1/2}\lambda)-\bm{\beta})$ w.r.t. $\sigma(W_{n, -k})$, for all $k$$\big)$}}\label{remark:measurabilitylassopath} 

Fix $k\in\{1,\dots,K\}$. Define
\(\hat{V}_{n,-k}:\mathbb{R}^p\times[0,\infty)\times\mathbb{R}^p
\longrightarrow
\mathbb{R}
\) by
\begin{align*}
\hat{V}_{n,-k}(\bm{W}_{n,-k},\lambda,\bm{u})
&=
\frac12\bm{u}^{\top}\bm{L}_{n,-k}\bm{u}
-\bm{u}^{\top}\bm{W}_{n,-k}+n^{1/2}\lambda
\sum_{j=1}^{p}
\left(
\left|
(n-m)^{-1/2}u_j+\beta_j
\right|
-
|\beta_j|
\right),
\end{align*}
Now equip
\(T_1=\mathbb{R}^p\times[0,\infty)\) with the Borel $\sigma$-field
$\mathcal{B}(\mathbb{R}^p\times[0,\infty))$.
For every fixed $\bm{u}\in\mathbb{R}^p$, the map
\[
(\bm{w},\lambda)
\longmapsto
\hat{V}_{n,-k}(\bm{w},\lambda,\bm{u})
\]
is continuous, and hence Borel measurable. Likewise, for every fixed
$(\bm{w},\lambda)\in T_1$, the map
\[
\bm{u}
\longmapsto
\hat{V}_{n,-k}(\bm{w},\lambda,\bm{u})
\]
is continuous since it is the sum of a quadratic form, a linear function
and an absolute value penalty. Therefore
$\hat{V}_{n,-k}$ is a Carath\'eodory integrand. By
Example 14.29 of \citet{rockafellar2009variational},
$\hat{V}_{n,-k}$ is a proper normal integrand. Further, we prove in Lemma \ref{lem:lassostrongconvex} that the function
$\bm{u}\mapsto
\hat{V}_{n,-k}(\bm{w},\lambda,\bm{u})$
is strictly convex for every fixed
$(\bm{w},\lambda)\in T_1$,
and hence $\bm{u}\mapsto\hat{V}_{n,-k}(\bm{w},\lambda,\bm{u})$ admits a unique minimizer  for every fixed
$(\bm{w},\lambda)\in T_1$. Define
\[
P(\bm{w},\lambda)
:=
\operatorname*{argmin}_{\bm{u}\in\mathbb{R}^p}
\hat{V}_{n,-k}(\bm{w},\lambda,\bm{u}).
\]
By Theorem 14.37 of \citet{rockafellar2009variational} or equivalently Theorem 18.19 of \citet{aliprantis2006infinite},
$P:T_1\rightrightarrows\mathbb{R}^p$ is a measurable correspondence. Since $P(\bm{w},\lambda)$ is a singleton
for every fixed  $(\bm{w},\lambda)\in T_1$, it follows that $P$ may be identified
with a Borel measurable function
\(\psi_k:
\mathbb{R}^p\times[0,\infty)
\longrightarrow
\mathbb{R}^p,\)
satisfying
\[
\psi_k(\bm{W}_{n,-k},\lambda)=\hat{\bm{u}}_{n,-k}(\lambda):
=
\operatorname*{argmin}_{\bm{u}\in\mathbb{R}^p}
\hat{V}_{n,-k}(\bm{W}_{n,-k},\lambda,\bm{u}).
\]
Hence, for fixed $\lambda\ge 0$, the random vector
$\hat{\bm{u}}_{n,-k}(\lambda)$ is $\sigma(\bm W_{n,-k})$-measurable for every $k\in\{1,\dots,K\}$. 
\end{remark}

\begin{remark}{\underline{(Measurability of $n^{-1/2}\hat{\lambda}_{n,K}$ w.r.t. $\sigma(\bm W_{n,-1},\ldots,\bm W_{n,-K})$)}}\label{remark:measurabilitywrtcv} 

Let
\(T_2=\underbrace{\mathbb R^p\times\cdots\times\mathbb R^p}_{K\ \text{copies}},
\) equipped with the Borel \(\sigma\)-field
\(\mathcal A=\mathcal B(T_2)\). For
\(\mathbf z=(\mathbf w_{-1},\ldots,\mathbf w_{-K})\in T_2,
\) define, for each \(k=1,\ldots,K\),
\[
\phi_k(\mathbf z)
=
\frac{1}{\sqrt{K-1}}
\left(
\sum_{\ell\neq k}\mathbf w_{-\ell}
-
(K-2)\mathbf w_{-k}
\right),
\]
so that \(\phi_k(\mathbf z)\) represents the vector
\(\mathbf W_{n,k}\) as a measurable function of
\((\mathbf W_{n,-1},\ldots,\mathbf W_{n,-K})\). Now define
\(\hat H'_{n,K}:T_2\times[0,\infty)\longrightarrow\mathbb R
\) by
\[
\hat H'_{n,K}(\mathbf z,\lambda)
=
\sum_{k=1}^{K}
\left[
\frac{m}{2(n-m)}
\psi_k(\mathbf w_{-k},\lambda)^\top
\mathbf L_{n,k}
\psi_k(\mathbf w_{-k},\lambda)
-
\sqrt{\frac{m}{n-m}}\,
\psi_k(\mathbf w_{-k},\lambda)^\top
\phi_k(\mathbf z)
\right].
\]
Since each \(\psi_k\) and \(\phi_k\) is Borel measurable, the map
\(\mathbf z
\longmapsto
\hat H'_{n,K}(\mathbf z,\lambda)
\) is Borel measurable for every fixed \(\lambda\ge0\). Moreover, by
Proposition~\ref{prop:eqicon}, the map
\(\lambda
\longmapsto
\psi_k(\mathbf w_{-k},\lambda)
\) is continuous for every fixed
\(\mathbf w_{-k}\in\mathbb R^p\). Consequently,
\(\lambda
\longmapsto
\hat H'_{n,K}(\mathbf z,\lambda)
\) is continuous for every fixed \(\mathbf z\in T_2\). Hence
\(\hat H'_{n,K}\) is a Carath\'eodory integrand and therefore a proper
normal integrand. Define the argmin correspondence
\[
P(\mathbf z)
=
\operatorname*{Argmin}_{\lambda\ge0}
\hat H'_{n,K}(\mathbf z,\lambda),
\qquad
\mathbf z\in T_2.
\]
By Theorem 14.37 of \citet{rockafellar2009variational} or Theorem 18.19 of \citet{aliprantis2006infinite},
\(P:T_2\rightrightarrows[0,\infty)\) is measurable correspondence and there exists a measurable
selection
\(\tau:T_2\longrightarrow[0,\infty)
\)
such that
\[
n^{-1/2}\hat\lambda_{n,K}
=
\tau
(\mathbf W_{n,-1},\ldots,\mathbf W_{n,-K}).
\]
Whenever we write \(n^{-1/2}\hat\lambda_{n,K}\) in the paper, we mean
such a measurable selection. Therefore,
\(n^{-1/2}\hat\lambda_{n,K}\) is measurable with respect to
\(
\sigma(\mathbf W_{n,-1}, \dots, W_{n, -K})
\) and we are done. 
\end{remark}

\begin{remark}{\underline{$\big($Measurablity of $\hat{\Lambda}_{\infty, K}$ w.r.t. $\sigma(\bm W_{\infty,-1},\ldots,\bm W_{\infty,-K})$$\big)$}}\label{reark:measurablelimit} 

Arguments similar to the Remark \ref{remark:measurabilitywrtcv} also apply to the limiting criterion $\hat{H}_{\infty, K}$, replacing
 $(\bm W_{n,-k}:k=1,..,K)^\top$ by its Gaussian limit $(\bm W_{\infty,-k}:k=1,...,K)^\top$ (see Lemma \ref{lem:gausswn-}). Thus we have $\hat{H}_{\infty, K}$ is a Carath\'eodory integrand and hence
 the measurability of $\hat{\Lambda}_{\infty,K}$ with respect to
 $\sigma(\mathbf W_{\infty,-1}, \dots, W_{\infty, -K})$ is immediate.
\end{remark}

\begin{remark}{\underline{$\big($Measurability of $\eta(\varepsilon)$ for any fixed $\varepsilon > 0$$\big)$}}\label{remark:measurableeta} 

Recall that
\[
\eta(\varepsilon)
:=
\inf_{\left|\lambda-\hat{\Lambda}_{\infty,K}\right|>\varepsilon}
\Big[
\hat H'_{\infty,K}(\lambda)
-
\hat H'_{\infty,K}(\hat{\Lambda}_{\infty,K})
\Big].
\]
First fix $\varepsilon > 0$. Now Remark~\ref{reark:measurablelimit} implies that there exists a measurable map
\(\tau:
\big(\mathbb{R}^{Kp},\mathcal B(\mathbb{R}^{Kp})\big)
\longrightarrow
\big([0,\infty),\mathcal B([0,\infty))\big)
\) such that
\[
\hat{\Lambda}_{\infty,K}
=
\tau(\mathbf W_{\infty,-1},\ldots,\mathbf W_{\infty,-K}).
\]
Let
\(\mathbf Z
:=
(\mathbf W_{\infty,-1},\ldots,\mathbf W_{\infty,-K}),\) and denote by
\(\mu:=\mathbf P\circ\mathbf Z^{-1}
\) the law of $\mathbf Z$ on $\mathbb R^{Kp}$.
Consider the measurable space
\((\mathbf T,\mathcal A)
:=
\bigl(\mathbb R^{Kp},
\overline{\mathcal B(\mathbb R^{Kp})}^{\,\mu}\bigr),\) where
\[
\overline{\mathcal B(\mathbb R^{Kp})}^{\,\mu}
:=
\left\{
A\cup N:
A\in\mathcal B(\mathbb R^{Kp}),
\;
N\subseteq Z
\text{ for some }
Z\in\mathcal B(\mathbb R^{Kp})
\text{ with }
\mu(Z)=0
\right\},
\]
is the completion of the Borel $\sigma$-field on
$\mathbb R^{Kp}$ with respect to $\mu$. Equivalently,
\(\overline{\mathcal{B}(\mathbb{R}^{Kp})}^{\,\mu}
\) is the smallest $\sigma$-algebra containing the elements of
$\mathcal{B}(\mathbb{R}^{Kp})$
and all subsets of $\mu$-null sets in
$\mathcal{B}(\mathbb{R}^{Kp})$. 
Let $G: T \rightrightarrows [0,\infty)$ be a correspondence defined by
\[
G(\mathbf z)
:=
\left\{
\lambda \in [0,\infty)
:\ |\lambda - \tau(\mathbf z)| > \varepsilon
\right\}.
\]
Thus, $G$ assigns to each $\mathbf z \in \mathbb{R}^{Kp}$ a subset of $[0,\infty)$ consisting of all
$\lambda$ satisfying the condition $|\lambda - \tau(\mathbf z)| > \varepsilon$. 
Note that the definition of $G$ implies that
\[
\begin{aligned}
G^{-1}(O)
&=
\Bigl\{
\mathbf z\in \mathbb R^{Kp}:
G(\mathbf z)\cap O\neq\varnothing
\Bigr\}=\Bigl\{
\mathbf z\in \mathbb R^{Kp}:
\exists\,\lambda\in O
\text{ such that }
|\lambda-\tau(\mathbf z)|>\varepsilon
\Bigr\}.
\end{aligned}
\]
We first verify that $G$ is measurable in the sense of
Definition~\ref{def:measurabilitycorrespondence}, i.e., we have to check if for every open set
$O \subset [0,\infty)$,
\[
G^{-1}(O)
:=
\{\mathbf z \in T : G(\bm z)\cap O \neq \emptyset\}
\in \mathcal{A}.
\]
Fix an open set $O\subset[0,\infty)$ and define
\[
A:=
\Bigl\{
(\mathbf z,\lambda):
\lambda\in O,\;
|\lambda-\tau(\mathbf z)|>\varepsilon
\Bigr\}
\subset
\mathbb R^{Kp}\times[0,\infty).
\]
Clearly, $G^{-1}(O) = \pi(A)$ where $\pi: \mathbb{R}^{Kp} \times [0, \infty) \rightarrow \mathbb{R}^{Kp}$ is the projection on first $Kp$ coordinates. We know that the projections of Borel sets in Polish spaces are analytic (see Proposition 4.1.1 of \citet{srivastava1998course} and exercise 14.3 of \citet{kechris2012classical}). Moreover, every analytic subset of $\mathbb R^{Kp}$ is universally
measurable (cf. Corollary 5.0.5 of \citet{muller1997polish}, Theorem 4.10.12 of \citet{srivastava1998course}), and hence is an element of the completion of the underlying Borel $\sigma$-field w.r.t. any probability measure. Therefore, to claim \(G^{-1}(O) \in \mathcal{A}\), it is enough to show that $A \in \mathcal{B}\big(\mathbb{R}^{Kp}\times [0, \infty)\big)$. 
Since \(\tau:(\mathbb R^{Kp},\mathcal B(\mathbb R^{Kp}))
\longrightarrow
([0,\infty),\mathcal B([0,\infty)))\) is measurable and any coordinate projection is continuous (hence Borel measurable),
the map
\[
f:\mathbb{R}^{Kp}\times[0,\infty)
\longrightarrow
\mathbb{R}, \;\; \text{where}\;
f(\mathbf z,\lambda)
:=
|\lambda-\tau(\mathbf z)|.
\]
is measurable with respect to $\mathcal{B}\big(\mathbb{R}^{Kp}\times
[0,\infty)\big)$. 
Thus,
\[
f^{-1}\big((\varepsilon,\infty)\big)
=
\Bigl\{
(\mathbf z,\lambda):
|\lambda-\tau(\mathbf z)|>\varepsilon
\Bigr\}
\in
\mathcal B\bigl(\mathbb R^{Kp}\times[0,\infty)\bigr),
\]
and therefore
\[
A
=
(\mathbb R^{Kp}\times O)
\cap
f^{-1}\big((\varepsilon,\infty)\big)
\in
\mathcal B\bigl(\mathbb R^{Kp}\times[0,\infty)\bigr).
\]
Therefore, for every open set $O \subset [0,\infty)$,
\[
G^{-1}(O) \in \overline{\mathcal B(\mathbb R^{Kp})}^{\,\mu}
=\mathcal A,
\]
implying that $G$ is a measurable correspondence with respect to
$\overline{\mathcal B(\mathbb R^{Kp})}^{\,\mu}$. Define
\[
\widetilde G(\mathbf z)
:=
\{\lambda\in[0,\infty):
|\lambda-\tau(\mathbf z)|\ge\varepsilon\}
=
\overline{G(\mathbf z)}.
\]
Since \(G\) is measurable, \(\widetilde G\) is also measurable. Indeed,
for every open set \(O\subset[0,\infty)\),
\[
\widetilde G^{-1}(O)
=
\{\mathbf z:\overline{G(\mathbf z)}\cap O\neq\varnothing\}
=
\{\mathbf z:G(\mathbf z)\cap O\neq\varnothing\}
=
G^{-1}(O),
\]
and hence \(\widetilde G^{-1}(O)\in\mathcal A\).
Moreover, \(\widetilde G(\mathbf z)\) is closed for every \(\mathbf z\). Define the indicator integrand
\[
\mathbf I_{\widetilde G}
:
\mathbb R^{Kp}\times[0,\infty)
\longrightarrow
\overline{\mathbb R}
\]
by
\[
\mathbf I_{\widetilde G(\mathbf z)}(\lambda)
:=
\begin{cases}
0,
&
\lambda\in\widetilde G(\mathbf z),
\\[1ex]
+\infty,
&
\lambda\notin\widetilde G(\mathbf z).
\end{cases}
\]
Since $\widetilde G$ is measurable and closed-valued,
Example~14.32 of \citet{rockafellar2009variational}
implies that
$\mathbf I_{\widetilde G}$
is a normal integrand. Note that by Remark~\ref{reark:measurablelimit}, $\hat H'_{\infty,K}$ is a proper normal integrand. Therefore,
\[
g(\mathbf z,\lambda)
:=
\hat H'_{\infty,K}(\mathbf z,\lambda)
+
\mathbf I_{\widetilde G(\mathbf z)}(\lambda)
\]
is a proper normal integrand. Define the map,
\[
v:\mathbb{R}^{Kp}\mapsto\bar{\mathbb R},\;\;v(\mathbf z)
:=
\inf_{\lambda\in[0,\infty)}
g(\mathbf z,\lambda).
\]
By Theorem~14.37 of \citet{rockafellar2009variational},
the function
$v$
is
$\mathcal A/\mathcal B(\overline{\mathbb R})$-measurable. Furthermore,
\[
\begin{aligned}
v(\mathbf z)
&=
\inf_{\lambda\in[0,\infty)}
\Bigl[
\hat H'_{\infty,K}(\mathbf z,\lambda)
+
\mathbf I_{\widetilde G}(\mathbf z,\lambda)
\Bigr]=
\inf_{\lambda\in\widetilde G(\mathbf z)}
\hat H'_{\infty,K}(\mathbf z,\lambda).
\end{aligned}
\]
We next show that \(v\) is in fact real-valued.
Since
\(\tau(\mathbf z)+\varepsilon\in \tilde G(\mathbf z),
\) we have
\(v(\mathbf z)
\le
\hat H'_{\infty,K}
\bigl(\mathbf z,\tau(\mathbf z)+\varepsilon\bigr).
\) Since \(\hat H'_{\infty,K}\) is a Carath\'eodory integrand, the right-hand side is finite, and therefore
\(v(\mathbf z)<\infty\). Moreover, since \(\tau(\mathbf z)\) minimizes
\(\hat H'_{\infty,K}(\mathbf z,\cdot)\),
\[
\hat H'_{\infty,K}(\mathbf z,\lambda)
\ge
\hat H'_{\infty,K}(\mathbf z,\tau(\mathbf z))
\qquad
\forall\,\lambda\in[0,\infty).
\]
Hence
\[
v(\mathbf z)
=
\inf_{\lambda\in G(\mathbf z)}
\hat H'_{\infty,K}(\mathbf z,\lambda)
\ge
\hat H'_{\infty,K}(\mathbf z,\tau(\mathbf z)).
\]
Again, since \(\hat H'_{\infty,K}\) is real-valued,
\(\hat H'_{\infty,K}(\mathbf z,\tau(\mathbf z))
>-\infty.
\) Therefore
\(v(\mathbf z)\in\mathbb R,\)\; for all \(\mathbf z\in\mathbb R^{Kp}.
\) Consequently,
\(v:
(\mathbb R^{Kp},\mathcal A)
\longrightarrow
(\mathbb R,\mathcal B(\mathbb R))
\) is measurable.\\

Now since
\(\{\lambda\in[0,\infty):
|\lambda-\tau(\mathbf z)|>\varepsilon\}
\subset
\widetilde G(\mathbf z),
\) we immediately have
\[
\inf_{\lambda\in\widetilde G(\mathbf z)}
\hat H'_{\infty,K}(\mathbf z,\lambda)
\le
\inf_{\{|\lambda-\tau(\mathbf z)|>\varepsilon\}}
\hat H'_{\infty,K}(\mathbf z,\lambda).
\]
Conversely, let
\(\lambda\in\widetilde G(\mathbf z)\).
Since
\(\{\lambda:
|\lambda-\tau(\mathbf z)|>\varepsilon\}
\) is dense in
\(\widetilde G(\mathbf z)\),
there exists a sequence
\(\{\lambda_m\}_{m\ge1}\)
satisfying
\[
|\lambda_m-\tau(\mathbf z)|>\varepsilon,
\qquad
\lambda_m\to\lambda.
\]
Since
\(\hat H'_{\infty,K}(\mathbf z,\cdot)\)
is continuous,
\(\hat H'_{\infty,K}(\mathbf z,\lambda_m)
\longrightarrow
\hat H'_{\infty,K}(\mathbf z,\lambda).
\) Therefore,
\[
\hat H'_{\infty,K}(\mathbf z,\lambda)
\ge
\inf_{\{|\lambda-\tau(\mathbf z)|>\varepsilon\}}
\hat H'_{\infty,K}(\mathbf z,\lambda).
\]
As this holds for every
\(\lambda\in\widetilde G(\mathbf z)\), thus \[
\inf_{\lambda\in\widetilde G(\mathbf z)}
\hat H'_{\infty,K}(\mathbf z,\lambda)
\ge
\inf_{\{|\lambda-\tau(\mathbf z)|>\varepsilon\}}
\hat H'_{\infty,K}(\mathbf z,\lambda).
\]
Combining the two inequalities yields
\[
\inf_{\lambda\in\widetilde G(\mathbf z)}
\hat H'_{\infty,K}(\mathbf z,\lambda)
=
\inf_{\{|\lambda-\tau(\mathbf z)|>\varepsilon\}}
\hat H'_{\infty,K}(\mathbf z,\lambda).
\]
Therefore we have, 
\[
v(\mathbf z)
:=
\inf_{\lambda \in G(\mathbf z)}
\hat H^\prime_{\infty, K}(\mathbf z,\lambda),
\]
and is measurable with respect to $\mathcal{A}$. Since $\hat H'_{\infty,K}$ is a Carath\'eodory integrand, it is
measurable with respect to
\(\mathcal B(\mathbb R^{Kp}
\times [0,\infty)).
\) Again
\(\tau:
(\mathbb R^{Kp},\mathcal A)
\longrightarrow
([0,\infty),\mathcal B([0,\infty)))
\) is measurable since
$\mathcal B(\mathbb R^{Kp})\subset\mathcal A$. Therefore, combining these two we obtain that 
the map
\(\mathbf z
\longmapsto
\hat H'_{\infty,K}(\mathbf z,\tau(\mathbf z))
\) is $\mathcal A$-measurable.
Consequently,
\[
\eta(\varepsilon)
:=
v(\bm W_{\infty,-1},...,\bm W_{\infty,-K})
-
\hat H^\prime_{\infty, K}(\bm W_{\infty,-1},...,\bm W_{\infty,-K};\tau(\bm W_{\infty,-1},...,\bm W_{\infty,-K}))
 \] is measurable, i.e., a random variable.

\end{remark}

\begin{remark}{\underline{$\big($Measurability of
$\sup_{\lambda\in \Psi}
|\hat H'_{n,K}(\lambda)-\hat H'_{\infty,K}(\lambda)|$ for any compact set $\Psi$
$\big)$}}
\label{remark:measurablesup}

Recall that from Remarks~\ref{remark:measurabilitywrtcv}
and~\ref{reark:measurablelimit}
that there exist deterministic maps
\(\Phi_n,\Phi_\infty:
\mathbb R^{Kp}\times[0,\infty)\longrightarrow\mathbb R
\) such that
\[
\hat H'_{n,K}(\lambda)
=
\Phi_n(\mathbf Z_n,\lambda),
\qquad
\hat H'_{\infty,K}(\lambda)
=
\Phi_\infty(\mathbf Z_\infty,\lambda),
\]
where
\[
\mathbf Z_n
=
(\mathbf W_{n,-1}^\top,\ldots,\mathbf W_{n,-K}^\top)^\top,
\qquad
\mathbf Z_\infty
=
(\mathbf W_{\infty,-1}^\top,\ldots,\mathbf W_{\infty,-K}^\top)^\top .
\]
Thus both criteria may be regarded as functions of the single random vector
\[
(\mathbf Z_n,\mathbf Z_\infty):
(\Omega,\mathcal F)
\longrightarrow
\bigl(
\mathbb R^{Kp}\times\mathbb R^{Kp},
\mathcal B(\mathbb R^{Kp}\times\mathbb R^{Kp})
\bigr),
\]
which is Borel measurable since both $\mathbf Z_n$ and $\mathbf Z_\infty$
are Borel measurable. Since $\Phi_n$ and $\Phi_\infty$ are proper normal integrands, they are
jointly measurable with respect to
\(
\mathcal B(\mathbb R^{Kp}
\times[0,\infty)).
\) Moreover, the coordinate projections
\[
\pi_1(\mathbf z_1,\mathbf z_2,\lambda)
=
(\mathbf z_1,\lambda),
\qquad
\pi_2(\mathbf z_1,\mathbf z_2,\lambda)
=
(\mathbf z_2,\lambda),
\]
are continuous, hence Borel measurable. Therefore,
\[
\Phi_n\circ\pi_1,
\qquad
\Phi_\infty\circ\pi_2,
\]
are Borel measurable on
$\mathbb R^{Kp}\times\mathbb R^{Kp}\times[0,\infty)$.
Consequently,
\((\mathbf z_1,\mathbf z_2,\lambda)
\longmapsto
\Bigl(
\Phi_n(\mathbf z_1,\lambda),
\Phi_\infty(\mathbf z_2,\lambda)
\Bigr)\) is an $\mathbb R^2$-valued Borel measurable map. Since
\((x,y)\longmapsto |x-y|
\) is continuous, the function
\[
F(\mathbf z_1,\mathbf z_2,\lambda)
:=
\left|
\Phi_n(\mathbf z_1,\lambda)
-
\Phi_\infty(\mathbf z_2,\lambda)
\right|
\]
is Borel measurable on
$\mathbb R^{Kp}\times\mathbb R^{Kp}\times[0,\infty)$. Define
\(v_{\Psi}(\mathbf z_1,\mathbf z_2)
:=
\sup_{\lambda\in\Psi}
F(\mathbf z_1,\mathbf z_2,\lambda).
\) Since
\(\lambda
\longmapsto
F(\mathbf z_1,\mathbf z_2,\lambda)
\) is continuous on the compact interval $\Psi\subset[0,\infty)$,
\[
v_{\Psi}(\mathbf z_1,\mathbf z_2)
=
\sup_{\lambda\in\mathbb Q\cap\Psi}
F(\mathbf z_1,\mathbf z_2,\lambda).
\]
As $\mathbb Q\cap\Psi$ is countable and each
$F(\cdot,\cdot,\lambda)$ is Borel measurable,
$v_{\Psi}$ is Borel measurable on
$\mathbb R^{Kp}\times\mathbb R^{Kp}$.
Therefore,
\[
v_{\Psi}(\mathbf Z_n,\mathbf Z_\infty)
=
\sup_{\lambda\in \Psi}
\left|
\hat H'_{n,K}(\lambda)
-
\hat H'_{\infty,K}(\lambda)
\right|
\]
is an
$\mathcal F/\mathcal B(\mathbb R)$-measurable, i.e.,
a real-valued random variable.

\end{remark}

\begin{remark}{\underline{$\big($Measurability of \(\sup_{\substack{\|\mathbf u'-\mathbf u\|<\delta\\
\max\{\|\mathbf u'\|,\|\mathbf u\|\}\le\psi}}
\Big|
V_n(\mathbf u',\hat\lambda_{n,K})
-
V_n(\mathbf u,\hat\lambda_{n,K})
\Big|
\)
$\big)$}}\label{remark:measurableequicontinuous} Fix $\delta,\psi>0$ and define
\[
A_{\delta,\psi}
:=
\Bigl\{
(\mathbf u,\mathbf u')
\in\mathbb R^p\times\mathbb R^p:
\|\mathbf u'-\mathbf u\|<\delta,\,
\max\{\|\mathbf u\|,\|\mathbf u'\|\}\le\psi
\Bigr\}.
\]

Let
\[
C_{\delta,\psi}
:=
\overline{A_{\delta,\psi}}
=
\Bigl\{
(\mathbf u,\mathbf u')
\in\mathbb R^p\times\mathbb R^p:
\|\mathbf u'-\mathbf u\|\le\delta,\,
\max\{\|\mathbf u\|,\|\mathbf u'\|\}\le\psi
\Bigr\}.
\]
By Remark~\ref{remark:measurabilitywrtcv}, there exists a measurable map
$\tau_n:\mathbb R^{Kp}\to[0,\infty)$ such that
\(\hat\lambda_{n,K}
=
\tau_n(\mathbf Z_n).
\) Moreover, by Remark~\ref{remark:measurabilitywrttraining},
there exists a continuous linear map
$\rho:\mathbb R^{Kp}\to\mathbb R^p$
satisfying
\(\mathbf W_n
=
\rho(\mathbf Z_n).
\) Define
\[
\widetilde V_n(\mathbf z,\mathbf u)
=
\frac12\mathbf u^\top\mathbf L_n\mathbf u
-
\rho(\mathbf z)^\top\mathbf u
+
\tau_n(\mathbf z)
\sum_{j=1}^{p}
\Bigl(
|\beta_j+n^{-1/2}u_j|
-
|\beta_j|
\Bigr),
\]
so that
\(V_n(\mathbf u,\hat\lambda_{n,K})
=
\widetilde V_n(\mathbf Z_n,\mathbf u).
\) Next define
\(F_n:
\mathbb R^{Kp}\times\mathbb R^{2p}
\longrightarrow
[0,\infty)
\) by
\[
F_n(\mathbf z,\mathbf u,\mathbf u')
=
\Big|
\widetilde V_n(\mathbf z,\mathbf u')
-
\widetilde V_n(\mathbf z,\mathbf u)
\Big|.
\]
Since $\rho$ is continuous and $\tau_n$ is measurable,
for every fixed $(\mathbf u,\mathbf u')\in\mathbb R^{2p}$,
the map
\(\mathbf z
\longmapsto
F_n(\mathbf z,\mathbf u,\mathbf u')
\) is measurable. Moreover, for every fixed
$\mathbf z\in\mathbb R^{Kp}$,
the map
\((\mathbf u,\mathbf u')
\longmapsto
F_n(\mathbf z,\mathbf u,\mathbf u')
\) is continuous. Hence $F_n$ is a Carath\'eodory integrand and therefore
a normal integrand. Define the constant correspondence
\[
S:\mathbb R^{Kp}\rightrightarrows\mathbb R^{2p},
\qquad
S(\mathbf z):=C_{\delta,\psi}\;\;\text{for all}\;\;\mathbf z\in\mathbb{R}^{Kp}.
\]
Since \(C_{\delta,\psi}\) is closed, \(S\) is closed-valued. Moreover,
for every open set \(O\subset\mathbb R^{2p}\),
\[
S^{-1}(O)
=
\begin{cases}
\mathbb R^{Kp}, & C_{\delta,\psi}\cap O\neq\varnothing,\\
\varnothing, & C_{\delta,\psi}\cap O=\varnothing,
\end{cases}
\]
and hence \(S\) is measurable. Therefore, by Example~14.32 of
\citet{rockafellar2009variational}, the indicator integrand
\[
(\mathbf z,\mathbf u,\mathbf u')
\mapsto
\mathbf I_{C_{\delta,\psi}}(\mathbf u,\mathbf u')
\]
is a normal integrand, where we define:
\[
\mathbf{I}_{C_{\delta,\psi}}(\mathbf u,\mathbf u')
:=
\begin{cases}
0,
&
(\mathbf u,\mathbf u')\in C_{\delta,\psi},
\\[1ex]
+\infty,
&
(\mathbf u,\mathbf u')\notin C_{\delta,\psi},
\end{cases}
\]

Consequently,
\[
G_n:\mathbb{R}^{Kp}\times \mathbb{R}^{2p}\mapsto \bar{\mathbb{R}},\;\;G_n(\mathbf z,\mathbf u,\mathbf u')
:=
-F_n(\mathbf z,\mathbf u,\mathbf u')
+
\mathbf{I}_{C_{\delta,\psi}}(\mathbf u,\mathbf u')
\]
is a normal integrand. Define
\(p_n(\mathbf z)
:=
\inf_{(\mathbf u,\mathbf u')\in\mathbb R^{2p}}
G_n(\mathbf z,\mathbf u,\mathbf u').
\) By Theorem~14.37 of \citet{rockafellar2009variational},
the optimal value function
\(p_n:\mathbb R^{Kp}\to\overline{\mathbb R}
\) is Borel measurable. Since
\[
-F_n(\mathbf z,\mathbf u,\mathbf u')
+
\mathbf I_{C_{\delta,\psi}}(\mathbf u,\mathbf u')
=
+\infty,\;\;\text{for every}\;\;(\mathbf u,\mathbf u')\notin C_{\delta,\psi},
\]
hence the points outside $C_{\delta,\psi}$ do not affect the infimum. Therefore,
\[
\begin{aligned}
p_n(\mathbf z)
&=
\inf_{(\mathbf u,\mathbf u')\in\mathbb R^{2p}}
\Bigl[
G_n(\mathbf z,\mathbf u,\mathbf u')\Bigr]=
\inf_{(\mathbf u,\mathbf u')\in C_{\delta,\psi}}
\bigl[
-F_n(\mathbf z,\mathbf u,\mathbf u')
\bigr]=
-\sup_{(\mathbf u,\mathbf u')\in C_{\delta,\psi}}
F_n(\mathbf z,\mathbf u,\mathbf u').
\end{aligned}
\]
Since $C_{\delta,\psi}$ is compact and
$F_n(\mathbf z,\cdot,\cdot)$ is continuous,
\(p_n(\mathbf z)
=
-\sup_{(\mathbf u,\mathbf u')\in C_{\delta,\psi}}
F_n(\mathbf z,\mathbf u,\mathbf u')
\in\mathbb R,
\) and hence \(p_n\) is in fact a
$\mathcal B(\mathbb R^{Kp})/\mathcal B(\mathbb R)$-measurable function.
Moreover,
\[
\sup_{(\mathbf u,\mathbf u')\in A_{\delta,\psi}}
F_n(\mathbf z,\mathbf u,\mathbf u')
=
\sup_{(\mathbf u,\mathbf u')\in C_{\delta,\psi}}
F_n(\mathbf z,\mathbf u,\mathbf u').
\]
Therefore,
\(\mathbf z
\longmapsto
\sup_{(\mathbf u,\mathbf u')\in A_{\delta,\psi}}
F_n(\mathbf z,\mathbf u,\mathbf u')
\) is Borel measurable. Define
\[
m_n:\mathbb R^{Kp}\to\mathbb R,
\qquad
m_n(\mathbf z)
:=
\sup_{(\mathbf u,\mathbf u')\in A_{\delta,\psi}}
F_n(\mathbf z,\mathbf u,\mathbf u').
\]
By the preceding argument,
$m_n$ is
$\mathcal B(\mathbb R^{Kp})/\mathcal B(\mathbb R)$-measurable. Since
\(\mathbf Z_n:
(\Omega,\mathcal F)
\longrightarrow
(\mathbb R^{Kp},\mathcal B(\mathbb R^{Kp}))
\) is measurable, the composition
\(m_n\circ \mathbf Z_n:
(\Omega,\mathcal F)
\longrightarrow
(\mathbb R,\mathcal B(\mathbb R))
\) is measurable. Moreover,
\[
\begin{aligned}
(m_n\circ \mathbf Z_n)
&=
m_n(\mathbf Z_n)=
\sup_{(\mathbf u,\mathbf u')\in A_{\delta,\psi}}
F_n(\mathbf Z_n,\mathbf u,\mathbf u')
\\
&=
\sup_{\substack{\|\mathbf u'-\mathbf u\|<\delta\\
\max\{\|\mathbf u'\|,\|\mathbf u\|\}\le\psi}}
\Big|
V_n(\mathbf u',\hat\lambda_{n,K})
-
V_n(\mathbf u,\hat\lambda_{n,K})
\Big|.
\end{aligned}
\]
Therefore,
\(\sup_{\substack{\|\mathbf u'-\mathbf u\|<\delta\\
\max\{\|\mathbf u'\|,\|\mathbf u\|\}\le\psi}}
\Big|
V_n(\mathbf u',\hat\lambda_{n,K})
-
V_n(\mathbf u,\hat\lambda_{n,K})
\Big|
\) is an
$\mathcal F/\mathcal B(\mathbb R)$-measurable, i.e., a random variable.
    
\end{remark}

\begin{remark}{\underline{$\big($Measurability of
\(\sup_{\substack{|\lambda_1-\lambda_2|<\delta\\
\max\{\lambda_1,\lambda_2\}\le\psi}}
\|\hat{\mathbf u}_n(\lambda_1)-\hat{\mathbf u}_n(\lambda_2)\|\)
$\big)$}}
\label{remark:measurabilityequicontinuousun}
Recall that we have, $\hat{\bm{u}}_n(\lambda):=n^{1/2}(\hat{\bm \beta}_{n}(n^{1/2}\lambda)-\bm \beta)$. Fix \(\delta,\psi>0\) and define
\[
A_{\delta,\psi}
:=
\Bigl\{
(\lambda_1,\lambda_2)\in[0,\infty)^2:
|\lambda_1-\lambda_2|<\delta,\,
\max\{\lambda_1,\lambda_2\}\le\psi
\Bigr\}.
\]
Let
\[
C_{\delta,\psi}
:=
\overline{A_{\delta,\psi}}
=
\Bigl\{
(\lambda_1,\lambda_2)\in[0,\infty)^2:
|\lambda_1-\lambda_2|\le\delta,\,
\max\{\lambda_1,\lambda_2\}\le\psi
\Bigr\}.
\]
By Remark~\ref{remark:measurabilitylassopath}, there exists a measurable map
\[
\xi_n:
\mathbb R^{Kp}\times[0,\infty)
\longrightarrow
\mathbb R^p\;\;\text{such that}\;\;
\hat{\mathbf u}_n(\lambda)
=
\xi_n(\mathbf Z_n,\lambda).
\]
Define
\(F_n:
\mathbb R^{Kp}\times[0,\infty)^2
\longrightarrow
[0,\infty)
\) by
\(
F_n(\mathbf z,\lambda_1,\lambda_2)
:=
\Bigl\|
\xi_n(\mathbf z,\lambda_1)
-
\xi_n(\mathbf z,\lambda_2)
\Bigr\|.
\) For every fixed
\((\lambda_1,\lambda_2)\in[0,\infty)^2\),
the map
\(\mathbf z
\longmapsto
F_n(\mathbf z,\lambda_1,\lambda_2)
\) is measurable. Moreover, for every fixed
\(\mathbf z\in\mathbb R^{Kp}\),
the map
\(
(\lambda_1,\lambda_2)
\longmapsto
F_n(\mathbf z,\lambda_1,\lambda_2)
\) is continuous. Hence \(F_n\) is a Carath\'eodory integrand and therefore
a normal integrand.
Define the constant correspondence
\[
S:\mathbb R^{Kp}\rightrightarrows\mathbb R^2,
\qquad
S(\mathbf z):=C_{\delta,\psi}
\quad\text{for all }\mathbf z\in\mathbb R^{Kp}.
\]

Exactly as in Remark~\ref{remark:measurableequicontinuous},
\(S\) is closed-valued and measurable. Therefore, by
Example~14.32 of \citet{rockafellar2009variational},
the indicator integrand
\((\mathbf z,\lambda_1,\lambda_2)
\mapsto
\mathbf I_{C_{\delta,\psi}}(\lambda_1,\lambda_2)
\) is a normal integrand, where
\[
\mathbf I_{C_{\delta,\psi}}(\lambda_1,\lambda_2)
:=
\begin{cases}
0,
&
(\lambda_1,\lambda_2)\in C_{\delta,\psi},
\\[1ex]
+\infty,
&
(\lambda_1,\lambda_2)\notin C_{\delta,\psi}.
\end{cases}
\]
Consequently,
\(G_n:
\mathbb R^{Kp}\times[0,\infty)^2
\longrightarrow
\overline{\mathbb R},
\) defined by
\(
G_n(\mathbf z,\lambda_1,\lambda_2)
:=
-F_n(\mathbf z,\lambda_1,\lambda_2)
+
\mathbf I_{C_{\delta,\psi}}(\lambda_1,\lambda_2),
\) is a normal integrand. Define
\(p_n(\mathbf z)
:=
\inf_{(\lambda_1,\lambda_2)\in[0,\infty)^2}
G_n(\mathbf z,\lambda_1,\lambda_2).
\) By Theorem~14.37 of \citet{rockafellar2009variational},
the function
\(p_n:\mathbb R^{Kp}\to\overline{\mathbb R}\)
is Borel measurable. Moreover,
\(p_n(\mathbf z)
=
-
\sup_{(\lambda_1,\lambda_2)\in C_{\delta,\psi}}
F_n(\mathbf z,\lambda_1,\lambda_2).
\) Since \(C_{\delta,\psi}\) is compact and
\(F_n(\mathbf z,\cdot,\cdot)\) is continuous,
\(p_n(\mathbf z)\in\mathbb R,
\) and hence \(p_n\) is in fact
\(\mathcal B(\mathbb R^{Kp})/\mathcal B(\mathbb R)
\text{-measurable}.
\) Since \(A_{\delta,\psi}\) is dense in \(C_{\delta,\psi}\) and
\(F_n(\mathbf z,\cdot,\cdot)\) is continuous,
\[
\sup_{(\lambda_1,\lambda_2)\in A_{\delta,\psi}}
F_n(\mathbf z,\lambda_1,\lambda_2)
=
\sup_{(\lambda_1,\lambda_2)\in C_{\delta,\psi}}
F_n(\mathbf z,\lambda_1,\lambda_2).
\]
Therefore,
\(m_n(\mathbf z)
:=
\sup_{(\lambda_1,\lambda_2)\in A_{\delta,\psi}}
F_n(\mathbf z,\lambda_1,\lambda_2)
\) is
\(\mathcal B(\mathbb R^{Kp})/\mathcal B(\mathbb R)\)-measurable. Since
\[
\mathbf Z_n:
(\Omega,\mathcal F)
\longrightarrow
(\mathbb R^{Kp},\mathcal B(\mathbb R^{Kp}))
\]
is measurable, the composition
\(m_n\circ\mathbf Z_n
:
(\Omega,\mathcal F)
\longrightarrow
(\mathbb R,\mathcal B(\mathbb R))
\) is measurable. Furthermore,
\[
\begin{aligned}
(m_n\circ\mathbf Z_n)
&=
\sup_{(\lambda_1,\lambda_2)\in A_{\delta,\psi}}
F_n(\mathbf Z_n,\lambda_1,\lambda_2)=
\sup_{\substack{|\lambda_1-\lambda_2|<\delta\\
\max\{\lambda_1,\lambda_2\}\le\psi}}
\Bigl\|
\hat{\mathbf u}_n(\lambda_1)
-
\hat{\mathbf u}_n(\lambda_2)
\Bigr\|.
\end{aligned}
\]
Therefore,
\(
\sup_{\substack{|\lambda_1-\lambda_2|<\delta\\
\max\{\lambda_1,\lambda_2\}\le\psi}}
\Bigl\|
\hat{\mathbf u}_n(\lambda_1)
-
\hat{\mathbf u}_n(\lambda_2)
\Bigr\|
\) is an
\(\mathcal F/\mathcal B(\mathbb R)\)-measurable random variable.

\end{remark}

\subsection{Requisite Lemmas related to Section \ref{sec:crossv}}\label{sec:appA}
For all $k\in \{1,\dots,K\}$, we recall the notations $\bm{S}_{n, -k},\bm{S}_{n, k},\bm{L}_{n, -k},\bm{L}_{n,k},\bm{W}_{n,-k},\bm{W}_{n,k}$ as in Table \ref{tab:unified}.
Denote that, $\Big(\hat{\bm{\beta}}_{n,-k}-\bm{\beta}\Big)=\; \operatorname{Argmin}_{\bm{u}}V_{n,-k}(\bm{u}),$
where,
\begin{align}\label{eqn:objectivelm}
 V_{n,-k}(\bm{u})&= 2^{-1}\bm{u}^\top \bm{L}_{n,-k}\bm{u}-n^{-1/2}(K-1)^{-1/2}K^{1/2}\bm{u}^\top \bm{W}_{n,-k}\nonumber\\
&\;\;\;\;\;\;\;\;\;\;\;\;\;\;\;\;\;\;\;\;\;\;\;\;\;\;\;\;+K(K-1)^{-1}(n^{-1}\lambda_n)\sum_{j=1}^{p}\Big(|\beta_j+u_j|-|\beta_j|\Big).   
\end{align}
\begin{remark}
Due to (C.1)-(C.3), $n^{-1/2}\sum_{i \in I_k}\varepsilon_i\bm{x}_i$ and $n^{-1/2}\sum_{i \notin I_k}\varepsilon_i\bm{x}_i$ are asymptotically normal and hence it's true that 
$\Bigg\{n^{-1/2}\sum_{i \in I_k}\varepsilon_i\bm{x}_i\Bigg\}_{n\geq 1}$ and $\Bigg\{n^{-1/2}\sum_{i \notin I_k}\varepsilon_i\bm{x}_i\Bigg\}_{n\geq 1}$ are tight sequences for all $k\in \{1,\dots, K\}$. Therefore using this observation, for any $\varepsilon>0$, there exists some $M_{\varepsilon}>0$ such that the set $A^\varepsilon$ has probability more than $ (1- \varepsilon)$. We are going to use this set frequently in this section, where we define,
\begin{align}\label{eqn:cvparentsetlm}
A^{\varepsilon}&=\Bigg\{\bigcap_{n \geq 1}\bigcap_{k=1}^{K}\Big\{\big\{\|n^{-1/2}\sum_{i \in I_k}\varepsilon_i\bm{x}_i\|\leq M_{\varepsilon}\big\}\cap\big\{\|n^{-1/2}\sum_{i \notin I_k}\varepsilon_i\bm{x}_i\|\leq M_{\varepsilon}\big\}\Big\}\Bigg\}
\end{align}
\end{remark}

\begin{lemma}\label{lem:crossvalidstronglm}
Suppose the set-up of Theorem \ref{prop:cv} is true. Then for all $\omega \in A^\varepsilon$ and for sufficiently large $n$,we have, $$\max_{k\in \{1,\dots, K\}}\big\|\hat{\bm{\beta}}_{n,-k}(\omega)-\bm{\beta}\big\|\leq 8\Tilde{{\gamma}}_0^{-1}K(K-1)^{-1}\big[n^{-1/2}M_{\varepsilon}+p^{1/2}(n^{-1}\lambda_n)\big].$$
    
\end{lemma}
Proof of Lemma \ref{lem:crossvalidstronglm}:  Recall that, $(\hat{\bm{\beta}}_{n,-k}-\bm{\beta})=\mbox{Argmin}_{\bm{v}}V_{{n,-k}}(\bm{v})$ where, $V_{{n,-k}}(\bm{v})$ is given by (\ref{eqn:objectivelm}). Therefore on the set $\big\{\bm{v}:\|\bm{v}\|>8\Tilde{{\gamma}}_0^{-1}K(K-1)^{-1}\big[n^{-1/2}M_{\varepsilon}+p^{1/2}(n^{-1}\lambda_n)\big]\big\}$ and from (\ref{eqn:objectivelm}) we have,
\begin{align*}
&V_{{n,-k}}(\bm{v})\geq 4^{-1}\Tilde{{\gamma}}_0 \|\bm{v}\|^2-K(K-1)^{-1}p^{1/2}(n^{-1}\lambda_n)\|\bm{v}\|-K(K-1)^{-1}\|\bm{v}\|\big\|n^{-1}\sum_{i \notin I_k}\varepsilon_i\bm{x}_i\big\|\\
&\geq 4^{-1}\Tilde{{\gamma}}_0 \|\bm{v}\|\Big\{\|\bm{v}\|-4\Tilde{{\gamma}}_0^{-1}K(K-1)^{-1}\big[n^{-1/2}M_{\varepsilon}+p^{1/2}(n^{-1}\lambda_n)\big]\Big\}\\
&>8^{-1}\Tilde{{\gamma}}_0\|\bm{v}\|^2>0,
\end{align*}
for sufficiently large $n$. Now since, $V_{{n,-k}}(\bm{0})=0$, therefore the minimizer can't lie in this set $\big\{\bm{v}:\|\bm{v}\|>8\Tilde{{\gamma}}_0^{-1}K(K-1)^{-1}\big[n^{-1/2}M_{\varepsilon}+p^{1/2}(n^{-1}\lambda_n)\big]\big\}$. Hence the proof is complete. \hfill $\square$

\begin{lemma}\label{lem:crossvalidstronginftylm}
Consider the same set-up as in Lemma \ref{lem:crossvalidstronglm}. Additionally assume that $n^{-1}\lambda_n>4\Tilde{\gamma}_0^{-1}$ where $\Tilde{\gamma}_0$ is the smallest eigen values of $\bm{L}$. Then for every $\omega \in A^{\varepsilon}$ as in (\ref{eqn:cvparentsetlm}) and for sufficiently large $n$ we have,
$$\max_{k\in \{1,\dots, K\}}\|\hat{\beta}_{{n,-k}}(\omega)\|_{\infty} \leq (n^{-1}\lambda_n)^{-1}.$$    
\end{lemma}
Proof of Lemma \ref{lem:crossvalidstronginftylm}: This lemma follows in the same line as in part (b) of Theorem 2.2 of \citet{chatterjee2011strong} by considering the set $A^\varepsilon$ as in equation (\ref{eqn:cvparentsetlm}). \hfill $\square$
\begin{lemma}\label{lem:crossvalidstrongboundawaylm}
Consider the same set-up as in Lemma \ref{lem:crossvalidstronglm}. Suppose that the sequence $\{n^{-1}\lambda_n\}_{n\geq 1}$ is such that $\tau<n^{-1}\lambda_n<M$ for some $0 < \tau < 1$
and $M> 1$. Then on the set $A^{\varepsilon}$, there exists some $\zeta>0$ (independent of $\varepsilon$ and $n$) such that for sufficiently large $n$,
$$\min_{k\in \{1,\dots, K\}}\big\|\hat{\bm{\beta}}_{n,-k}-\bm{\beta}\big\| > \zeta.$$
\end{lemma}

Proof of Lemma \ref{lem:crossvalidstrongboundawaylm}: Fix $k\in \{1,2,..,K\}$ and $\omega \in A^{\varepsilon}$. Note that $(\hat{\bm{\beta}}_{n,-k}-\bm{\beta})=\mbox{Argmin}_{\bm{v}}V_{{n,-k}}(\bm{v})$ where, $V_{{n,-k}}(\bm{v})$ as in (\ref{eqn:objectivelm}). Again for any $\omega \in A^{\varepsilon}$, $\big\|n^{-1}\sum_{i \notin I_k}\varepsilon_i\bm{x}_i\big\|\leq \min\{\frac{\tau}{2},\frac{Mp^{1/2}}{2}\}$, for sufficiently large $n$. Now on the set $\big\{\bm{v}:\|\bm{v}\| \leq \zeta \big\}$ and with the choice of $\zeta$ given as,
$$\zeta= \min\Big\{\frac{\|\bm{\beta}\|K\tau}{[K\tau+4(K-1)\tilde{\gamma}_1\|\bm{\beta}\|]}\ ,\ \frac{\|\bm{\beta}\|K^2\tau^3}{3Mp^{1/2}\big[K\tau+4(K-1)\tilde{\gamma}_1\|\bm{\beta}\|\big]^2}\Big\},$$ we have for sufficiently large $n$,
\begin{align*}
&V_{{n,-k}}(\bm{v})\\
&\geq 4^{-1}\Tilde{{\gamma}}_0 \|\bm{v}\|^2-K(K-1)^{-1}p^{1/2}(n^{-1}\lambda_n)\|\bm{v}\|-K(K-1)^{-1}\|\bm{v}\|\big\|n^{-1}\sum_{i \notin I_k}\varepsilon_i\bm{x}_i\big\|\\
&\geq \|\bm{v}\|\Big[4^{-1}\Tilde{\gamma}_0\|\bm{v}\|-2^{-1}MK(K-1)^{-1}p^{1/2}-MK(K-1)^{-1}p^{1/2}\Big]\\
&>-\frac{K^{3}\tau^{3}\|\bm{\beta}\|}{2(K-1)\Big[K\tau+4(K-1)\Tilde{\gamma}_1\|\bm{\beta}\|\Big]^2}.
\end{align*}
Hence we have
\begin{align}\label{eqn:cvlowerlm}
\inf_{\{\bm{v}:\|\bm{v}\| \leq \zeta \}}V_{{n,-k}}(\bm{v}) \geq -\frac{K^{3}\tau^{3}\|\bm{\beta}\|}{2(K-1)\Big[K\tau+4(K-1)\Tilde{\gamma}_1\|\bm{\beta}\|\Big]^2}  
\end{align}
Now define $\bm{v}_0=-\frac{K\tau}{K\tau+4(K-1)\Tilde{\gamma}_1\|\bm{\beta}\|}\bm{\beta}$. From assumption on $\tau$, it's easy to see that, $\|\bm{v}_0\|>\zeta$ and hence from (\ref{eqn:objectivelm}) for sufficiently large $n$ we have
\begin{align}\label{eqn:cvupperlm}
&\inf_{\{\bm{v}:\|\bm{v}\|>\zeta\}}V_{{n,-k}}(\bm{v})\nonumber\\
&\leq V_{{n,-k}}(\bm{v}_0)\nonumber\\
&\leq\; 2\Tilde{\gamma}_1\|\bm{v}_0\|^2+K(K-1)^{-1}\|\bm{v}_0\|\Big(\|n^{-1}\sum_{i \notin I_k}\varepsilon_i\bm{x}_i\|\Big)\nonumber\\
&\;\;\;\;\;\;\;\;\;\;\;\;\;\;\;\;\;\;\;\;\;\;\;\;\;\;\;\;\;\;\;\;\;\;\;\;\;\;\;+K(K-1)^{-1}(n^{-1}\lambda_n)\sum_{j=1}^{p}[|\beta_{{0j}}+v_{0j}|-|\beta_{{0j}}|]\nonumber\\
&\leq\; \frac{2\Tilde{\gamma}_1K^{2}\tau^{2}\|\bm{\beta}\|^2}{\Big[K\tau+4(K-1)\Tilde{\gamma}_1\|\bm{\beta}\|\Big]^2}+\frac{K^2\tau^2(K-1)^{-1}\|\bm{\beta}\|}{2\Big[K\tau+4(K-1)\Tilde{\gamma}_1\|\bm{\beta}\|\Big]}-\frac{K^{2}\tau^{2}(K-1)^{-1}\|\bm{\beta}\|}{\Big[K\tau+4(K-1)\Tilde{\gamma}_1\|\bm{\beta}\|\Big]}\nonumber\\
&< -\frac{K^{3}\tau^{3}\|\bm{\beta}\|}{2(K-1)\Big[K\tau+4(K-1)\Tilde{\gamma}_1\|\bm{\beta}\|\Big]^2}.
\end{align}
Now comparing (\ref{eqn:cvupperlm}) and (\ref{eqn:cvlowerlm}), the proof is now complete. \hfill $\square$
\begin{lemma}\label{lem:crossvalidboundlm}
Consider the same set up as in Lemma \ref{lem:crossvalidstronglm} and assume that
$n^{-1/2}\lambda_n \leq \eta$ for all $n$, for some $\eta  \in [0,\infty)$. Then on the set $A^{\varepsilon}$ for sufficiently large $n$ we have,
$$\max_{k\in\{1,\dots, K\}}(n-m)^{1/2}\big\|\hat{\bm{\beta}}_{n,-k}-\bm{\beta}\big\| \leq (\Tilde{\gamma}_0/8)^{-1}\big\{M_{\varepsilon}+K^{1/2}(K-1)^{-1/2} \eta p^{1/2}\big\}.$$
 \end{lemma}
Proof of Lemma \ref{lem:crossvalidboundlm}: Fix $k \in \{1,\dots, K\}$. Then note that, $\hat{\bm{u}}_{n,-k}=(n-m)^{1/2}\Big(\hat{\bm{\beta}}_{n,-k}-\bm{\beta}\Big)= \operatorname*{Argmin}_{\bm{u}} V_{n,-k}(\bm{u})$ where 
\begin{align*}
V_{n,-k}(\bm{u})=\;&2^{-1}\bm{u}^\top \bm{L}_{n, -k}\bm{u}-\bm{u}^\top \bm{W}_{n, -k} +\lambda_n\sum_{j=1}^{p}\Big\{|\beta_{j}+(n-m)^{-1/2}u_j|-|\beta_{j}|\Big\},
\end{align*}
Then on the set $A^\varepsilon$ we have,
\begin{align*}
V_{n,-k}(\bm{u})\;&\geq (\Tilde{\gamma}_0/4)\|\bm{u}\|^2-\|\bm{u}\|\|\bm{W}_{n,-k}\|-[(n-m)^{-1/2}\lambda_n]p^{1/2}\|\bm{u}\|\\
&\geq (\Tilde{\gamma}_0/4)\|\bm{u}\|\big[\|\bm{u}\|-(\Tilde{\gamma}_0/4)^{-1}\big(\|\bm{W}_{n,-k}\|+ K^{1/2}(K-1)^{-1/2} \eta p^{1/2}\big)\big]\\
&\geq (\Tilde{\gamma}_0/8)\|\bm{u}\|^2 > 0,
\end{align*}
for sufficiently large $n$, provided $\|\bm{u}\|>(\Tilde{\gamma}_0/8)^{-1}\big\{M_{\varepsilon}+ K^{1/2}(K-1)^{-1/2} \eta p^{1/2}\big\}$. Now since $V_{n,-k}(\bm{0})=0$, $\hat{\bm{u}}_{n,-k}$ can't lie in the set $\big\{\bm{u}:\|\bm{u}\|>(\Tilde{\gamma}_0/8)^{-1}\big[M_{\varepsilon}+K^{1/2}(K-1)^{-1/2} \eta p^{1/2}\big]\big\}$ for sufficiently large $n$. Therefore the proof is complete. \hfill $\square$
\begin{lemma}\label{lem:crossvalidunboundlm}
Consider the same set-up as in Lemma \ref{lem:crossvalidstronglm} and assume that first $p_0$ components of $\bm{\beta}$ is non-zero. If $n^{-1}\lambda_n\rightarrow 0$ and $n^{-1/2}\lambda_n \rightarrow \infty$, as $n\rightarrow \infty$, then for all $\omega \in A^{\varepsilon}$, $(n-m)^{1/2}\big(\hat{\bm{\beta}}_{n,-k}(\omega) -\bm{\beta}\big) \in \Tilde{B}_{1n}^c\cap \Tilde{B}_{2n}$, for sufficiently large $n$, where
\begin{align*}
\Tilde{B}_{1n}=\;&\Big\{\bm{u}: \|u\| > (8\Tilde{\gamma}_{0}^{-1}p^{1/2})K^{1/2}(K-1)^{-1/2}(n^{-1/2}\lambda_n)\Big\}\\
& \text{and}\;\;\Tilde{B}_{2n}=\Big\{\bm{u}: \operatorname*{\max}_{j=1(1)p_0}|u_j| > (n^{-1/2}\lambda_n)^{3/4} \Big\}.
\end{align*}
\end{lemma}
Proof of Lemma \ref{lem:crossvalidunboundlm}: We will consider everything on the set $A^{\varepsilon}$ and for sufficiently large $n$. Then first of all note that,
\begin{align*}
&\inf_{\bm{u}\in \Tilde{B}_{1n}}V_{n,-k}(\bm{u})\\
&=
\inf_{\bm{u}\in \Tilde{B}_{1n}}\Big[2^{-1}\bm{u}^\top \bm{L}_{n, -k}\bm{u}-\bm{u}^\top \bm{W}_{n, -k} +\lambda_n\sum_{j=1}^{p}\Big\{|\beta_{j}+(n-m)^{-1/2}u_j|-|\beta_{j}|\Big\}\Big]\\
&\geq \inf_{\bm{u}\in \Tilde{B}_{1n}}\|\bm{u}\|\big\{(\Tilde{\gamma}_0/4)\|\bm{u}\|-\|\bm{W}_{n,-k}\| - \lambda_n(n-m)^{-1/2} p^{1/2}\big\}\\
&\geq (8\Tilde{\gamma}_{0}^{-1}p^{1/2})K^{1/2}(K-1)^{-1/2}(n^{-1/2}\lambda_n)\big\{K^{1/2}(K-1)^{-1/2}p^{1/2}(n^{-1/2}\lambda_n)-M_{\varepsilon}\big\},
\end{align*}
which goes to $\infty$ as $n\rightarrow \infty$. Again since $V_{n,-k}(\bm{0})=0, \hat{\bm{u}}_{n,-k}\in \Tilde{B}_{1n}^c$. Now it is left to show that $\inf_{\bm{u}\in \Tilde{B}^c_{2n}}V_{n,-k}(\bm{u})\geq \inf_{\bm{u}\in \Tilde{B}_{2n}}V_{n,-k}(\bm{u})$. To that end, note that,
\begin{align*}
\inf_{\bm{u}\in \Tilde{B}^c_{2n}}V_{n,-k}(\bm{u})\geq \inf_{\bm{u}\in \Tilde{B}^c_{2n}}\bigg[\sum_{j=1}^{p_0}\Big\{(\Tilde{\gamma}_0/4)u_j^2-|u_j|\big(K^{1/2}(K-1)^{-1/2}(n^{-1/2}\lambda_n)+M_{\varepsilon}\big)\Big\}\bigg].
\end{align*}
Now consider the function, $g(y)=c_1y^2-c_2y,\ y\geq 0,\ c_1,c_2>0.$ This function is strictly decreasing on $(0,\frac{c_2}{2c_1})$, strictly increasing on $(\frac{c_2}{2c_1},\infty)$ and attains minimum at $y^*=\frac{c_2}{2c_1}=2\Tilde{\gamma}_0^{-1}\Big[K^{1/2}(K-1)^{-1/2}(n^{-1/2}\lambda_n)+M_{\varepsilon}\Big]$. Again note that, $y_0=(n^{-1/2}\lambda_n)^{3/4}\in \big(0,y^*\big)$ since $n^{-1/2}\lambda_n \rightarrow \infty$ as $n \rightarrow \infty$. Therefore we have, 
\begin{align}\label{eqn:cvshwlm}
\inf_{\bm{u}\in \Tilde{B}^c_{2n}}V_{n,-k}(\bm{u}) & \geq p_0\Big(\frac{\lambda_n}{n^{1/2}}\Big)^{3/4}\Big[(\Tilde{\gamma}_0/4)\Big(\frac{\lambda_n}{n^{1/2}}\Big)^{3/4}-K^{1/2}(K-1)^{-1/2}\Big(\frac{\lambda_n}{n^{1/2}}\Big)-M_{\varepsilon}\Big]\nonumber\\
& \geq -p_0 K^{1/2}(K-1)^{-1/2}(n^{-1/2}\lambda_n)^{7/4}.
\end{align}
Now due to the assumption that $n^{-1}\lambda_n \rightarrow 0$ as $n \rightarrow \infty$, Lemma \ref{lem:crossvalidstronglm} implies that $\|\hat{\bm{\beta}}_{n, -k}-\bm{\beta}\| = o(1)$. Hence we can assume that $n^{-1/2}u_j = o(1)$ for all $j\in \{1,\dots, p_0\}$, which in turn implies that $$\big|\beta_{j}+K^{1/2}(K-1)^{-1/2}n^{-1/2}u_j\big|-\big|\beta_{j}\big|=K^{1/2}(K-1)^{-1/2}n^{-1/2}u_jsgn(\beta_{j}),$$ for large enough $n$. Now consider the vector $$\bm{u}_0=2\Big(-sgn(\beta_{1}),....,-sgn(\beta_{p_0}),0,.....,0\Big)^\top (n^{-1/2}\lambda_n)^{3/4} $$ which clearly lies in $\Tilde{B}_{2n}$. Then denoting the largest eigen value of the leading $p_0\times p_0$ sub-matrix of $\bm{L}$ by $\Tilde{\gamma}_1^*$, we have for sufficiently large $n$,\\
\begin{align}\label{eqn:cvneglm}
&\inf_{\bm{u}\in \Tilde{B}_{2n}}V_{n,-k}(\bm{u})\leq V_{n,-k}(\bm{u}_0)\nonumber\\
&=2^{-1}\bm{u}_0^\top \bm{L}_{n,-k}\bm{u_0}-\bm{u}_0^\top \bm{W}_{n,-k}+ \lambda_n\sum_{j=1}^{p}\Big[|\beta_{j}+(n-m)^{-1/2}u_{0j}|-|\beta_{j}|\Big]\nonumber\\
&\leq 2^{-1}\bm{u}_0^\top \bm{L}_{n,-k}\bm{u}_0+\sum_{j=1}^{p_0}u_{0j}\big[sgn(\beta_{j})K^{1/2}(K-1)^{-1/2}(n^{-1/2}\lambda_n)-W_{n,-k}^{(j)}\big]\nonumber\\
&\leq \Tilde{\gamma}_1^*\|\bm{u}_0\|^2-2p_0(n^{-1/2}\lambda_n)^{3/4}K^{1/2}(K-1)^{-1/2}(n^{-1/2}\lambda_n)+2p_0M_{\varepsilon}(n^{-1/2}\lambda_n)^{3/4}\nonumber\\
&\leq \Tilde{\gamma}_1^*p_0(n^{-1/2}\lambda_n)^{3/2}-2p_0(n^{-1/2}\lambda_n)^{3/4}K^{1/2}(K-1)^{-1/2}(n^{-1/2}\lambda_n)+2p_0M_{\varepsilon}(n^{-1/2}\lambda_n)^{3/4}\nonumber\\
&\leq p_0(n^{-1/2}\lambda_n)^{3/4}\Big[2M_{\varepsilon}+\Tilde{\gamma}_1^*(n^{-1/2}\lambda_n)^{3/4}-2K^{1/2}(K-1)^{-1/2}(n^{-1/2}\lambda_n)\Big]\nonumber\\
& \leq - 1.5p_0K^{1/2}(K-1)^{-1/2}(n^{-1/2}\lambda_n)^{7/4}.
\end{align}
Now comparing (\ref{eqn:cvshwlm}) and (\ref{eqn:cvneglm}), we can conclude that $\hat{\bm{u}}_{n,-k} \in \Tilde{B}_{1n}^c\bigcap \Tilde{B}_{2n}.$ and hence the proof is complete. \hfill $\square$

\subsection{Requisite Lemmas related to Section \ref{sec:crossvconv}}\label{sec:appB}
Before holding on to the lemmas, we first define the stochastic equicontinuity of a process (cf. \citet{newey1991uniform}) and strong convexity (cf. \citet{rockafellar2009variational}) of a function.
\begin{definition}\label{def:steqicon}
A sequence of $\mathbb{R}^p-$valued stochastic processes $\{\bm{Y}_n(t)\}_{n\geq 1}$ with the index set $T$, a metric space with metric $d(\cdot, \cdot)$, is said to be \underline{stochastically equicontinuous} if for any fixed $t \in T$ and every $\eta,\varepsilon>0$, there exist a $\delta:=\delta(\eta,\varepsilon, t)>0$ and a natural number $N_0:=N_0(\eta,\varepsilon)$, such that
$$\mathbf{P}\Big[\sup_{\{t^\prime \in T:d(t^\prime,t)<\delta\}}\|\bm{Y}_n(t^\prime)-\bm{Y}_n(t)\|>\eta\Big]<\varepsilon\;\;\text{for all}\; n\geq N_0.$$
\end{definition}
If in some contexts, $t,t^\prime\in \mathcal{K}$, a compact set, we will mention it as stochastic equicontinuity of the process on compact sets. The above probability should ideally be written based on the notion of outer probability, as in \citet{kim1990cube}. However, we will essentially work with measurable maps as discussed in Remark \ref{remark:measurabilitywrttraining}-\ref{remark:measurabilityequicontinuousun} in details. Hence we can keep on using the usual probability and expectation operations. \begin{definition}\label{def:strongconvexity}
 An objective function $G(\bm{u}):\mathbb{R}^p\mapsto \mathbb{R}$ is said to be \underline{strongly convex} if there exists a constant $\mu>0$ such that for any $\bm{u_1},\bm{u_2}\in \mathbb{R}^p$ and for all $\alpha\in [0,1]$, we have,
 $$G\big[\alpha \bm{u}_1+(1-\alpha)\bm{u}_2\big]\leq \alpha G(\bm{u}_1)+(1-\alpha)G(\bm{u}_2)-\frac{\mu}{2}\alpha(1-\alpha)\|\bm{u}_1-\bm{u}_2\|^2.$$
\end{definition}

Now note that, for all $k\in\{1,\dots,K\}$, 
\begin{equation}
   \hat{\bm{u}}_{n, -k}(\lambda) = (n-m)^{1/2}\big(\hat{\bm{\beta}}_{n,-k}\big(n^{1/2}\lambda\big)-\bm{\beta}\big)= \operatorname*{Argmin}_{\bm{u}} \hat{V}_{n,-k}(\bm{u},\lambda)
   \label{eqn:2}
\end{equation}
where,
\begin{equation}
\hat{V}_{n,-k}(\bm{u},\lambda)=(1/2)\bm{u}^\top\bm{L}_{n,-k}\bm{u} -\bm{u}^\top \bm{W}_{n,-k} + n^{1/2}\lambda\Big\{\sum_{j=1}^{p}\big(|(n-m)^{-1/2}u_j+\beta_j|-|\beta_j|\big)\Big\}
\label{eqn:3}
\end{equation}

Now we are ready to state and prove the lemmas.

\begin{lemma}\label{lem:jointstrongconvex}
Suppose, for a fixed $\lambda\in\mathbb{K}\subset (0,\infty)$,
$G_1(\bm{u},\lambda):\mathbb{R}^p\times \mathbb{K}\mapsto \mathbb{R}$ be strongly convex in $\bm{u}$ and 
$G_2(\bm{u},\lambda):\mathbb{R}^p\times \mathbb{K}\mapsto \mathbb{R}$ be convex in $\bm{u}$. Then $F(\bm{u},\lambda):=G_1(\bm{u},\lambda)+G_2(\bm{u},\lambda)$ is strongly convex in $\bm{u}$ for a fixed $\lambda$.
\end{lemma}
Proof of Lemma \ref{lem:jointstrongconvex}: It will follow easily from the definition \ref{def:strongconvexity} and convexity of a function. \hfill $\square$

\begin{lemma}\label{lem:lassostrongconvex}
Under the assumption (C.1), the objective function $\hat{V}_{n,-k}(\bm{u},\lambda)$ as in (\ref{eqn:3}), is strongly convex in $\bm{u}$ for all $k\in\{1,\dots,K\}$ and fixed $\lambda$. 
\end{lemma}
Proof of Lemma \ref{lem:lassostrongconvex}: Note that the second term in RHS of (\ref{eqn:3}), $G_{2_{n,-k}}(\bm{u},\lambda)=n^{1/2}\lambda\Big\{\sum_{j=1}^{p}\big(|(n-m)^{-1/2}u_j+\beta_j|-|\beta_j|\big)\Big\}$ is convex in $\bm{u}$ for fixed $\lambda$. Denote the first term there as $G_{1_{n,-k}}(\bm{u},\lambda)=(1/2)\bm{u}^\top\bm{L}_{n,-k}\bm{u} -\bm{u}^\top \bm{W}_{n,-k}.$ Also let, $\tilde{\gamma}_0$ be the smallest eigen value of the positive definite matrix $L$. Now for fixed $\lambda$, for any $\alpha\in[0,1]$ and $\bm{u}_1,\bm{u}_2\in \mathbb{R}^p$,
\begin{align}\label{eqn:lassostrong}
& \alpha G_{1_{n,-k}}(\bm{u}_1,\lambda)+(1-\alpha)G_{1_{n,-k}}(\bm{u}_2,\lambda)-G_{1_{n,-k}}\big[\alpha\bm{u}_1+(1-\alpha)\bm{u}_2,\lambda\big]\nonumber\\
&=(1/2)\alpha(1-\alpha)\Big\{(\bm{u}_1-\bm{u}_2)^\top \bm{L}_{n,-k}(\bm{u}_1-\bm{u}_2)\Big\}\geq (\tilde{\gamma}_0/4)\alpha(1-\alpha)||\bm{u}_1-\bm{u}_2||^2
\end{align}
Then by definition \ref{def:strongconvexity}, $G_{1_{n,-k}}(\bm{u},\lambda)$ is strongly convex in $\bm{u}$ with $\mu=\tilde{\gamma}_0/2>0$ for fixed $\lambda.$ Now the proof is complete due to Lemma \ref{lem:jointstrongconvex}. \hfill $\square$

\begin{lemma}\label{lem:strongconvexquadgrowth}
Under the assumption (C.1), for any $\bm{u}\in\mathbb{R}^p$, for all $k\in\{1,\dots,K\}$ and fixed $\lambda$, we have;
\begin{align}\label{eqn:strongquadgrowth}
\hat{V}_{n,-k}(\bm{u},\lambda)\geq \hat{V}_{n,-k}(\hat{\bm{u}}_{n,-k}(\lambda),\lambda)+\frac{\tilde{\gamma}_0}{8}||\bm{u}-\hat{\bm{u}}_{n,-k}(\lambda)||^2  
\end{align}
\end{lemma}
Proof of Lemma \ref{lem:strongconvexquadgrowth}: Fix $\bm{u}$ such that $\bm{u}\neq \hat{\bm{u}}_{n,-k}(\lambda)$. Under the existing assumptions, $\hat{\bm{u}}_{n,-k}(\lambda)$ is the unique minimizer of $\hat{V}_{n,-k}(\bm{u},\lambda)$ for fixed $\lambda$. This fact together with Lemma \ref{lem:lassostrongconvex} will give us;
\begin{align*}
\hat{V}_{n,-k}(\hat{\bm{u}}_{n,-k}(\lambda),\lambda)&<\hat{V}_{n,-k}(\frac{1}{2}\hat{\bm{u}}_{n,-k}(\lambda)+\frac{1}{2}\bm{u},\lambda)\\&\leq \frac{1}{2}\hat{V}_{n,-k}(\hat{\bm{u}}_{n,-k}(\lambda),\lambda)+ \frac{1}{2}\hat{V}_{n,-k}(\bm{u},\lambda)-\frac{\tilde{\gamma}_0}{16}||\bm{u}-\hat{\bm{u}}_{n,-k}(\lambda)||^2    
\end{align*}
Therefore, the proof is complete. \hfill $\square$

 \begin{lemma}\label{lem:stcontwrtu}
For $\bm{u}\in\mathbb{R}^p$, define $V_n(\bm{u},\hat{\lambda}_{n,K})=(1/2)\bm{u}^\top \bm{L}_n\bm{u}-\bm{W}_n^\top \bm{u}+\hat{\lambda}_{n,K}\sum_{j=1}^{p}\big(|\beta_j+n^{-1/2}u_j|-|\beta_j|\big)$ with $\bm{L}_n=n^{-1}\sum_{i=1}^{n}\bm{x}_i{x}_i^\top$, $\bm{W}_n=n^{-1/2}\sum_{i=1}^{n}\varepsilon_i\bm{x}_i$ and $\bm{\beta}$ being a fixed quantity. Then under the assumption (C.1), (C.2) and (C.3), $\{V_n(\cdot,\hat{\lambda}_{n,K})\}_{n\geq 1}$ is stochastically equicontinuous on compact set. More precisely, for every $\varepsilon,\eta,\psi>0$, there exists a $\delta:=\delta(\varepsilon,\eta,\psi)>0$ such that
$$\mathbf{P}\Big(\sup_{\{\|\bm{u}^\prime-\bm{u}\|<\delta\},\{\max\{\|\bm{u}^\prime\|,\|\bm{u}\|\}\leq \psi\}}|V_n(\bm{u}^\prime,\hat{\lambda}_{n,K})-V_n(\bm{u},\hat{\lambda}_{n,K})|>\eta\Big)<\varepsilon,$$
for large enough $n$.
\end{lemma}
Proof of Lemma \ref{lem:stcontwrtu}: Fix $\bm{u}\in \mathbb{K}(\bm{u})\subset \mathbb{R}^p$, a compact set. For some $\varepsilon,\eta>0$, choose a $\delta:=\delta(\varepsilon,\eta,\bm{u})$ such that we consider all such $\bm{u}^\prime\in\mathbb{K}(\bm{u})$ for which $\|\bm{u}^\prime-\bm{u}\|<\delta$. Now due to assumptions (C.2) and (C.3), it's easy to prove that $\bm{W}_n\xrightarrow{d} \bm{W}_\infty$, with $\bm{W}_\infty\sim N_p(\bm{0},\bm{S})$. Therefore for every $\varepsilon>0$, there exists $0<M^\prime:=M^\prime(\varepsilon)<\infty$ such that, 
\begin{align}\label{eqn:stcontwrtu}
 \mathbf{P}\big(\|\bm{W}_n\|>M^\prime(\varepsilon)\big)<\varepsilon/2.   
\end{align}
Similarly, due to Theorem 4.1, there exists $0<M^{\prime\prime}:=M^{\prime\prime}(\varepsilon)<\infty$ such that, 
\begin{align}\label{eqn:tightcv}
   \mathbf{P}(n^{-1/2}\hat{\lambda}_{n,K}>M^{\prime\prime}(\varepsilon))<\varepsilon/2. 
\end{align}
Now we see that,
\begin{align}\label{eqn:difference}
&V_n(\bm{u}^\prime,\hat{\lambda}_{n,K})-V_n(\bm{u},\hat{\lambda}_{n,K})\nonumber\\
&=(1/2)\Big[(\bm{u}^\prime-\bm{u})^\top \bm{L}_n(\bm{u}^\prime-\bm{u})+2(\bm{u}^\prime-\bm{u})^\top \bm{L}_n\bm{u}\Big]-\bm{W}_n^\top [\bm{u}^\prime-\bm{u}]\nonumber\\
&\;\;\;\;\;\;\;\;\;\;\;\;\;\;\;\;\;\;\;\;\;\;\;\;\;\;\;\;\;\;\;\;\;\;\;\;\;\;\;\;\;\;\;\;\;\;\;\;\;\;\;\;\;\;+\hat{\lambda}_{n,K}\sum_{j=1}^{p}\Big[|\beta_j+n^{-1/2}u^\prime_j|-|\beta_j+n^{-1/2}u_j|\Big].
\end{align}
Since $\bm{u}$ lies in a compact set, assume there exists a $\psi>0$ such that $\|\bm{u}\|\leq \psi$. Therefore from (\ref{eqn:difference}) we have,
\begin{align}\label{eqn:towardstcont}
&\sup_{\{\|\bm{u}^\prime-\bm{u}\|<\delta\},\{\max\{\|\bm{u}^\prime\|,\|\bm{u}\|\}\leq \psi\}}|V_n(\bm{u}^\prime,\hat{\lambda}_{n,K})-V_n(\bm{u},\hat{\lambda}_{n,K})|\nonumber\\
&\leq \tilde{\gamma}_1\delta(\delta+2\psi)+\delta\|\bm{W}_n\|+(n^{-1/2}\hat{\lambda}_{n,K})p^{1/2}\delta
\end{align}
Now considering, $\delta=\frac{1}{2}\min\Big\{\Big(\frac{\eta}{2\tilde{\gamma}_1}\Big)^{1/2},\frac{\eta}{4\tilde{\gamma}_1\psi},\frac{\eta}{2M^\prime(\varepsilon)},\frac{\eta}{2p^{1/2}M^{\prime\prime}(\varepsilon)}\Big\}$ and from (\ref{eqn:stcontwrtu}) and (\ref{eqn:tightcv}), for large enough $n$ we have,
\begin{align}\label{eqn:final1232}
\mathbf{P}\Big(\sup_{\{\|\bm{u}^\prime-\bm{u}\|<\delta\},\{\max\{\|\bm{u}^\prime\|,\|\bm{u}\|\}\leq \psi\}}|V_n(\bm{u}^\prime,\hat{\lambda}_{n,K})-V_n(\bm{u},\hat{\lambda}_{n,K})|>\eta\Big)<\varepsilon. 
\end{align}
Therefore our proof is complete. \hfill $\square$

\begin{lemma}\label{lem:argmin}
Suppose that $\{\bm{U}_n(\cdot)\}_{n\geq 1}$ and $\bm{U}_{\infty}(\cdot)$ are convex stochastic processes on $\mathbb{R}^p$ such that $\bm{U}_{\infty}(\cdot)$ has almost surely unique minimum $\xi_{\infty}$. Also assume that\\
(a) every finite dimensional distribution of $\bm{U}_n(\cdot)$ converges to that of $\bm{U}_{\infty}(\cdot)$, that is, for any natural number $k$ and for any $\{\bm{t}_1,..,\bm{t}_k\}\subset \mathbb{R}^p$, we have  $(\bm{U}_n(\bm{t}_1),...\bm{U}_n(\bm{t}_k))\xrightarrow{d}(\bm{U}_{\infty}(\bm{t}_1),...\bm{U}_{\infty}(\bm{t}_k)).$\\ 
(b) $\{\bm{U}_n(\cdot)\}_{n\geq 1}$ is stochastically equicontinuous on compacta, i.e., for every $\varepsilon, \eta, M>0$, there exists a $\delta:=\delta(\varepsilon,\eta,M)>0$ such that for large enough $n$,
$$\mathbf{P}\Bigg(\sup_{\{\|\bm{r}-\bm{s}\|<\delta, \max \{\|\bm{r}\|,\|\bm{s}\|\}<M\}}\big|\bm{U}_n(r)-\bm{U}_n(s)\big|>\eta\Bigg)<\varepsilon.$$
Then we have $\operatorname*{Argmin}_{\bm{t}}\bm{U}_n(\bm{t}) \xrightarrow{d} \xi_{\infty}$.
\end{lemma}
Proof of Lemma \ref{lem:argmin}. This lemma is stated and proved in \citet{choudhury2024bootstrapping}. \hfill $\square$

\subsection{Requisite Lemmas related to Section \ref{sec:bootstrap}}\label{sec:appCBoot}
\begin{lemma}\label{lem:F-N}
Suppose $Y_1,\dots,Y_n$ are zero mean independent random variables with $\mathbf{E}(|Y_i|^t)< \infty$ for $i\in \{1,\dots,n\}$ and $S_n = \sum_{i = 1}^{n}Y_i$. Let $\sum_{i = 1}^{n}\mathbf{E}(|Y_i|^t) = \sigma_t$, $c_t^{(1)}=\big(1+\frac{2}{t}\big)^t$ and $c_t^{(2)}=2(2+t)^{-1}e^{-t}$. Then, for any $t\geq 2$ and $x>0$,
\begin{equation*}
\mathbf{P}[|S_n|>x]\leq c_t^{(1)}\sigma_t x^{-t} + exp(-c_t^{(2)}x^2/\sigma_2)
\end{equation*}
\end{lemma}
Proof of Lemma \ref{lem:F-N}.
This inequality was proved in \citet{fuk1971probability}. \hfill $\square$

\begin{lemma}\label{lem:pointargmin}
Let $C\subseteq \mathbb{R}^p$ be open convex set and let $f_n:C\rightarrow \mathbb{R}$, $n\geq 1$, be a sequence of convex functions such that $\lim_{n\rightarrow \infty}f_n(x)$ exists for all $x\in C_0$ where $C_0$ is a dense subset of $C$. Then $\{f_n\}_{n\geq 1}$ converges pointwise on $C$ and the limit function $$f(x)=\lim_{n\rightarrow \infty}f_n(x) $$ is finite and convex on $C$. Moreover, $\{f_n\}_{n\geq 1}$ converges to $f$ uniformly over any compact subset $K$ of $C$, i.e. $$\sup_{x\in K}|f_n(x)-f(x)|\rightarrow 0,\;\;\; \text{as}\;\; n \rightarrow \infty.$$
\end{lemma}
Proof of Lemma \ref{lem:pointargmin}. This lemma is stated as Theorem 10.8 of \citet{rockafellar1997convex}. \hfill $\square$

\begin{lemma}\label{lem:nearness}
Suppose that  $\{f_n\}_{n\geq 1}$ and $\{g_n\}_{n\geq 1}$ are random convex functions on $\mathbb{R}^p$. The sequence of minimizers are $\{\alpha_n\}_{n\geq 1}$ and $\{\beta_n\}_{n\geq 1}$ respectively, where the sequence $\{\beta_n\}_{n\geq 1}$ is unique. For some $\delta>0$, define the quantities $$\Delta_n(\delta)=\sup_{\|s-\beta_n\|\leq \delta}|f_n(s)-g_n(s)|\;\;\text{and}\;\; h_n(\delta)=\inf_{\|s-\beta_n\|=\delta}g_n(s)-g_n(\beta_n).$$ Then we have, $\Big\{\|\alpha_n-\beta_n\|\geq \delta\Big\}\subseteq \Big\{\Delta_n(\delta)\geq \frac{1}{2}h_n(\delta)\Big\}.$
\end{lemma}
Proof of Lemma \ref{lem:nearness}. This lemma follows from Lemma 2 of \citet{hjort1993asymptotics}. \hfill $\square$

\begin{lemma}\label{lem:asargmin}
Consider the sequence of convex functions $\{f_n:\mathbb{R}^p\rightarrow\mathbb{R}\}_{n\geq 1}$ having the form $$f_n(\bm{u})=\bm{u}^\top\bm{\Sigma}_n \bm{u} + R_n(\bm{u}),$$
where $\bm{\Sigma}_n$ converges almost surely to a positive definite matrix $\bm{\Sigma}$ and $\mathbf{P}\big[\lim_{n\rightarrow \infty}\|R_n(\bm{u})\|=0\big]=1$ for any $u\in \mathbb{R}^p$. Let $\{\alpha_n\}_{n\geq 1}$ be the sequence of minimizers of $\{f_n\}_{n\geq 1}$ over $\mathbb{R}^p$. Then,
\begin{align}\label{eqn:5.5.1}
\mathbf{P}\big(\lim_{n\rightarrow \infty}\|\alpha_n\|=0\big)=1.
\end{align}
\end{lemma}
Proof of Lemma \ref{lem:asargmin}.  This lemma is stated and proved in \citet{choudhury2024bootstrapping}.

\begin{lemma}\label{lem:lil}
Under the assumptions (C.2) and (C.3), we have
\begin{align*}
\|\bm{W}_n\| = o(\log n)\;\; w.p\; 1.
\end{align*}
where $\bm{W}_n=n^{-1/2}\sum_{i=1}^{n}\varepsilon_i\bm{x}_i$.
\end{lemma}
Proof of Lemma \ref{lem:lil}. This lemma follows from Lemma 4.1 of \citet{chatterjee2010asymptotic}. \hfill $\square$

\begin{lemma}\label{lem:asconcentration}
Under the assumptions (C.1)-(C.3), we have
\begin{align}\label{eqn:6.0}
\mathbf{P}\Big[\|(\hat{\bm{\beta}}_n(\hat{\lambda}_{n,K})-\bm{\beta})\|=o\big(n^{-1/2}\log n\big)\Big]=1.
\end{align}
\end{lemma}
Proof of Lemma \ref{lem:asconcentration}.
Note that 
\begin{align}\label{eqn:6.1}
(\log n)^{-1}n^{1/2}(\hat{\bm{\beta}}_n(\hat{\lambda}_{n,K})-\bm{\beta}) = \mbox{Argmin}_{\bm{u}} \Big\{ w_{1n}(\bm{u}) + w_{2n}(\bm{u},\hat{\lambda}_{n,K}) \Big\}
\end{align}
where we define,
$w_{1n}(\bm{u})= (1/2)\bm{u}^\top\bm{L}_n\bm{u} - (\log n)^{-1}\bm{W}_n^\top \bm{u}$ with $\bm{L}_n=n^{-1}\sum_{i=1}^{n}\bm{x}_i\bm{x}_i^\top$ and also we denote $w_{2n}(\bm{u},\hat{\lambda}_{n,K})=(\log n)^{-2}\hat{\lambda}_{n,K}\sum_{j=1}^{p}\left( |\beta_{j}+\frac{u_{j} \log n}{n^{1/2}}|-|\beta_{j}|\right)$. Now observe the following: 
(i) Due to Lemma \ref{lem:lil}, $(\log n)^{-1}\bm{W}_n^\top \bm{u}=o(1)$ almost surely. (ii) Due to assumption (C.1), $\bm{L}_n$ converges pointwise to positive definite matrix $\bm{L}$. (iii) Due Theorem 4.1, $w_{2n}(\bm{u},\hat{\lambda}_{n,K})=o(1)$ almost surely. Therefore we write,
\begin{align}\label{eqn:6.2}
(\log n)^{-1}n^{1/2}(\hat{\bm{\beta}}_{n}- \bm{\beta}) = \mbox{Argmin}_{\bm{u}} \Big[ (1/2)\bm{u}^\top \bm{L}_n \bm{u} + \bm{r}_{n}(\bm{u},\hat{\lambda}_{n,K}) \Big ],  
\end{align}
where $\bm{r}_{n}(\bm{u},\hat{\lambda}_{n,K})=o(1)\;\;w.p\; 1$. Therefore, (\ref{eqn:6.2}) is in the setup of Lemma \ref{lem:asargmin} and hence (\ref{eqn:6.0}) follows. \hfill $\square$

\begin{lemma}\label{lem:var2}
Under the assumptions (C.1)-(C.3), we have
$$\|\tilde{\bm{S}}_n^*-\bm{S}\|=o(1)\;\; w.p\; 1,$$
where $\tilde{\bm{S}}_n^*=n^{-1}\sum_{i=1}^{n}\bm{x}_i\bm{x}^\top_i\Big\{y_i-\bm{x}_i^\top\tilde{\bm{\beta}}_n(\hat{\lambda}_{n,K})\Big\}^2$
\end{lemma}
Proof of Lemma \ref{lem:var2} This result is stated and proved in \citet{choudhury2024bootstrapping} \hfill $\square$

\begin{lemma}\label{lem:bnormal}
Under the assumptions (C.1)-(C.3), we have
\[
\mathcal{L}\big(\bm{W}_n^{*}\mid \mathcal{E}\big)
\xrightarrow{d_*} N\big(0,\bm{S}\big)
\quad \text{almost surely}.
\]
where $\bm{W}_n^{*}=n^{-1/2}\sum_{i=1}^{n}\Big\{y_i-\bm{x}_i^\top\tilde{\bm{\beta}}_n(\hat{\lambda}_{n,K})\Big\}\bm{x}_i\Big(\frac{G_i^*}{\mu_{G^*}}-1\Big)$, $\mathcal{E}$ is the sigma-field generated by $\{\varepsilon_1,..,\varepsilon_n\}$ and $\mathcal{L}(\cdot|\mathcal{E})$ denotes the conditional law or distribution of the underlined random vector given the data.
\end{lemma}
Proof of Lemma \ref{lem:bnormal} This result follows in the similar line of Lemma 4.4 of \citet{chatterjee2010asymptotic}. \hfill $\square$

\subsection{Requisite Lemmas regarding Proposition \ref{prop:C.4}}\label{sec:app4.c}
\begin{lemma}\label{lem:gausswn-}
Recall the notation, $\bm{W}_{n,-k} = (n-m)^{-1/2} \sum_{i \notin I_k} \varepsilon_i \bm{x}_i$ for all $k=1,...,K.$  Then under assumption (C.1),
$$\left(
\bm W_{n,-1},\dots,\bm W_{n,-K}
\right)
\xrightarrow{d}
N\!\left(0,\bm\Sigma^{(-)}\right),$$
where, the positive definite matrix $\bm\Sigma^{(-)}$ has block structure: \(\Sigma^{(-)}_{kk}=\bm S\;\;\text{and}\;\;
\Sigma^{(-)}_{kl}=\frac{K-2}{K-1}\bm S,\quad k\neq l\) with $K\ge 2$.
\end{lemma}
Proof of Lemma \ref{lem:gausswn-}: Note that due to assumption (C.1),
$$\mathrm{Cov}(\bm{W}_{n,-k}, \bm{W}_{n,-k})= \frac{1}{n-m} \sum_{i \notin I_k}\bm{x}_i \bm{x}_i^\top \mathbf{E}[\varepsilon_i^2 ] \to \bm{S}\;\;\text{as}\;n\to\infty.$$
And for $k\neq l$, we have  $$\mathrm{Cov}(\bm{W}_{n,-k}, \bm{W}_{n,-l})= \frac{1}{n-m} \sum_{i \in I_k^c \cap I_l^c} \bm{x}_i \bm{x}_i^\top \mathbf{E}[\varepsilon_i^2 ]\;\;\text{and}\;|I_k^c \cap I_l^c| = n - |I_k \cup I_l| = n - 2m.$$
Next due to assumption (C.1) and since $I_k\cap I_l=\phi$ for all $k\neq l$, we can verify,
\begin{align*}
\frac{1}{n-m} \sum_{i \in I_k^c \cap I_l^c} \bm{x}_i \bm{x}_i^\top\mathbf{E}[\varepsilon_i^2 ]\to \frac{K-2}{K-1}\bm{S}\;\quad\text{as}\;n\to\infty.
\end{align*}

Define the stacked vector in $\mathbb{R}^{Kp}$:
\[
\bm{\mathcal W}_n^{(-)}
:=\Big(\bm{W}_{n,-1}^\top,....,\bm{W}_{n,-K}^\top\Big)^\top
\]

The $(k,l)$ block ($p\times p$) of covariance matrix $\bm \Sigma_n^{(-)}$ is
\[
\mathrm{Cov}(\bm{W}_{n,-k}, \bm{W}_{n,-l}):=\frac{1}{n-m}
\sum_{i\in I_k^c\cap I_l^c}
\bm x_i\bm x_i^\top \mathbf E[\varepsilon_i^2]\;\quad\text{for all}\; k\neq l.
\]
From the earlier derivations,
\(
\bm \Sigma_n^{(-)} \longrightarrow \bm \Sigma^{(-)},\;\;\text{as}\;n\to\infty,
\)
where
\[
\Sigma^{(-)}_{kk}=\bm S,
\qquad
\Sigma^{(-)}_{kl}=\frac{K-2}{K-1}\bm S,\quad k\neq l.
\]

To prove joint Gaussian convergence with a Berry--Esseen rate, it suffices by the
Cram\'er--Wold device to study, for an arbitrary fixed
\(\bm t=(\bm t_1^\top,\dots,\bm t_K^\top)^\top\in\mathbb R^{Kp}\),
the scalar projection
\(
T_n^{(-)} := \bm t^\top \bm{\mathcal W}_n^{(-)}.
\) Hence
\begin{align*}
T_n^{(-)}
&= \sum_{k=1}^K \bm t_k^\top \bm W_{n,-k}= (n-m)^{-1/2}\sum_{i=1}^n \varepsilon_i
\sum_{k=1}^K
\bm t_k^\top \bm x_i \mathbf 1_{\{i\notin I_k\}}= \sum_{i=1}^n \varepsilon_i a_{n,i},
\end{align*}
where
\(
a_{n,i}
:=
(n-m)^{-1/2}
\sum_{k=1}^K
\bm t_k^\top \bm x_i \mathbf 1_{\{i\notin I_k\}}.
\) Thus \(T_n^{(-)}\) is a sum of independent mean-zero scalars. It's easy to establish that,
\[
T_n^{(-)}
\xrightarrow{d}
N\!\left(0,\bm t^\top \bm\Sigma^{(-)} \bm t\right).
\]
Thus by Cramér--Wold,
\[
\left(
\bm W_{n,-1},\dots,\bm W_{n,-K}
\right)
\xrightarrow{d}
N\!\left(0,\bm\Sigma^{(-)}\right).
\]
Therefore the proof is complete. \hfill $\square$

\begin{lemma}\label{lem:equiobj}
Let $U$ be a set. For each $u\in U$, let $T(u)$ be a (possibly $u$-dependent and non-empty) feasible set, and let
\(F:\{(u,t): u\in U,\; t\in T(u)\}\to \mathbb{R}.\) Define
\(\phi(u):=\inf_{t\in T(u)} F(u,t).\) Then
\begin{equation}
\label{eq:minmin-eq}
\inf_{u\in U}\ \inf_{t\in T(u)} F(u,t)
\;=\;
\inf_{\substack{u\in U\\ t\in T(u)}} F(u,t).
\end{equation}
\end{lemma}
Proof of Lemma \ref{lem:equiobj}: Define the joint feasible set
\(\mathcal{S}
:=
\{(u,t): u\in U,\ t\in T(u)\}.\) Fix any $u\in U$. For every $t\in T(u)$,
\[
\inf_{(u',t')\in\mathcal{S}} F(u',t')
\;\le\;
F(u,t).
\]
Taking infimum over $t\in T(u)$ gives:
\[
\inf_{(u',t')\in\mathcal{S}} F(u',t')
\;\le\;
\inf_{t\in T(u)} F(u,t)
=
\phi(u).
\]
Since this holds for every $u$,
\begin{align}\label{eqn:knb}
\inf_{(u',t')\in\mathcal{S}} F(u',t')
\;\le\;
\inf_{u\in U} \phi(u).
\end{align}
We will prove the reverse direction. Let $\varepsilon>0$. By definition of infimum over $\mathcal{S}$, there exists
$(\bar u,\bar t)\in\mathcal{S}$ such that
\[
F(\bar u,\bar t)
\le
\inf_{(u,t)\in\mathcal{S}} F(u,t) + \varepsilon.
\]
By definition of $\phi(\bar u)$,
\[
\phi(\bar u)
=
\inf_{t\in T(\bar u)} F(\bar u,t)
\le
F(\bar u,\bar t)\implies 
\phi(\bar u)
\le
\inf_{(u,t)\in\mathcal{S}} F(u,t) + \varepsilon.
\]
Taking infimum over $u$ and letting $\varepsilon\to 0$,
\begin{align}\label{eqn:mkjl}
\inf_{u\in U} \phi(u)
\le
\inf_{(u,t)\in\mathcal{S}} F(u,t).
\end{align}
Now (\ref{eq:minmin-eq}) is true combining (\ref{eqn:knb}) and (\ref{eqn:mkjl}). \hfill $\square$

\begin{lemma}\label{lem:equivminimizer}
Suppose $\bm W\in\mathbb R^p$ is an absolutely continuous random vector, $L$ is a $p\times p$ symmetric positive definite matrix and $\lambda\ge 0$ is a tuning parameter. Let $\bm u\in \mathbb R^p$ and $\bm t\in\mathbb R^{p-p_0}$ with $p_0<p$. Then as in the notations of Lemma \ref{lem:equiobj}, define:
\begin{align}\label{eqn:xxxc}
\phi(\bm u)
:=
\frac12 \bm u^\top L \bm u - \bm u^\top \bm W
+
\lambda\!\left(
\sum_{j\le p_0} c_j u_j
+
\sum_{j>p_0} |u_j|
\right),\;\;c_j=sgn(\beta_j)\;\text{with}\;sgn(\beta_j)\in\{\pm 1\},
\end{align}
and
\begin{align}\label{eqn:yyyyc}
F(\bm u,\bm t)
:=
\frac12 \bm u^\top L \bm u - \bm u^\top \bm W
+
\lambda\!\left(
\sum_{j\le p_0} c_j u_j
+
\sum_{j>p_0} t_j
\right),\;\text{subject to}\; t_j \ge |u_j|,\;t_j\ge 0\;\text{for all}\;j>p_0.
\end{align}
Then following hold true.
\begin{itemize}
    \item[(a)] If \(\hat{\bm{u}}\) minimizes the  $\phi(\bm u)$, then it induces a minimizer of  $F(\bm u,\bm t)$.
    \item[(b)] If \((\hat u,\hat t)\) minimizes  $F(\bm u,\bm t)$, then \(\hat u\) minimizes $\phi(\bm u)$ and $\hat t_j=|\hat u_j|$. 
\end{itemize}

\end{lemma}

Proof of Lemma \ref{lem:equivminimizer}: If $\lambda=0$, both optimization problems have the same unique minimizer in the $\bm u$-variable, namely $\hat{\bm u}=L^{-1}\bm W.$ Hence the conclusions of parts (a) and (b) follow immediately. Therefore, it suffices to consider the case $\lambda>0$. We recall a basic fact that, for any $x\in\mathbb R$:
\begin{align}\label{eqn:mnbv}
|x|
=
\min_{t\in\mathbb R}
\left\{
t :\;
t\ge 0,\;
-\,t \le x \le t
\right\}.
\end{align}
Fix a $\bm u$ and define $T(\bm u):=\{\bm t\in\mathbb{R}^{p-p_0}:t_j\ge|u_j|,t_j\ge0;j>p_0\}$. Then following Lemma \ref{lem:equiobj} we can write, 
\begin{equation}
\label{eq:key-identity}
\min_{\bm t\in T(\bm u)} F(\bm u,\bm t) \;=\; \phi(\bm u).
\end{equation}
\emph{\bf{Regarding case (a)}}: Assume \(\hat{\bm{u}}\) minimizes   $\phi(\bm u)$. Then: $\hat{\bm{u}}= \arg\min_{\bm{u}} \phi(\bm u).$ Define
\(\hat t_j := |\hat u_j|,\; j>p_0.
\) Then \((\hat {\bm{u}},\hat{\bm{t}})\) is feasible for   $F(\bm u,\bm t)$ in the sense that it satisfies the constraints. Now observing (\ref{eqn:mnbv}) and (\ref{eq:key-identity}) we have, $F(\hat{\bm{u}},\hat{\bm{t}})=\phi(\hat{\bm{u}}).$ For any feasible pair \((\bm u,\bm t)\),
since \(t_j\ge |u_j|\), hence $F(\bm u,\bm t) \ge \phi(\bm u).$ Since $\hat{\bm{u}}$ minimizes $\phi(\bm u)$ we have;
\[
\phi(\bm u)\ge \phi(\hat{\bm{u}})\implies F(\bm u,\bm t)\ge \phi(\hat{\bm{u}})=F(\hat{\bm{u}},\hat{\bm{t}}).
\]
Hence \((\hat{\bm{u}},\hat{\bm{t}})\) is a minimizer of   $F(\bm u,\bm t)$.\\

\emph{\bf{Regarding case (b)}}: Assume \((\hat{\bm{u}},\hat{\bm{t}})\) is optimal for   $F(\bm u,\bm t)$. First, we show that
\(\hat t_j = |\hat u_j|,\; j>p_0.\) Then constraints imply
$t_j \ge |\hat u_j|.$ In the objective function $F(\bm u,\bm t)$, the variable \(t_j\) appears only through the linear term
\(\lambda t_j\), where \(\lambda>0\).
Thus, for fixed \(\hat{\bm{u}}\) and all other \(t_k\) (\(k\neq j\)) fixed,
\(F(\hat{\bm{u}},\bm{t})\) is a strictly increasing function of \(t_j\).  Suppose, for the contradiction assume, that for one such index $j$,
\(\hat t_j > |\hat u_j|.\) Define a new vector \(\tilde t\) by
\[
\tilde t_j := |\hat u_j|,
\quad
\tilde t_k := \hat t_k \quad (k\neq j).
\]
By feasibility condition, \(\tilde{\bm{t}}\) is still feasible. Since only the \(j\)-th component changed,
\[
F(\hat{\bm{u}},\tilde{\bm{t}})
=
F(\hat{\bm{u}},\hat{\bm{t}})
-
\lambda\big(\hat t_j-|\hat u_j|\big)
<
F(\hat{\bm{u}},\hat{\bm{t}}),
\]
because \(\hat t_j>|\hat u_j|\) and \(\lambda>0\). This contradicts the optimality of \((\hat{\bm{u}},\hat{\bm{t}})\). Hence necessarily,
\[
\hat t_j = |\hat u_j|,\; j>p_0.
\] and by \eqref{eq:key-identity},
\(F(\hat{\bm{u}},\hat{\bm{t}})=\phi(\hat{\bm{u}}).\) Now take any \(\bm u\) and set \(t_j=|u_j|\). Then \((\bm u,\bm t)\) is feasible,
so optimality gives
\(F(\hat{\bm{u}},\hat{\bm{t}}) \le F(\bm{u},\bm t).\) Using \eqref{eq:key-identity} again,
\[
\phi(\hat{\bm{u}}) \le \phi(\bm{u}).
\]
Thus \(\hat{\bm{u}}\) minimizes  . Hence the proof is complete. \hfill $\square$

\begin{lemma}\label{lem:moore}
Let $M \in \mathbb{R}^{n \times n}$ be any real matrix, and let $M^\dagger$ denote its Moore--Penrose pseudo inverse. Define
\(P := I - M M^\dagger.\) Then
\[
P \text{ is the orthogonal projection onto } \operatorname{col}(M)^\perp.
\]
Equivalently,
\(\operatorname{Im}(P) = \operatorname{col}(M)^\perp.\)
\end{lemma}

Proof of Lemma \ref{lem:moore}: By definition, the Moore--Penrose pseudo inverse $M^\dagger$ is the unique matrix satisfying:
\begin{align*}
(1)M M^\dagger M = M,\;(2)M^\dagger M M^\dagger = M^\dagger,\;(3)(M M^\dagger)^\top = M M^\dagger,\;(4)(M^\dagger M)^\top = M^\dagger M.
\end{align*}
Note that using property (3),
\[
P^\top = (I - M M^\dagger)^\top = I - (M M^\dagger)^\top = I - M M^\dagger = P.
\]
Now we compute using (1),
\[
(M M^\dagger)^2 = M M^\dagger M M^\dagger= M M^\dagger.
\]
Hence we have:
\[
P^2 = (I - M M^\dagger)^2
= I - 2 M M^\dagger + (M M^\dagger)^2
= I - M M^\dagger = P.
\]
so $P$ is an orthogonal projection matrix. Now we prove:
\(\operatorname{Im}(P) = \operatorname{col}(M)^\perp.\)\\

\emph{\bf{Forward step:} $\operatorname{Im}(P) \subseteq \operatorname{col}(M)^\perp$} Let $y \in \operatorname{Im}(P)$. Then for some $x$,
\[
y = Px = x - M M^\dagger x.
\]
Take any vector in $\operatorname{col}(M)$, say $Mz$. Then
\(\langle Mz, y \rangle = (Mz)^\top y = z^\top M^\top y.
\) This gives rise to,
\[
M^\top y = M^\top(x - M M^\dagger x)
= M^\top x - M^\top M M^\dagger x.
\]

Now use the identity (transpose of (1) and using (3)):
\[
(M M^\dagger M)^\top = M^\top
\;\Rightarrow\;
M^\top M M^\dagger = M^\top \implies 
M^\top y = M^\top x - M^\top x = 0.
\]
Hence:
\(\langle Mz, y \rangle = 0\; \forall z,\)
so
\(y \in \operatorname{col}(M)^\perp.\)\\

\emph{\bf{Backward step}: $\operatorname{col}(M)^\perp \subseteq \operatorname{Im}(P)$.} Let $v \in \operatorname{col}(M)^\perp$. Then
\(v \perp \operatorname{col}(M)
\;\Longleftrightarrow\;
M^\top v = 0.\)\\

\textbf{Fact:}
\(M M^\dagger \text{ is the orthogonal projection onto } \operatorname{col}(M).\)\\

\emph{\bf{Justification of the fact}:} From earlier step, $M M^\dagger$ is symmetric and idempotent, hence an orthogonal projector. For any $x$,
\(M M^\dagger x \in \operatorname{col}(M).\) Conversely, for any $y = Mz$ using (1),
\[
M M^\dagger y = M M^\dagger M z = M z = y.
\]
Thus:
\(\operatorname{Im}(M M^\dagger) = \operatorname{col}(M).\) Now come back to the proof of backward step.\\

Since $v \perp \operatorname{col}(M)$ and $M M^\dagger$ projects onto $\operatorname{col}(M)$, we have:
\(M M^\dagger v = 0.
\) Therefore:
\[
Pv = (I - M M^\dagger)v = v - 0 = v\implies 
v \in \operatorname{Im}(P).
\]
Hence we have,
\[
\operatorname{Im}(P) = \operatorname{col}(M)^\perp.
\]
The proof is complete. \hfill $\square$

\begin{lemma}\label{lem:P-kernel}
Let \(L\in\mathbb R^{p\times p}\) be positive definite and let
\(A\in\mathbb R^{m\times p}\). Define
\(M:=AL^{-1}A^\top\) (not necessarily invertible) with $M^\dagger$ be its usual Moore-Penrose inverse. Also we define that \(P
:=
L^{-1}
-
L^{-1}A^\top M^\dagger AL^{-1}.\)
Then the following are true:
$$(i)\;P^\top=P,\;(ii)\;AP=0,\;(iii)\;PLP=P,\;(iv)\;\operatorname{Im}(P)\subseteq\ker(A),\;(v)\;\operatorname{rank}(P)=\dim\ker(A).$$
Consequently,
\(\operatorname{Im}(P)=\ker(A).\)
\end{lemma}

Proof of Lemma \ref{lem:P-kernel}: Since \(M\) is symmetric positive semi-definite, its Moore--Penrose
inverse \(M^\dagger\) is symmetric. Therefore
\[
P^\top
=
\Bigl(
L^{-1}
-
L^{-1}A^\top M^\dagger AL^{-1}
\Bigr)^\top
=
P,\;\;\text{proving (i)}.
\]
Next,
\[
AP
=
AL^{-1}
-
AL^{-1}A^\top M^\dagger AL^{-1}
=
(I-MM^\dagger)AL^{-1}.
\]
Since \(M=AL^{-1}A^\top\) and \(L^{-1}\) is invertible, we have \(\operatorname{col}(M)=\operatorname{col}(A).\) Hence, for every \(x\),
\[
AL^{-1}x\in\operatorname{col}(M).
\]
Because \(I-MM^\dagger\) is the orthogonal projector onto
\(\operatorname{col}(M)^\perp\) due to Lemma \ref{lem:moore}, we must have, \((I-MM^\dagger)AL^{-1}=0.\) Thus
\(AP=0,\) which proves (ii).\\

For (iii),
\begin{align*}
PLP
&=
\Bigl(
L^{-1}
-
L^{-1}A^\top M^\dagger AL^{-1}
\Bigr)
L
\Bigl(
L^{-1}
-
L^{-1}A^\top M^\dagger AL^{-1}
\Bigr)
\\
&=
L^{-1}
-2L^{-1}A^\top M^\dagger AL^{-1}+
L^{-1}A^\top M^\dagger
(AL^{-1}A^\top)
M^\dagger AL^{-1}.
\end{align*}
Using \(M=AL^{-1}A^\top\) and the Penrose identity
\(M^\dagger M M^\dagger=M^\dagger\),
\[
PLP
=
L^{-1}
-
L^{-1}A^\top M^\dagger AL^{-1}
=
P.
\]
From (ii), for every \(x\),
\(A(Px)=0.\) Hence
\[
\operatorname{Im}(P)\subseteq\ker(A),\;\;\text{proving (iv)}.
\]
To establish (v), let \(B:=AL^{-1/2}.\) Then
\(P
=
L^{-1/2}
\Bigl(
I-B^\top(BB^\top)^\dagger B
\Bigr)
L^{-1/2}.\) Set
\(Q:=B^\top(BB^\top)^\dagger B.\) Since \(BB^\top\) and \((BB^\top)^\dagger\) are symmetric, hence \(Q^\top=Q.\) Moreover,
\[
Q^2
=
B^\top(BB^\top)^\dagger
BB^\top
(BB^\top)^\dagger B
=
Q,
\]
by the Moore-Penrose identity. Therefore \(Q\) is an orthogonal projector. We claim that
\[
\operatorname{Im}(Q)=\operatorname{col}(B^\top).
\]
Indeed, for every \(x\),
\[
Qx
=
B^\top(BB^\top)^\dagger(Bx)
\in
\operatorname{col}(B^\top)\;\implies \operatorname{Im}(Q)\subseteq\operatorname{col}(B^\top).
\]
Conversely, let \(y\in\operatorname{col}(B^\top)\) implying $y=B^\top z$ for some $z$. Then
\[
Qy
=
B^\top(BB^\top)^\dagger(BB^\top)z.
\]
Since \((BB^\top)^\dagger(BB^\top)\) is the orthogonal projector onto
\(\operatorname{col}(BB^\top)=\operatorname{col}(B)\) (similarly as in Lemma \ref{lem:moore}), we write
\[
z=z_\parallel+z_\perp,
\qquad
z_\parallel\in\operatorname{col}(B),
\quad
z_\perp\in\ker(B^\top).
\]
Then
\((BB^\top)^\dagger(BB^\top)z=z_\parallel,\) and
\(B^\top z_\parallel
=
B^\top(z_\parallel+z_\perp)
=
B^\top z
=
y.
\) Hence \(Qy=y\), proving the claim. Therefore
\(I-Q\) is the orthogonal projector onto
\(\operatorname{col}(B^\top)^\perp
=
\ker(B).\) Thus
\[
\operatorname{rank}(I-Q)
=
\dim\ker(B).
\]
Since \(L^{-1/2}\) is invertible,
\(\operatorname{rank}(P)
=
\operatorname{rank}(I-Q)
=
\dim\ker(B).\) Finally,
\(\ker(B)=\ker(A),
\) because \(L^{-1/2}\) is invertible. Hence
\(\operatorname{rank}(P)
=
\dim\ker(A),\) which proves (v).\\

Combining (iv) and (v) we have that,
\[
\operatorname{Im}(P)=\ker(A).
\]
The proof is complete. \hfill$\square$

\begin{lemma}\label{lem:unique-dual-image}
Let \(L\in\mathbb R^{p\times p}\) be positive definite and let
\(A\in\mathbb R^{m\times p}\). Define
\(M:=AL^{-1}A^\top\) (not necessarily invertible). Then
\[
\ker(M)=\ker(A^\top).
\]
Consequently, if \(b\in\operatorname{Im}(M)\) and
\(\mu_1,\mu_2\) satisfy
\(M\mu_1=b,\;\;
M\mu_2=b,\) then
\(A^\top\mu_1=A^\top\mu_2.\) In particular, the vector \(A^\top\mu\) is uniquely determined by the
linear system \(M\mu=b\), even though the solution \(\mu\) need not be
unique.
\end{lemma}

Proof of Lemma \ref{lem:unique-dual-image}: 
Let \(v\in\ker(M)\). Then
\[
0
=
v^\top Mv
=
v^\top AL^{-1}A^\top v
=
(A^\top v)^\top L^{-1}(A^\top v).
\]
Since \(L^{-1}\) is positive definite, it follows that
\(A^\top v=0\). Hence
\(\ker(M)\subseteq\ker(A^\top).\) The reverse inclusion is immediate.
If \(A^\top v=0\), then
\[
Mv
=
AL^{-1}A^\top v
=
0.
\]
Therefore
\(\ker(M)=\ker(A^\top).\) Now let \(\mu_1,\mu_2\) be solutions of \(M\mu=b\). Then
\[
M(\mu_1-\mu_2)=0\;\implies \mu_1-\mu_2\in\ker(M)=\ker(A^\top)\]. Hence
\[
A^\top(\mu_1-\mu_2)=0\;\implies A^\top\mu_1=A^\top\mu_2.
\]
Thus the proof is complete. \hfill $\square$

\begin{lemma}\label{lem:degeneracy-probability-zero}
Suppose $(\bm x_{1},\ldots,\bm x_{K})^\top\in\mathbb R^{Kp}$ be a deterministic vector such that the polynomial
\(Q:\mathbb R^{Kp}\to\mathbb R\)
is not identically zero. Assume that $\bm{Z}=(\bm Z_1,..,\bm Z_K)^\top\in\mathbb R^{Kp}$ has a absolutely continuous distribution with respect to the Lebesgue measure on $\mathbb R^{Kp}$. Then
\[
\mathbf P\!\left[Q(\bm Z)=0
\right]
=
0.
\]
\end{lemma}

Proof of Lemma \ref{lem:degeneracy-probability-zero}: Define the algebraic zero set
\[
\mathcal Z
:=
\left\{
\bm x\in\mathbb R^{Kp}
:
Q(\bm x_1,...,\bm{x}_K)=0
\right\}.
\]
Since $Q$ is a nonzero polynomial identically, we have
\[
\lambda_{Kp}(\mathcal Z)=0,
\]
where $\lambda_{Kp}$ denotes Lebesgue measure on $\mathbb R^{Kp}$. See \citet{caron2005zero} for a proof and \citet{Federer69} for general result in case of analytic functions. Since $\bm Z$ has a absolutely continuous distribution with respect to the Lebesgue measure, it admits a density $f_{\bm Z}$ such that,
\[
\mathbf P\!\bigl(\bm Z\in\mathcal Z\bigr)
=
\int_{\mathcal Z}
f_{\bm Z}(\bm z)\,d\bm z
=
0.
\]
This completes the proof. \hfill $\square$

\subsection{Proofs of Main Results}\label{sec:appC}
In this section, we provide the the proofs of main results viz. Theorem \ref{prop:cv}, Theorem \ref{thm:cvthm}, Proposition \ref{prop:eqicon}, Proposition \ref{prop:C.4},  Theorem \ref{thm: lambdaCVconv}, Theorem \ref{thm:distconvcv} and Theorem \ref{thm:bootconsistency}. 
\subsubsection{Proof of Theorem \ref{prop:cv}}\label{sec:prop3.1}

Note that we need to establish $\mathbf{P}\Big(n^{-1}\hat{\lambda}_{n,K} \rightarrow 0\ \text{as}\ n\ \to\ \infty \Big)=1.$ Recall that, $\hat{\lambda}_{n,K}\in\operatorname*{Argmin}_{\lambda_n} H_{n,K}$ where
\begin{align}\label{eqn:cvpropobjlm}
n^{-1}H_{n,K}&=\; K^{-1}\sum_{k=1}^{K}\Bigg[2^{-1}\Big(\hat{\bm{\beta}}_{n,-k}-\bm{\beta}\Big)^\top \bm{L}_{{n,k}}\Big(\hat{\bm{\beta}}_{n,-k}-\bm{\beta}\Big)-K\Big(\hat{\bm{\beta}}_{n,-k}-\bm{\beta}\Big)^\top \Big[n^{-1}\sum_{i \in I_k}\varepsilon_i\bm{x}_i\Big]\Bigg].
\end{align}
Now for any sequence of penalty parameters $\{\lambda_n\}_{n\geq 1}$, there are three possible cases: (a) When $n^{-1}\lambda_n\rightarrow 0,\; \text{as}\; n \rightarrow \infty$, (b) When $n^{-1}\lambda_n\rightarrow \infty,\; \text{as}\; n \rightarrow \infty$ and (c) When $\{n^{-1}\lambda_n\}_{n\geq 1}$ does not satisfy (a) or (b).\\

Let us first consider the case (a). Then for any $0<\varepsilon^*_1< 1$, for sufficiently large $n$ we have $n^{-1}\lambda_n < \varepsilon^*_1$.
Hence due to Lemma \ref{lem:crossvalidstronglm}, from equation (\ref{eqn:cvpropobjlm}) we have for sufficiently large $n$,
\begin{align}\label{eqn:cvupperzerolm}
n^{-1}H_{n,K}&\leq K^{-1}\sum_{k=1}^{K}\Big[\Tilde{\gamma}_1\Big\|\Big(\hat{\bm{\beta}}_{n,-k}-\bm{\beta}\Big)\Big\|^2+K\Big\|\Big(\hat{\bm{\beta}}_{n,-k}-\bm{\beta}\Big)\Big\| \Big\|n^{-1}\sum_{i \in I_k}\varepsilon_i\bm{x}_i\Big\|\Big]\nonumber\\
&\leq K^{-1}\sum_{k=1}^{K}\Big\{64\Tilde{\gamma}_0^{-2}\Tilde{\gamma}_1K^2(K-1)^{-2}\big(n^{-1/2}M_{\varepsilon}+p^{1/2}\varepsilon^*_1\big)^2\nonumber\\
&\;\;\;\;\;\;\;\;\;\;\;\;\;\;\;\;+8\Tilde{\gamma}_0^{-1}K^2(K-1)^{-1}\big(n^{-1/2}M_{\varepsilon}+p^{1/2}\varepsilon^*_1\big)(n^{-1/2}M_{\varepsilon})\Big\}\nonumber\\
&\leq 8\Tilde{\gamma}_0^{-1}K^2(K-1)^{-1}(2+p^{1/2})\Big\{1+8\Tilde{\gamma}_0^{-1}\Tilde{\gamma}_1(K-1)^{-1}(2+p^{1/2})\Big\}\varepsilon^{*2}_1.
\end{align}
Now consider case (b) i.e. when $n^{-1}\lambda_n\rightarrow\infty$. Then due to Lemma \ref{lem:crossvalidstronginftylm}, from (\ref{eqn:cvpropobjlm}) we have for sufficiently large $n$,
\begin{align}\label{eqn:cvlowerinflm}
n^{-1}H_{n,K}&\geq K^{-1}\sum_{k=1}^{K}\Big[4^{-1}\Tilde{\gamma}_0\Big\|\Big(\hat{\bm{\beta}}_{n,-k}-\bm{\beta}\Big)\Big\|^2-K\Big\|\Big(\hat{\bm{\beta}}_{n,-k}-\bm{\beta}\Big)\Big\| \Big\|n^{-1}\sum_{i \in I_k}\varepsilon_i\bm{x}_i\Big\|\Big]\nonumber\\
&\geq K^{-1}\sum_{k=1}^{K}\Big\{8^{-1}\Tilde{\gamma}_0\|\bm{\beta}\|^2 -K\Big(\frac{3\|\bm{\beta}\|}{2}\Big)(n^{-1/2}M_{\varepsilon})\Big\}\geq \frac{\Tilde{\gamma}_0\|\bm{\beta}\|^2}{16}.
\end{align}
Lastly consider case (c). Then without loss of generality we can assume that $\tau < n^{-1}\lambda_n < M$ for all $n$ for some $0< \tau < 1$ and $M> 1$. Otherwise, we can argue through a sub-sequence in the same line. Hence due to Lemma \ref{lem:crossvalidstrongboundawaylm} and from (\ref{eqn:cvpropobjlm}) we have for sufficiently large $n$,
\begin{align}\label{eqn:cvlowerboundawaylm}
n^{-1}H_{n,K}&\geq\; K^{-1}\sum_{k=1}^{K}\Big[4^{-1}\Tilde{\gamma}_0\Big\|\Big(\hat{\bm{\beta}}_{n,-k}-\bm{\beta}\Big)\Big\|^2-K\Big\|\Big(\hat{\bm{\beta}}_{n,-k}-\bm{\beta}\Big)\Big\| \Big\|n^{-1}\sum_{i \in I_k}\varepsilon_i\bm{x}_i\Big\|\Big]\geq \frac{\Tilde{\gamma}_0\zeta^2}{16}   ,
\end{align}
where $\zeta= \min\Big\{\frac{\|\bm{\beta}\|K\tau}{[K\tau+4(K-1)\gamma_1\|\bm{\beta}\|]}\ ,\ \frac{\|\bm{\beta}\|K^2\tau^3}{3Mp^{1/2}\big[K\tau+4(K-1)\gamma_1\|\bm{\beta}\|\big]^2}\Big\}$ as defined in the proof of Lemma \ref{lem:crossvalidstrongboundawaylm}. Now since $\varepsilon^*_1$ can be arbitrarily small, hence comparing (\ref{eqn:cvupperzerolm}), (\ref{eqn:cvlowerinflm}) and (\ref{eqn:cvlowerboundawaylm}) corresponding to cases (a), (b) and (c),
we can claim that the sequence $\{\hat{\lambda}_{n, K}\}_{n\geq 1}$ should belong to case (a) on the set $A^{\varepsilon}$. Since $\varepsilon > 0$ is arbitrary, the proof is now complete. \hfill $\square$

\subsubsection{Proof of Theorem \ref{thm:cvthm}}\label{sec:thm3.1}

We now fix an $\varepsilon > 0$ and consider our analysis on $A^{\varepsilon}$ of (\ref{eqn:cvparentsetlm}). Recall that, $\hat{\lambda}_{n,K}\in\operatorname*{Argmin}_{\lambda_n} H_{n,K}$ where 
\begin{align*}
H_{n,K}=\;& \sum_{k=1}^{K}\Bigg[2^{-1}(n-m)^{1/2}\Big(\hat{\bm{\beta}}_{n,-k}-\bm{\beta}\Big)^\top \Big\{(n-m)^{-1}m\bm{L}_{n,k}\Big\}(n-m)^{1/2}\Big(\hat{\bm{\beta}}_{n,-k}-\bm{\beta}\Big)\\
&-(n-m)^{1/2}\Big(\hat{\bm{\beta}}_{n,-k}-\bm{\beta}\Big)^\top(n-m)^{-1/2}m^{1/2}\bm{W}_{n,k}\Bigg].
\end{align*}
Note that we have to establish, $\mathbf{P}\Big(\limsup_{n\rightarrow\infty}n^{-1/2}\hat{\lambda}_{n,K}< \infty\Big)= 1.$
Now for any  sequence of penalty parameters $\{\lambda_n\}_{n\geq 1}$, either the sequence $\{n^{-1/2}\lambda_n\}_{n\geq 1}$ is bounded or  $\{n^{-1/2}\lambda_n\}_{n\geq 1}$ diverges to $\infty$ through a sub-sequence. Consider the first situation, i.e. $\{n^{-1/2}\lambda_n\}_{n\geq 1}$ is bounded, say by $\eta \in [0, \infty)$. Then assuming $M_{\varepsilon}^{(1)}=(\Tilde{\gamma}_0/8)^{-1}\big\{M_{\varepsilon}+K^{1/2}(K-1)^{-1/2} \eta p^{1/2}\big\}$, due to Lemma \ref{lem:crossvalidboundlm}, we have,
\begin{align}\label{eqn:cvboundlm}
 H_{n,K}&\leq \frac{m}{(n-m)} \sum_{k=1}^{K}\Big\{ \|(n-m)^{1/2}\Big(\hat{\bm{\beta}}_{n,-k}-\bm{\beta}\Big)\|^2\Tilde{\gamma}_1\Big\}\nonumber\\
 &\;\;\;\;+\big\{\frac{m}{(n-m)}\big\}^{1/2}\sum_{k=1}^{K}\Big\{\|(n-m)^{1/2}\Big(\hat{\bm{\beta}}_{n,-k}-\bm{\beta}\Big)\|\|\bm{W}_{n,k}\|\Big\} \nonumber\\
&\leq K\Big\{(K-1)^{-1} (M_\varepsilon^{(1)})^2\Tilde{\gamma}_1+(K-1)^{-1/2}M_\varepsilon^{(1)}M_\varepsilon\Big\}.
\end{align}
Now consider the second situation i.e. when $\{n^{-1/2}\lambda_n\}_{n\geq 1}$ diverges to $\infty$ through a sub-sequence. Here without loss of generality we can consider the sequence $\{n^{-1/2}\lambda_n\}_{n\geq 1}$ itself diverges since otherwise the remaining argument can be carried out through a sub-sequence. Therefore, due to Lemma \ref{lem:crossvalidunboundlm} and Theorem  \ref{prop:cv} we have,
\begin{align}\label{eqn:cvinftylm}
 H_{n,K}&\geq \frac{m}{4(n-m)} \sum_{k=1}^{K}\Big\{ \|(n-m)^{1/2}\Big(\hat{\bm{\beta}}_{n,-k}-\bm{\beta}\Big)\|^2\Tilde{\gamma}_0\Big\}\nonumber\\
&\;\;\;\;\;\;\;\;\;\;\;\;-\big\{\frac{m}{(n-m)}\big\}^{1/2}\sum_{k=1}^{K}\Big\{\|(n-m)^{1/2}\Big(\hat{\bm{\beta}}_{n,-k}-\bm{\beta}\Big)\|\|\bm{W}_{n,k}\|\Big\} \nonumber\\
&\geq K\Big\{(4(K-1))^{-1} (n^{-1/2}\lambda_n)^{3/2}\Tilde{\gamma}_0-K^{1/2}(K-1)^{-1}(8\Tilde{\gamma}_{0}^{-1}p^{1/2})(n^{-1/2}\lambda_n)(M_\varepsilon)\Big\}, 
\end{align}
which may be arbitrarily large as $n$ increases. Therefore, comparing (\ref{eqn:cvboundlm}) and (\ref{eqn:cvinftylm}), it is evident that the sequence $\{n^{-1/2}\hat{\lambda}_{n,K}(\omega)\}_{n\geq 1}$ must be bounded for any $\omega \in A^{\varepsilon}$, which implies that
$$\mathbf{P}\Big(\limsup_{n\rightarrow\infty}n^{-1/2}\hat{\lambda}_{n,K} < \infty \Big)= 1.$$ Hence, the proof is now complete. \hfill $\square$

\subsubsection{Proof of Proposition \ref{prop:eqicon}}\label{sec:prop4.1}
Fix any $\varepsilon>0$, choose a $\delta:=\delta(\varepsilon)>0$ such that $|\lambda_1-\lambda_2|<\delta.$ Now define the following quantities:

\begin{align*}
    &\hat{V}_{n}(\bm{u},\lambda)=(1/2)\bm{u}^\top\bm{L}_{n}\bm{u} -\bm{u}^\top \bm{W}_{n} + n^{1/2}\lambda\Big\{\sum_{j=1}^{p}\big(|n^{-1/2}u_j+\beta_j|-|\beta_j|\big)\Big\}\\
   &\;\text{and}\; \hat{\bm{u}}_{n}(\lambda)=n^{1/2}(\hat{\bm{\beta}}_n(n^{1/2}\lambda)-\bm{\beta})=argmin_{\bm{u}}\hat{V}_{n}(\bm{u},\lambda).
\end{align*}
Now due to Lemma \ref{lem:strongconvexquadgrowth}, we can write;
\begin{align}\label{eqn:againstrongquadgrowth}
&\frac{\tilde{\gamma}_0}{8}||\hat{\bm{u}}_{n}(\lambda_1)-\hat{\bm{u}}_{n}(\lambda_2)||^2\leq \hat{V}_{n}(\hat{\bm{u}}_{n}(\lambda_1),\lambda_2)- \hat{V}_{n}(\hat{\bm{u}}_{n}(\lambda_2),\lambda_2)\nonumber\\
&\leq \underbrace{\Big[\hat{V}_{n}(\hat{\bm{u}}_{n}(\lambda_1),\lambda_1)-\hat{V}_{n}(\hat{\bm{u}}_{n}(\lambda_2),\lambda_1)\Big]}_{\le 0\;\text{since }\;\hat{\bm{u}}_{n}(\lambda_1):\;\text{argmin of}\;\hat{V}_n(\cdot,\lambda_1) }+\Big[\hat{V}_{n}(\hat{\bm{u}}_{n}(\lambda_2),\lambda_1)-\hat{V}_{n}(\hat{\bm{u}}_{n}(\lambda_2),\lambda_2)\Big]\nonumber\\
&\qquad\qquad\qquad-\Big[\hat{V}_{n}(\hat{\bm{u}}_{n}(\lambda_1),\lambda_1)-\hat{V}_{n}(\hat{\bm{u}}_{n}(\lambda_1),\lambda_2)\Big]\nonumber\\
&\le \Big[\hat{V}_{n}(\hat{\bm{u}}_{n}(\lambda_2),\lambda_1)-\hat{V}_{n}(\hat{\bm{u}}_{n}(\lambda_2),\lambda_2)\Big]-\Big[\hat{V}_{n}(\hat{\bm{u}}_{n}(\lambda_1),\lambda_1)-\hat{V}_{n}(\hat{\bm{u}}_{n}(\lambda_1),\lambda_2)\Big]
\end{align}

Now,
\begin{align}\label{eqn:100}
& \hat{V}_{n}(\hat{\bm{u}}_{n}(\lambda_2),\lambda_1)-\hat{V}_{n}(\hat{\bm{u}}_{n}(\lambda_2),\lambda_2)=n^{1/2}(\lambda_1-\lambda_2)\Bigg[\Big\|n^{-1/2}\hat{\bm{u}}_{n}(\lambda_2)+\bm{\beta}\Big\|_1-\Big\|\bm{\beta}\Big\|_1\Bigg]
\end{align}
Similarly we can obtain expression for $\Big[\hat{V}_{n}(\hat{\bm{u}}_{n}(\lambda_1),\lambda_1)-\hat{V}_{n}(\hat{\bm{u}}_{n}(\lambda_1),\lambda_2)\Big]$. Combining these two will give,
\begin{align}\label{eqn:101}
 & \Big[\hat{V}_{n}(\hat{\bm{u}}_{n}(\lambda_2),\lambda_1)-\hat{V}_{n}(\hat{\bm{u}}_{n}(\lambda_2),\lambda_2)\Big]-\Big[\hat{V}_{n}(\hat{\bm{u}}_{n}(\lambda_1),\lambda_1)-\hat{V}_{n}(\hat{\bm{u}}_{n}(\lambda_1),\lambda_2)\Big]\nonumber\\
 &=n^{1/2}(\lambda_1-\lambda_2)\Bigg[\Big\|n^{-1/2}\hat{\bm{u}}_{n}(\lambda_2)+\bm{\beta}\Big\|_1-\Big\|n^{-1/2}\hat{\bm{u}}_{n}(\lambda_1)+\bm{\beta}\Big\|_1\Bigg]\nonumber\\
 &\leq n^{1/2}|\lambda_1-\lambda_2|\;\Big\|n^{-1/2}\Big(\hat{\bm{u}}_{n}(\lambda_1)-\hat{\bm{u}}_{n}(\lambda_2)\Big)\Big\|_1\leq p^{1/2}|\lambda_1-\lambda_2|\; \|\hat{\bm{u}}_{n}(\lambda_1)-\hat{\bm{u}}_{n}(\lambda_2)\|.
\end{align}
Therefore combining (\ref{eqn:againstrongquadgrowth}) and (\ref{eqn:101}) we get,
\begin{align}\label{eqn:102}
\Big\|\hat{\bm{u}}_{n}(\lambda_1)-\hat{\bm{u}}_{n}(\lambda_2)\Big\| \le 8\tilde \gamma_0^{-1}p^{1/2}|\lambda_1-\lambda_2|.
\end{align}
Now for any $\varepsilon>0$, choosing $\delta:=\frac{\varepsilon\tilde \gamma_0}{16p^{1/2}}$, will complete the proof. \hfill $\square$

\begin{remark}\label{remark:uinftyeqicon}
  In addition to the proof of Proposition \ref{prop:eqicon}, similar steps will lead us to stochastic continuity of $\hat{\bm{u}}_{\infty}(\lambda)$. We define $\hat{\bm{u}}_{\infty}(\lambda)=argmin_{\bm{u}}\hat{V}_{\infty}(\bm{u},\lambda)$, where:
  \begin{align}\label{eqn:objinftyu}
  \hat{V}_{\infty}(\bm{u},\lambda)=(1/2)\bm{u}^\top\bm{L}\bm{u} -\bm{u}^\top \bm{W}_{\infty} + \lambda\Big\{\sum_{j=1}^{p_0}u_jsgn({{\beta}_{j}})+\sum_{j=p_0+1}^{p}|u_j|\Big\},   \end{align}
  and $\bm{L},\bm{S}$ are two positive definite matrices and assumed to be the respective limits of $\bm{L}_n$ and $\bm{S}_n$ as defined in the statement of Proposition \ref{prop:eqicon}. $\bm{W}_{\infty}$ is a copy of $N(\bm{0},\bm{S})$. $\bm{\beta}$ is a fixed quantity in this context. Note that $\hat{V}_{\infty}(\bm{u},\lambda)$ is strongly convex in $\bm{u}$ for fixed $\lambda$. Similarly we can conclude the rest as in the proof of Proposition \ref{prop:eqicon}.
\end{remark}

\subsubsection{Proof of Proposition  \ref{prop:C.4}}\label{sec:appBprop3.3}
We will establish that, for every $\varepsilon>0$ and $\omega\in \Omega^\star$ with $\mathbf{P}(\Omega^\star)=1$, there exists $\eta:=\eta(\varepsilon,\omega)>0$ such that,
\begin{align}\label{eqn:kmljnk}
 \mbox{inf}_{\big\{\lambda:|\lambda - \hat{\Lambda}_{\infty,K}(\omega)| > \varepsilon\big\}}\hat{H}_{\infty, K}^\prime(\lambda,\omega)>\hat{H}_{\infty, K}^\prime(\hat{\Lambda}_{\infty,K},\omega)+\eta 
\end{align}
Fix such $\omega$.  Therefore the notations reduce to,
\begin{align}\label{eqn:bvcf}
\hat V_{\infty,-k}(\bm u,\lambda,\omega)
=
\frac12 \bm u^\top L\bm u
-
\bm u^\top \bm W_{\infty,-k}(\omega)
+
\lambda\Bigg(
\sum_{j=1}^{p_0} u_j \,\mathrm{sgn}(\beta_j)
+
\sum_{j=p_0+1}^{p} |u_j|
\Bigg),
\end{align}
and define $\hat{\bm u}_{\infty,-k}(\lambda,\omega)
=
\arg\min_{\bm u} \hat V_{\infty,-k}(\bm u,\lambda,\omega)$ and $\hat{\Lambda}_{\infty,K}(\omega)\in \mbox{argmin}_{\lambda}\hat H'_{\infty,K}(\lambda,\omega)$ where,
\[
\hat H'_{\infty,K}(\lambda,\omega)
=
\sum_{k=1}^K
\left[
\frac12\hat{\bm u}_{\infty,-k}(\lambda,\omega)^\top L\hat{\bm u}_{\infty,-k}(\lambda,\omega)
-
\hat{\bm u}_{\infty,-k}(\lambda,\omega)^\top \bm W_{\infty,k}(\omega)
\right]
\]
 Now note that, the objective function $\hat V_{\infty,-k}(\bm u,\lambda,\omega)$ in (\ref{eqn:bvcf}) is in the form (\ref{eqn:xxxc}) of Lemma \ref{lem:equivminimizer}. Then due to Lemma \ref{lem:equiobj} and \ref{lem:equivminimizer} it's enough to handle the equivalent minimization problem:
\begin{align}\label{eqn:asdews}
 F_{\infty,-k}(\bm u,\bm t,\lambda,\omega)&=\frac12 \bm u^\top L\bm u
-
\bm u^\top \bm W_{\infty,-k}(\omega)
+
\lambda\Bigg(
\sum_{j=1}^{p_0} u_j \,\mathrm{sgn}(\beta_j)
+
\sum_{j=p_0+1}^{p} t_j\Bigg),\nonumber\\
&\qquad\qquad\text{subject to the constraints}\; t_j \ge |u_j|,\;t_j\ge 0\;\text{for all}\;j>p_0. 
\end{align}
For the ease of notation, we drop the index $\omega$ from everywhere. Next for this problem, we will derive the KKT conditions. Now for each $j>p_0$,
\begin{align}
g_{1,j}(\bm u,\bm t):=u_j - t_j &\le 0 \quad \text{(constraint 1)} \label{c1}\\
g_{2,j}(\bm u,\bm t):=-\,u_j - t_j &\le 0 \quad \text{(constraint 2)} \label{c2}\\
g_{3,j}(\bm u,\bm t):=-\,t_j &\le 0 \quad \text{(constraint 3)} \label{c3}
\end{align}

For each $j>p_0$, associate the Lagrangian multipliers:
\(\alpha_j \ge 0\) with constraint $1$,\; \(\theta_j \ge 0 \) with constraint $2$ and \(\gamma_j \ge 0\) with constraint $3$. Write the Lagrangian form of (\ref{eqn:asdews}) as: 
\begin{align}\label{eqn:ytuy}
\mathcal{L}(\bm u,\bm t,\bm\alpha,\bm\theta,\bm\gamma)
&=
\frac12 \bm u^\top L\bm u
-
\bm u^\top \bm W_{\infty,-k}
+
\lambda\Bigg(
\sum_{j=1}^{p_0} u_j \,\mathrm{sgn}(\beta_j)
+
\sum_{j=p_0+1}^{p} t_j\Bigg)\nonumber\\
&+ \sum_{j>p_0} \alpha_j (u_j - t_j)
+ \sum_{j>p_0} \theta_j (-u_j - t_j)
+ \sum_{j>p_0} \gamma_j (-t_j).
\end{align}
Next we examine the stationarity conditions. This will give that,
\begin{align}
(L\bm u-\bm W_{\infty,-k})_j+\lambda c_j &=0,
\qquad j\le p_0,\label{eqn:bh1}\\
(L\bm u-\bm W_{\infty,-k})_j+\alpha_j-\theta_j &=0,
\qquad j>p_0,\label{eqn:bh2}\\
\lambda-\alpha_j-\theta_j-\gamma_j &=0,
\qquad j>p_0.\label{eqn:bh3}
\end{align}
Then the first two sets of equations can be written compactly as:
\begin{align}
\label{eq:stationarity_ut}
&L\bm{u} - \bm{W}_{\infty,-k} + \lambda \bm{d} = 0,
\end{align}
where,
\[
d_j =
\begin{cases}
c_j=sgn(\beta_j), & j \le p_0, \\
\frac{1}{\lambda}(\alpha_j - \theta_j), & j > p_0,
\end{cases}
\]
For each $j>p_0$, complementary slackness will give:
\begin{align}
\alpha_j (u_j - t_j) &= 0 \label{cs1}\\
\theta_j (-u_j - t_j) &= 0 \label{cs2}\\
\gamma_j (-t_j) &= 0 \label{cs3}
\end{align}
    Let $(\hat{\bm{u}}, \hat{\bm{t}}, \hat{\bm{\alpha}}, \hat{\bm{\theta}},\hat{\bm{\gamma}})$ be the optimal primal--dual solution satisfying the KKT conditions. Any constraint among (\ref{c1}), (\ref{c2}) and (\ref{c3}) is said to be active if for any $j>p_0$,  $g_i(\bm u,\bm t):i=1,2,3$ is exactly equal to $0$ and otherwise inactive if it is strictly less than $0$. We handle this case wise.\\
    
\emph{\bf{Case 1}: $\hat u_j > 0$.} Under this case, due to part (b) of Lemma \ref{lem:equivminimizer}, we have that $\hat{t}_j = |\hat{u}_j| = \hat{u}_j.$ This will eventually give us that for all $j>p_0$,
\begin{align}
g_{1,j}(\hat{\bm{u}},\hat{\bm{t}}):=\hat u_j - \hat t_j &= 0 \quad \text{(active)} \label{c1a}\\
g_{2,j}(\hat{\bm{u}},\hat{\bm{t}}):=-\hat u_j - \hat t_j&=-2\hat u_j< 0 \quad \text{(inactive)} \label{c2a}\\
g_{3,j}(\hat{\bm{u}},\hat{\bm{t}}):=-\,\hat{t}_j &=-\hat{u}_j< 0 \quad \text{(inactive)} \label{c3a}
\end{align}
By the complementary slackness conditions (\ref{cs1}), (\ref{cs2}), (\ref{cs3}) and active-inactive behaviour of the constraints (\ref{c1a}), (\ref{c2a}) and (\ref{c3a}), it's easy to conclude that, 
\begin{align}\label{eqn:oolop}
\hat{\theta}_j = 0,\; \hat{\gamma}_j = 0,\;\text{and}\; \hat{\alpha}_j\ge 0.
\end{align}
From the stationarity condition,
\[
\lambda - \hat{\alpha}_j - \hat{\theta}_j - \hat{\gamma}_j = 0\implies \hat{\alpha}_j= \lambda.
\]

Thus for $j>p_0$,
\[
d_j(\hat{u}_j(\lambda)) = \frac{1}{\lambda}\bigl(\hat{\alpha}_j- \hat{\theta}_j\bigr)=1=\text{sgn}(\hat{u}_j(\lambda)).
\]
\emph{\bf{Case 2}: $\hat{u}_j< 0$.} Following exact similar steps will give us, $\hat t_j=|\hat{u}_j|=-\hat{u}_j$. This will eventually give us that for all $j>p_0$,
\begin{align}
g_{1,j}(\hat{\bm{u}},\hat{\bm{t}}):=\hat u_j - \hat t_j &= 2\hat u_j<0  \quad \text{(inactive)} \label{c1b}\\
g_{2,j}(\hat{\bm{u}},\hat{\bm{t}}):=-\hat u_j - \hat t_j&= 0 \quad \text{(active)} \label{c2b}\\
g_{3,j}(\hat{\bm{u}},\hat{\bm{t}}):=-\,\hat{t}_j &=\hat{u}_j< 0 \quad \text{(inactive)} \label{c3b}
\end{align}
By the complementary slackness conditions (\ref{cs1}), (\ref{cs2}), (\ref{cs3}) and active-inactive behaviour of the constraints (\ref{c1b}), (\ref{c2b}) and (\ref{c3b}), it's easy to conclude that, 
\begin{align}\label{eqn:oolopxcd}
\hat{\theta}_j\ge 0,\; \hat{\gamma}_j = 0,\;\text{and}\; \hat{\alpha}_j= 0.
\end{align}
From the stationarity condition,
\[
\lambda - \hat{\alpha}_j - \hat{\theta}_j - \hat{\gamma}_j = 0\implies \hat{\theta}_j= \lambda.
\]
Thus for $j>p_0$,
\[
d_j(\hat{u}_j(\lambda)) = \frac{1}{\lambda}\bigl(\hat{\alpha}_j- \hat{\theta}_j\bigr)=-1=\text{sgn}(\hat{u}_j(\lambda)).
\]
\emph{\bf{Case 3: $\hat{u}_j = 0$}.} In this case, $\hat t_j=0$. This will eventually give us that for all $j>p_0$,
\begin{align}
g_{1,j}(\hat{\bm{u}},\hat{\bm{t}}):=\hat u_j - \hat t_j &=0  \quad \text{(active)} \label{c1c}\\
g_{2,j}(\hat{\bm{u}},\hat{\bm{t}}):=-\hat u_j - \hat t_j&= 0 \quad \text{(active)} \label{c2c}\\
g_{3,j}(\hat{\bm{u}},\hat{\bm{t}}):=-\,\hat{t}_j &=0 \quad \text{(active)} \label{c3c}
\end{align}
By the complementary slackness conditions (\ref{cs1}), (\ref{cs2}), (\ref{cs3}) and active-inactive behaviour of the constraints (\ref{c1c}), (\ref{c2c}) and (\ref{c3c}), it's easy to conclude that, 
\begin{align}\label{eqn:oolopyycd}
\hat{\theta}_j\ge 0,\; \hat{\gamma}_j \ge 0,\;\text{and}\; \hat{\alpha}_j\ge 0.
\end{align}
From the stationarity condition,
\[
\lambda - \hat{\alpha}_j - \hat{\theta}_j - \hat{\gamma}_j = 0\implies \hat{\alpha}_j+\hat{\theta}_j+\hat{\gamma}_j = \lambda.
\]
This will imply that, $\hat{\alpha}_j,\hat{\theta}_j\le\lambda$, $\hat{\alpha}_j - \hat{\theta}_j\le\hat{\alpha}_j\le\lambda$ and $\hat{\alpha}_j - \hat{\theta}_j\ge-\hat{\theta}_j\ge-\lambda$. Hence $\hat{\alpha}_j - \hat{\theta}_j\in[-\lambda,\lambda]$. Thus for $j>p_0$,
\[
d_j(\hat{u}_j(\lambda)) = \frac{1}{\lambda}\bigl(\hat{\alpha}_j- \hat{\theta}_j\bigr)\in[-1,1].
\]
Combining Cases 1--3, for every $j>p_0$ we obtain
\[
d_j\bigl(\hat u_j(\lambda)\bigr)
\in
\partial \bigl|\hat u_j(\lambda)\bigr|
=
\begin{cases}
\{1\}, & \hat u_j(\lambda)>0,\\[4pt]
[-1,1], & \hat u_j(\lambda)=0,\\[4pt]
\{-1\}, & \hat u_j(\lambda)<0.
\end{cases}
\]
Now denote for a fixed $k\in\{1,...,K\}$,
\[
\hat{\bm{x}}_{-k}(\lambda)
:=
\begin{pmatrix}
\hat{\bm{u}}_{\infty,-k}(\lambda) \\
\hat{\bm{t}}(\lambda)
\end{pmatrix},
\quad
\hat{\bm{u}}_{\infty,-k}(\lambda)\in\mathbb{R}^p,\;\;
\hat{\bm{t}}(\lambda)\in\mathbb{R}^{p-p_0}.
\] 
Now based on the KKT equations mentioned above, we will argue that whenever, 
\begin{itemize}
\item[(i)] a particular non-zero component $\hat{u}_{\infty,-k,j}(\lambda)$ for $j\in\{p_0+1,..,p\}$ becomes $\hat{u}_{\infty,-k,j}(\lambda)=0$,
\item[(ii)] and/or a zero component $\hat{u}_{\infty,-k,j^\prime}(\lambda)$ becomes non zero,
\end{itemize}
it will result in alteration in active-inactive constraints $(\ref{c1})-(\ref{c3})$. As a consequence, this will govern existence of some finitely many points (will be referred as break points) on $[0,\infty)$ that will create a partition of the domain $[0,\infty)$. We will now discuss these arguments sequentially.\\

Note that, situation (i) refers to ``sign change of a non-zero coordinate''. Suppose for a fixed index $j$ and fixed $\lambda_1<\lambda_2$, we have $\hat{u}_{\infty,-k,j}(\lambda_1)>0$ and $\hat{u}_{\infty,-k,j}(\lambda_2)<0$. Then there exists a point $\lambda_{-k,j}^{(1)}\in (\lambda_1,\lambda_2)$ such that $\hat{u}_{\infty,-k,j}(\lambda_{-k,j}^{(1)})=0$ since $\hat{\bm u}_{\infty,-k}(\lambda)$ is stochastically continuous in $\lambda$ (see remark 8.1 of main paper). Similar argument can handle negative to positive. This essentially tells us that we have to cross $0$ intermediately for a sign change to happen. Eventually this transition from ``non-zero'' to ``zero'' will alter the index of the active constraints. For example, $\hat{u}_{\infty,-k,j}(\lambda)<0$ corresponds to $g_{2,j}(\hat{u}_{\infty,-k,j}(\lambda),\hat{t}_{j}(\lambda))=0$ (i.e only one constraint is active) but $\hat{u}_{\infty,-k,j}(\lambda)=0$ corresponds to $g_{i,j}(\hat{u}_{\infty,-k,j}(\lambda),\hat{t}_{j}(\lambda))=0$ for all $i=1,2,3$ (i.e all three are active). Thus $\lambda_{-k,j}^{(1)}$ is a break point pertaining to situation (i), typically where sign(s) of $\hat{u}_{\infty,-k,j}(\lambda)$ for one or more $j>p_0$, change(s).\\

Now situation (ii) refers to ``violation of sub-gradient condition at a zero coordinate''. Suppose for a fixed $j$ and $\lambda$, we have $\hat{u}_{\infty,-k,j}(\lambda)=0$. Then KKT sub-gradient condition requires that, $\text{sgn}(\hat{u}_{\infty,-k,j}(\lambda))\in[-1,1]$. Now, for $\hat{u}_{\infty,-k,j}(\lambda)$ to be non-zero, we require that the $\text{sgn}(\hat{u}_{\infty,-k,j}(\lambda))$ should hit the boundaries of $[-1,1]$. Whenever that happens, it will imply existence of a quantity $\lambda_{-k,j}^{(2,\pm)}$ such that $\text{sgn}(\hat{u}_{\infty,-k,j}\Big(\lambda_{-k,j}^{(2,\pm)})\Big)=\pm 1$. So, from this context we will term $\lambda_{-k,j}^{(2,\pm)}$ as another possible break point.\\

\emph{\bf{Why break points should be finite in number}?} So far, we have discussed which scenarios can possibly result in break points. Now we will argue that the collection of break points has finite cardinality. For that it's important to know how the break points look like. Now we define the active set at a fixed $\lambda$ as:
\[
\tilde{\mathcal{A}}_{-k}(\lambda) :=
\left\{
(i,j) \in \{1,2,3\}\times \{p_0+1,..,p\}:
g_{i,j}(\hat{u}_{\infty,-k,j}(\lambda),\hat{t}_{j}(\lambda)) = 0
\right\},
\]
which has finite cardinality. For each active set $\mathcal{A}$, define
\[
\Lambda_{-k}(\mathcal{A})
:=
\{\lambda>0 : \tilde{\mathcal{A}}_{-k}(\lambda)=\mathcal{A}\}.
\]
Let $e_j\in\mathbb R^p$ and
$f_j\in\mathbb R^{p-p_0}$ denote the standard basis vectors of the
$\bm u$- and $\bm t$-blocks, respectively.
For each $j>p_0$, the three inequality constraints
(\ref{c1})--(\ref{c3}) at optimal values may be written as
\[
a_{i,j}^{\top}\hat{\bm x}_{-k}(\lambda) \le 0,
\qquad i=1,2,3,
\]
where
\[
a_{1,j}
=
\begin{pmatrix}
e_j\\
-f_{j-p_0}
\end{pmatrix},
\qquad
a_{2,j}
=
\begin{pmatrix}
-\,e_j\\
-f_{j-p_0}
\end{pmatrix},
\qquad
a_{3,j}
=
\begin{pmatrix}
\bm 0_p\\
-f_{j-p_0}
\end{pmatrix}.
\]
Observe that, since $j>p_0$, the first $p_0$ coordinates of
$e_j$ are zero. Consequently, the first $p_0$ coordinates of
$a_{1,j}$ and $a_{2,j}$ are also zero. Define the active constraint matrix for a fixed active set $\mathcal A$:
\[
A_{\mathcal A}
:=
\begin{bmatrix}
a_{i,j}^\top
\end{bmatrix}_{(i,j)\in\mathcal A}
\in
\mathbb R^{\,|\mathcal A|
\times (2p-p_0)}.
\]
Then for every $\lambda\in\Lambda_{-k}(\mathcal A)$,
\[
A_{\mathcal A}\,\hat{\bm x}_{-k}(\lambda)=\bm 0.
\]
Let the full multiplier vector be: $$\hat{\bm\mu}_{\mathcal A}(\lambda)
: =\Big(\hat{\alpha}_{1,p_0+1}(\lambda),\hat{\theta}_{2,p_0+1}(\lambda),\hat{\gamma}_{3,p_0+1}(\lambda)\;\vdots\;....\;\vdots\;\hat{\alpha}_{1,p}(\lambda),\hat{\theta}_{2,p}(\lambda),\hat{\gamma}_{3,p}(\lambda)\Big)^\top$$ which has dimension $3(p-p_0)$. We use the identification
\[
\hat{\mu}_{1,j}(\lambda)=\hat{\alpha}_{1,j}(\lambda),\quad
\hat{\mu}_{2,j}(\lambda)=\hat{\theta}_{2,j}(\lambda),\quad
\hat{\mu}_{3,j}(\lambda)=\hat{\gamma}_{3,j}(\lambda),
\qquad j>p_0.
\] For each $j>p_0$, each triplet $(\hat{\alpha}_{1,j}(\lambda),\hat{\theta}_{2,j}(\lambda),\hat{\gamma}_{3,j}(\lambda))$ is associated with $3$ complementary slackness equations (\ref{cs1}),(\ref{cs2}) and (\ref{cs3}) respectively.  By complementary slackness, a multiplier is zero whenever the corresponding constraint is inactive. Hence, for any \((i,j)\notin \mathcal A\), the corresponding multiplier satisfies
\(
\hat{\mu}_{i,j}(\lambda)=0.
\) Therefore, we define the restricted multiplier vector as the subvector of active coordinates:
\[
\hat{\bm\mu}_{\mathcal A}(\lambda)
:=
\bigl(\hat{\mu}_{i,j}(\lambda)\bigr)_{(i,j)\in \mathcal A}.
\]
Next define
\[
Q :=
\begin{pmatrix}
L & \bm 0\\
\bm 0 & \bm 0
\end{pmatrix},
\quad
\bar{\bm c}
:=(c_1,\ldots  c_{p_0},0\ldots,0)^\top
\in\mathbb R^p,
\]
and
\[
c_{0,-k}(\lambda)
:=
\begin{pmatrix}
-\bm W_{\infty,-k}
+\lambda \bar{\bm c}
\\[2mm]
\lambda \bm 1
\end{pmatrix},
\]
where \(\bm 1\in\mathbb R^{p-p_0}\) denotes the vector of all ones. Since inactive constraints have zero multipliers, the stationarity conditions may be written using only the active multipliers. First note that
\[
Q\hat{\bm x}_{-k}(\lambda)+c_{0,-k}(\lambda)
=
\begin{pmatrix}
L\hat{\bm u}_{\infty,-k}(\lambda)
-\bm W_{\infty,-k}
+\lambda \bar{\bm c}
\\[2mm]
\lambda \bm 1
\end{pmatrix}.
\]
By construction,
\(A_{\mathcal A}^{\!\top}
\hat{\bm\mu}_{\mathcal A}(\lambda)
=
\sum_{(i,j)\in\mathcal A}
\hat\mu_{i,j}(\lambda)\,a_{i,j}.\) Hence we write, 
\[
\begin{aligned}
A_{\mathcal A}^{\!\top}
\hat{\bm\mu}_{\mathcal A}(\lambda)
&=
\sum_{(1,j)\in\mathcal A}
\hat{\alpha}_{1,j}(\lambda)
\begin{pmatrix}
e_j\\
-f_{j-p_0}
\end{pmatrix}
+
\sum_{(2,j)\in\mathcal A}
\hat{\theta}_{2,j}(\lambda)
\begin{pmatrix}
-\,e_j\\
-f_{j-p_0}
\end{pmatrix}
+
\sum_{(3,j)\in\mathcal A}
\hat{\gamma}_{3,j}(\lambda)
\begin{pmatrix}
\bm 0_p\\
-f_{j-p_0}
\end{pmatrix}\\
&=
\begin{pmatrix}
\displaystyle
\sum_{j=p_0+1}^{p}
\bigl(
\hat{\alpha}_{1,j}(\lambda)
-
\hat{\theta}_{2,j}(\lambda)
\bigr)e_j
\\[4mm]
\displaystyle
-\sum_{j=p_0+1}^{p}
\bigl(
\hat{\alpha}_{1,j}(\lambda)
+
\hat{\theta}_{2,j}(\lambda)
+
\hat{\gamma}_{3,j}(\lambda)
\bigr)
f_{j-p_0}
\end{pmatrix}
\end{aligned}
\]
Equivalently, if
\(\bm\alpha
=
\bigl(\hat{\alpha}_{1,j}(\lambda)\bigr)_{j=p_0+1}^{p},\;
\bm\theta
=
\bigl(\hat{\theta}_{2,j}(\lambda)\bigr)_{j=p_0+1}^{p},\;
\bm\gamma
=
\bigl(\hat{\gamma}_{3,j}(\lambda)\bigr)_{j=p_0+1}^{p},\)
then
\[
A_{\mathcal A}^{\!\top}
\hat{\bm\mu}_{\mathcal A}(\lambda)
=
\begin{pmatrix}
\bm 0_{p_0}
\\[1mm]
\bm\alpha-\bm\theta
\\[2mm]
-(\bm\alpha+\bm\theta+\bm\gamma)
\end{pmatrix},
\]
with inactive coordinates set equal to zero. Hence
\[
Q\hat{\bm x}_{-k}(\lambda)
+c_{0,-k}(\lambda)
+A_{\mathcal A}^{\!\top}
\hat{\bm\mu}_{\mathcal A}(\lambda)
=
\begin{pmatrix}
L\hat{\bm u}_{\infty,-k}(\lambda)
-\bm W_{\infty,-k}
+\lambda \bar{\bm c}
+\begin{pmatrix}
\bm 0_{p_0}\\
\bm\alpha-\bm\theta
\end{pmatrix}
\\[4mm]
\lambda\bm 1
-(\bm\alpha+\bm\theta+\bm\gamma)
\end{pmatrix}.
\]
Therefore, the stationarity condition,
\begin{align}\label{eqn:stat}
Q\hat{\bm x}_{-k}(\lambda)
+c_{0,-k}(\lambda)
+A_{\mathcal A}^{\!\top}
\hat{\bm\mu}_{\mathcal A}(\lambda)
=\bm 0
\end{align}
is equivalent to
\[
\begin{aligned}
L\hat{\bm u}_{\infty,-k}(\lambda)
-\bm W_{\infty,-k}
+\lambda \bar{\bm c}
+\begin{pmatrix}
\bm 0_{p_0}\\
\bm\alpha-\bm\theta
\end{pmatrix}
&=\bm 0,\quad
\bm\alpha+\bm\theta+\bm\gamma=\lambda\bm 1.
\end{aligned}
\]
More precisely if we write,
\[
L\hat{\bm u}_{\infty,-k}(\lambda)
-\bm W_{\infty,-k}
+\lambda \bar{\bm c}
+
\begin{pmatrix}
\bm 0_{p_0}\\
\bm\alpha-\bm\theta
\end{pmatrix}
=
\begin{pmatrix}
\bigl(
L\hat{\bm u}_{\infty,-k}(\lambda)
-\bm W_{\infty,-k}
\bigr)_{1:p_0}
+\lambda(c_1,\ldots,c_{p_0})^\top
\\[3mm]
\bigl(
L\hat{\bm u}_{\infty,-k}(\lambda)
-\bm W_{\infty,-k}
\bigr)_{p_0+1:p}
+\bm\alpha-\bm\theta
\end{pmatrix}.
\]
Then equation (\ref{eqn:stat}) is just the matrix analogue of (\ref{eqn:bh1})-(\ref{eqn:bh3}) together with the primal feasibility
\(A_{\mathcal A}\hat{\bm x}_{-k}(\lambda)=\bm 0.\) We now partition the active constraint matrix as
\[
A_{\mathcal{A}}
=
\begin{pmatrix}
A_{u,\mathcal{A}} & A_{t,\mathcal{A}}
\end{pmatrix},\;A_{u,\mathcal{A}} \in \mathbb{R}^{|\mathcal{A}| \times p},
\quad
A_{t,\mathcal{A}} \in \mathbb{R}^{|\mathcal{A}| \times (p-p_0)}
\]
Now splitting the stationary conditions block wise for $\hat{\bm{u}}_{\infty,-k}(\lambda)$ we get:
\[L\hat{\bm u}_{\infty,-k}(\lambda)
-\bm W_{\infty,-k}
+\lambda \bar{\bm c}
+A_{u,\mathcal{A}}^\top \hat{\bm\mu}_{\mathcal A}(\lambda)=\bm 0,\;
A_{u,\mathcal{A}}\hat{\bm u}_{\infty,-k}(\lambda)=\bm 0,\;
\hat{\bm{\mu}}_{\mathcal{A}}(\lambda) \ge \bm 0,
\]
The KKT system reads
\begin{align}
L \hat{\bm u}_{\infty,-k}(\lambda)
+
A_{u,\mathcal{A}}^\top \hat{\bm\mu}_{\mathcal A}(\lambda)
&=
\bm W_{\infty,-k} - \lambda \bar{\bm c},
\label{eq:kkt_stationarity}\\
A_{u,\mathcal{A}}\hat{\bm u}_{\infty,-k}(\lambda)&= \bm 0.
\label{eq:kkt_feas}
\end{align}
From \eqref{eq:kkt_stationarity},
\begin{align}\label{eqn:d3}
\hat{\bm u}_{\infty,-k}(\lambda)
=
L^{-1}(\bm W_{\infty,-k}-\lambda \bar{\bm c})
-
L^{-1}A_{u,\mathcal{A}}^\top \hat{\bm\mu}_{\mathcal A}(\lambda).
\end{align}
Substitute into \eqref{eq:kkt_feas}:
\[
M_{\mathcal A}\hat{\bm\mu}_{\mathcal A}(\lambda)
=
A_{u,\mathcal{A}} L^{-1}(\bm W_{\infty,-k}-\lambda \bar{\bm c}),
\quad
M_{\mathcal A}:=A_{u,\mathcal{A}} L^{-1}A_{u,\mathcal{A}}^\top.
\] 
Since $M_{\mathcal A}$ may be singular due to redundant active constraints, the
multiplier vector need not be unique. Note that,
\[
A_{u,\mathcal{A}} L^{-1}(\bm W_{\infty,-k}-\lambda\bar{\bm c})
\in \operatorname{col}(M_\mathcal{A}).
\]
Moreover, for any \(b\in\operatorname{col}(M_\mathcal{A})\),
the vector \(M_\mathcal{A}^\dagger b\) (where $M_\mathcal{A}^\dagger$ is the usual Moore-Penrose inverse of $M_\mathcal{A}$) satisfies
\[
M_\mathcal{A}(M_\mathcal{A}^\dagger b)
=
M_\mathcal{A} M_\mathcal{A}^\dagger b
=
b,
\]
due to Lemma \ref{lem:moore}. Therefore, we choose the Moore--Penrose solution
\begin{align}\label{eqn:hghg}
\hat{\bm\mu}_{\mathcal A}(\lambda)
=
M_{\mathcal A}^\dagger
A_{u,\mathcal{A}} L^{-1}(\bm W_{\infty,-k}-\lambda \bar{\bm c}).
\end{align}
Although this quantity may not be unique, the term \(
A_{u,\mathcal{A}}^\top \hat{\bm\mu}_\mathcal{A}(\lambda)
\)
appearing in the stationarity equation is unique (see Lemma \ref{lem:unique-dual-image}). Note that at a fixed $\lambda$, $\hat{\bm u}_{\infty,-k}(\lambda)$ is unique although $\hat{\bm\mu}_{\mathcal A}(\lambda)$ may not be unique. The non-uniqueness of $\hat{\bm\mu}_{\mathcal A}(\lambda)$ is due to the fact that the sub-gradient under Case 3 (mentioned before) may be anything inside $[-1, 1]$. 
Substituting this (\ref{eqn:hghg}) back to equation (\ref{eqn:d3}):
\begin{align*}
\hat{\bm u}_{\infty,-k}(\lambda)
=
P_{\mathcal A}\bm W_{\infty,-k}
-\lambda P_{\mathcal A}\bar{\bm c},
\end{align*}
where
\(P_{\mathcal A}
:=
L^{-1}
-
L^{-1}A_{u,\mathcal{A}}^\top (A_{u,\mathcal{A}} L^{-1}A_{u,\mathcal{A}}^\top)^\dagger A_{u,\mathcal{A}} L^{-1}.
\) Thus for all $\lambda\in\Lambda_{-k}(\mathcal A)$,
\[
\hat{\bm u}_{\infty,-k}(\lambda)
=
\bm b_{\mathcal A,-k}
+\lambda \bm a_{\mathcal A,-k},\;\;\text{with}\;\bm b_{\mathcal A,-k}:=P_{\mathcal A}\bm W_{\infty,-k}\;\;\text{and}\;\bm a_{\mathcal A,-k}:=-P_{\mathcal A}\bar{\bm c}. 
\]

\emph{\bf{ Characterization of breakpoints within $\Lambda_{-k}(\mathcal A)$}:}

We now finally derive the expressions for the break points according to situation (i) and (ii).
\medskip

\noindent
\textbf{(i) Sign change of a nonzero coordinate.} For $j$ such that $\hat u_{\infty,-k,j}(\lambda)\neq 0$ on $\Lambda_{-k}(\mathcal A)$, we will have alteration in sign if
\[
b_{\mathcal A,-k,j}+a_{\mathcal A,-k,j}\lambda=0
\quad\Rightarrow\quad
\lambda^{(1)}_{\mathcal A,-k,j}
=
-\frac{b_{\mathcal A,-k,j}}{a_{\mathcal A,-k,j}}.
\]
\textbf{(ii) Violation of the subgradient condition at a zero coordinate.} For a coordinate with $\hat u_{\infty,-k,j}(\lambda)=0$ on $\Lambda_{-k}(\mathcal A)$,
the KKT condition requires
\(\frac{(\bm W_{\infty,-k}-L\hat{\bm u}_{\infty,-k}(\lambda))_j}{\lambda}
\in[-1,1].\) Substituting the affine form gives
\[
\frac{(\bm W_{\infty,-k}-L\bm b_{\mathcal A,-k})_j}{\lambda}
-(L\bm a_{\mathcal A,-k})_j
\in[-1,1].
\]
Hence boundary points satisfy
\[
\frac{(\bm W_{\infty,-k}-L\bm b_{\mathcal A})_j}{\lambda}
-(L\bm a_{\mathcal A})_j
=\pm 1,
\]
yielding a break point: 
\[
\lambda^{(2,\pm)}_{\mathcal A,-k,j}
=
\frac{(\bm W_{\infty,-k}-L\bm b_{\mathcal A,-k})_j}
{(L\bm a_{\mathcal A,-k})_j\pm 1}.
\]

\medskip

Consequently we define the collection of break points as:
\[
\left\{
\lambda:
\lambda=\lambda^{(1)}_{\mathcal A,-k,j}
\ \text{or}\
\lambda=\lambda^{(2,\pm)}_{\mathcal A,-k,j}
\ \text{for some }j
\right\}.
\]
We now show that the piecewise affine solution path $\lambda \mapsto \hat{\bm u}_{\infty,-k}(\lambda)$ has finitely many linear segments (and hence break points). Now we argue that for any fixed active set $\mathcal A$, the set
\[
\Lambda_{-k}(\mathcal A)
:=
\{\lambda>0 : \tilde{\mathcal A}_{-k}(\lambda)=\mathcal A\}
\]
is either empty or an interval. Towards that fix an active set $\mathcal A$ and suppose that 
$\lambda_1 < \lambda_2$ are two elements of $\Lambda_{-k}(\mathcal A)$. 
We show that $(\lambda_1,\lambda_2)\subset \Lambda_{-k}(\mathcal A)$. For any $\lambda \in \Lambda_{-k}(\mathcal A)$, the KKT system associated with $\mathcal A$ yields the affine representation
\[
\hat{\bm u}_{\infty,-k}(\lambda)
=
\bm b_{\mathcal A,-k}
+
\lambda \bm a_{\mathcal A,-k},
\]
and similarly the corresponding multiplier vector $\hat{\bm\mu}_{\mathcal A}(\lambda)$ is affine in $\lambda$. The condition $\tilde{\mathcal A}_{-k}(\lambda)=\mathcal A$ is equivalent to the following three requirements:

\begin{itemize}
\item[(i)] Active constraints hold as equalities:
\(A_{\mathcal A}\hat{\bm x}_{-k}(\lambda)=\bm 0,
\)
\item[(ii)] Inactive constraints remain strictly feasible:
\(a_{i,j}^\top \hat{\bm x}_{-k}(\lambda) < 0, \quad (i,j)\notin \mathcal A,
\)
\item[(iii)] Dual feasibility:
\(\hat{\bm\mu}_{\mathcal A}(\lambda)\ge 0.\)
\end{itemize}
Each of the above conditions can be written as either an affine equality or an affine inequality in $\lambda$, since both $\hat{\bm x}_{-k}(\lambda)$ and $\hat{\bm\mu}_{\mathcal A}(\lambda)$ depend affinely on $\lambda$. Now choose a $\lambda \in (\lambda_1,\lambda_2)$. Then there exists $\theta \in (0,1)$ such that
\[
\lambda = \theta \lambda_1 + (1-\theta)\lambda_2.
\]
Since affine functions preserve convex combinations, any affine function $\phi(\lambda)$ satisfies
\[
\phi(\lambda)
=
\theta \phi(\lambda_1)
+
(1-\theta)\phi(\lambda_2).
\]
Therefore:
\begin{itemize}
\item if $\phi(\lambda_1)=0$ and $\phi(\lambda_2)=0$, then $\phi(\lambda)=0$,
\item if $\phi(\lambda_1)<0$ and $\phi(\lambda_2)<0$, then $\phi(\lambda)<0$,
\item if $\phi(\lambda_1)> 0$ and $\phi(\lambda_2)> 0$, then $\phi(\lambda)> 0$.
\end{itemize}
Applying this to all conditions (i)--(iii), we conclude that they remain satisfied at $\lambda$. Hence,
\[
\tilde{\mathcal A}_{-k}(\lambda)=\mathcal A,
\]
which shows that $\lambda \in \Lambda_{-k}(\mathcal A)$. Therefore,
\((\lambda_1,\lambda_2)\subset \Lambda_{-k}(\mathcal A),
\) and hence $\Lambda_{-k}(\mathcal A)$ is an interval. If no such $\lambda$ exists, it is empty. Hence, for all $\lambda \in (\lambda_1,\lambda_2)$,
\[
\hat{\bm u}_{\infty,-k}(\lambda)
=
\bm b_{\mathcal A,-k}
+
\lambda \bm a_{\mathcal A,-k},
\]
that is, the solution path is affine on $(\lambda_1,\lambda_2)$. Therefore, whenever two solutions 
$\hat{\bm u}_{\infty,-k}(\lambda_1)$ and 
$\hat{\bm u}_{\infty,-k}(\lambda_2)$ correspond to the same active set, the regularization path between $\lambda_1$ and $\lambda_2$ is a linear segment. As a consequence, the number of distinct linear segments of the path is bounded by the number of possible active sets. Since for each coordinate $j>p_0$, the component $\hat u_{\infty,-k,j}(\lambda)$ can be in one of three states (positive, negative, or zero), the total number of possible active configurations is bounded by
\(3^{\,p-p_0}.\) Hence, the number of break points is finite and at most $3^{\,p-p_0}$. This phenomenon is similar to the analysis of number of break points obtained for piece wise affine solution of Lasso (cf. Lemma 2 of \citet{mairal2012complexity}, Lemma 6 of \citet{tibshirani2013lasso}). \\

\emph{\bf{Finite partition of $[0,\infty)$ induced by break points}:} Let $\mathcal B \subset [0,\infty)$ denote the (finite) set of break points, i.e., values of $\lambda$ at which the active set $\tilde{\mathcal A}_{-k}(\lambda)$ changes. Since the number of break points is finite, we may enumerate them as
\[
\mathcal B = \{\lambda_1,\dots,\lambda_M\}, 
\qquad 0 < \lambda_1 < \cdots < \lambda_M < \infty.
\]
Define the extended sequence: \(\lambda_0 := 0,\;
\lambda_{M+1} := \infty.\) Then the complement $[0,\infty)\setminus \mathcal B$ can be written as:
\[[0,\infty)\setminus \mathcal B
=
\bigcup_{k=0}^{M} T_k,
\quad
\text{where }
T_0 := [\lambda_0,\lambda_1),\;
T_k := (\lambda_k,\lambda_{k+1}) \text{ for } k\ge 1.\]
Fix any interval in the above decomposition. We claim that the active set $\tilde{\mathcal A}_{-k}(\lambda)$ is constant on each such interval. Suppose, for contradiction, that there exist $\lambda',\lambda''$ in the same interval such that
\[
\tilde{\mathcal A}_{-k}(\lambda') \neq \tilde{\mathcal A}_{-k}(\lambda'').
\]
Then, since the active set can change only when one of the defining KKT conditions becomes tight or violated (i.e., through a sign change or a subgradient boundary event), there must exist some $\lambda^\star$ strictly between $\lambda'$ and $\lambda''$ at which such a transition occurs. By definition, this $\lambda^\star$ is a break point, which contradicts the fact that the interval contains no elements of $\mathcal B$. Hence, on each interval there exists an active set $\mathcal A$ such that
\[
\tilde{\mathcal A}_{-k}(\lambda) = \mathcal A
\quad \text{for all } \lambda \text{ in that interval}.
\]
By the KKT characterization derived earlier, it follows that on each such interval the solution admits the affine representation
\(\hat{\bm u}_{\infty,-k}(\lambda)
=
\bm b_{\mathcal A,-k}
+
\lambda \bm a_{\mathcal A,-k}.
\) Finally, since changes in the active set can occur only at points in $\mathcal B$, we obtain the partition
\[
[0,\infty)
=
\bigcup_{k=0}^{M}
[\lambda_k,\lambda_{k+1}),
\quad \lambda_0=0,\;\lambda_{M+1}=\infty,
\]
such that on each interval the active set is constant (except possibly at the left endpoint). Therefore, the solution path $\lambda \mapsto \hat{\bm u}_{\infty,-k}(\lambda)$ is piecewise affine with finitely many affine segments. \medskip

From this point onward, it suffices to carry out the analysis on a fixed interval
\[
T_\ell := [\lambda_\ell,\lambda_{\ell+1}),
\]
since the active set remains constant on $T_\ell$ (except possibly at the left endpoint $\lambda_\ell$). Accordingly, for a fixed $k \in \{1,\ldots,K\}$, we simplify the notation by replacing the suffix $(\mathcal A,-k)$ with $(\ell,-k)$, and write
\begin{align}\label{eqn:mkj}
\hat{\bm u}_{\infty,\ell,-k}(\lambda)
=
\bm b_{\ell,-k}
+
\lambda \bm a_{\ell,-k},
\qquad \lambda \in T_\ell.
\end{align}

\emph{\bf{Proposition (Existence of a global minimizer)}:}

Let $\hat{H}'_{\infty,K}(\lambda)$ be defined as
\begin{align}\label{eqn:picw}
\hat{H}'_{\infty,K}(\lambda)
=
\sum_{k=1}^{K}
\left[
\frac12 \hat{\bm{u}}_{\infty,-k}(\lambda)^\top L \hat{\bm{u}}_{\infty,-k}(\lambda)
-
\hat{\bm{u}}_{\infty,-k}(\lambda)^\top W_{\infty,k}
\right],
\quad \lambda \ge 0,
\end{align}
where $L$ is positive definite and $\hat{\bm{u}}_{\infty,-k}(\lambda)$ is the unique solution of (\ref{eqn:bvcf}) as a function of $\lambda$. Then $\hat{H}'_{\infty,K}(\lambda)$ admits a global unique  minimizer on $[0,\infty)$.

\medskip

\textbf{Proof.} We have established that $[0,\infty)$ admits a finite partition through disjoint intervals governed by finitely many break points and the optimal solution path $\hat{\bm{u}}_{\infty,-k}(\lambda)$ is piece wise affine in $\lambda$, i.e if we fix any interval $T_\ell=[\lambda_\ell,\lambda_{\ell+1})$ from that partition, we will have $\hat{\bm u}_{\infty,\ell,-k}(\lambda)$ as in (\ref{eqn:mkj}). Consequently, it is obvious that the objective function $\hat{H}'_{\infty,K}(\lambda)$ in (\ref{eqn:picw}) is piece wise quadratic in $\lambda$ provided the coefficient of the quadratic term is non-zero (which we will argue later). Mathematically, on that same fixed interval $T_\ell$, we should get something as:
$$\hat{H}'_{\infty,\ell,K}(\lambda)=\frac12\bar{A}_\ell \lambda^2 + B_\ell \lambda + C_\ell,\;\;\text{with}\;\;\bar{A}_\ell\neq 0,$$
and these coefficients $\bar{A}_\ell,B_\ell,C_\ell$ are determined through the index of the interval fixed by break point $\lambda_\ell$. Now to arrive at the conclusion of this proposition, we will proceed in following steps.
\begin{itemize}
\item Since number of partitions are finite say $M+1$, we will get $\{\hat{H}'_{\infty,\ell,K}(\lambda)\}_{\ell=0}^{M}$ many quadratic representations of $\hat{H}'_{\infty,K}(\lambda)$ on $M+1$ many intervals $\{T_\ell\}_{\ell=0}^M$ respectively.\\[1pt]
\item On each $T_\ell$ for $\ell=0,..,M$, we will establish that $\hat{H}'_{\infty,\ell,K}(\lambda)$ is indeed a strictly convex quadratic function in $\lambda$. We will argue that $\bar A_\ell>0$.\\[1pt]
\item This will imply that each fixed interval $T_\ell=[\lambda_\ell,\lambda_{\ell+1})$, can accommodate at most one (hence locally unique) minimizer $\lambda_\ell^{opt}$ restricted to that interval. That is:
\[
\lambda_\ell^{\mathrm{opt}}
=
\begin{cases}
-\dfrac{B_\ell}{\bar A_\ell}, & \text{if } -\dfrac{B_\ell}{\bar A_\ell} \in [\lambda_\ell,\lambda_{\ell+1}), \\[8pt]
\lambda_\ell, & \text{if } -\dfrac{B_\ell}{\bar A_\ell} < \lambda_\ell, \\[4pt]
\lambda_{\ell+1}, & \text{if } -\dfrac{B_\ell}{\bar A_\ell} \ge \lambda_{\ell+1}.
\end{cases}
\]
\item Now, consider the last interval $T_M = [\lambda_M,\infty)$. On this interval, the objective function takes the quadratic form
\[
\hat{H}'_{\infty,M,K}(\lambda)
=
\frac12 \bar{A}_M \lambda^2 + B_M \lambda + C_M,
\qquad \bar{A}_M > 0.
\]
Hence,
\(\hat{H}'_{\infty,M,K}(\lambda) \to +\infty\;\text{as } \lambda \to \infty,
\) which shows that the objective is coercive on $[0,\infty)$. In particular, the minimum cannot be attained at $\lambda_{M+1}=\infty$, and therefore must occur at a finite value of $\lambda$.\\[1pt]
\item As seen before, these boundaries are nothing but the break points and given by union of the following collections on $T_\ell$:\\
\emph{(1) Sign change of nonzero coordinates:}
\(\Lambda_\ell^{(1)}
:=
\left\{
-\frac{b_{\ell,-k,j}}{a_{\ell,-k,j}}
:
j \notin Z_\ell,\; a_{\ell,-k,j} \neq 0
\right\}.\)

\medskip

\noindent
\emph{(2) Violation of sub-gradient condition:}
\(\Lambda_\ell^{(2)}
:=
\left\{
\frac{(\bm{W}_{\infty,-k} - L \bm b_{\ell,-k})_j}{(L \bm a_{\ell,-k})_j \pm 1}
:
j \in Z_\ell
\right\}\),\\

where $Z_\ell:=\{j>p_0:\hat{u}_{\infty,\ell,-k,j}(\lambda)=0\;\text{for}\;\lambda\in T_\ell\}$.\\[1pt]
\item Define the breakpoint set on the interval $T_\ell$ by: \(
\Lambda_\ell := \Lambda_\ell^{(1)} \cup \Lambda_\ell^{(2)}\) and the global break point set
\(\Lambda := \bigcup_{\ell=0}^{M} \Lambda_\ell.\) Collecting all candidates, the minimizer of $\hat H'_{\infty,K}(\lambda)$ over $[0,\infty)$ belongs to the finite set
\[\mathcal C
=
\{0\}
\;\cup\;
\Lambda
\;\cup\;
\left\{
-\frac{B_\ell}{\bar A_\ell}
:
\ell = 0,\dots,M,\;
-\frac{B_\ell}{\bar A_\ell} \in [\lambda_\ell,\lambda_{\ell+1})
\right\}.
\]
Therefore, \(\min_{\lambda \ge 0} \hat H'_{\infty,K}(\lambda)
=
\min_{\lambda \in \mathcal C}
\hat H'_{\infty,K}(\lambda).\)\\[1pt]
\item Then we prove that the global unique minimizer, $\hat{\Lambda}_{\infty,K}:=\min_{\lambda \in \mathcal C}
\hat H'_{\infty,K}(\lambda).$ exists. Now towards contradiction,  assume that there does not exist any such unique global minimizer $\hat{\Lambda}_{\infty,K}$. This will imply that any one of the following scenarios is plausible:\\ 

\begin{itemize}
\item $(\nu.1)$\;\textbf{Two Interior Candidates:} For any two disjoint intervals  $T_\ell \neq T_t$, $\hat{\lambda}_{\infty,\ell}=-\frac{B_{\ell}}{\bar A_{\ell}}\neq-\frac{B_{t}}{\bar A_t}=\hat{\lambda}_{\infty,t}$ implies $\hat{H}_{\infty,\ell,K}^\prime\Big(\frac{B_{\ell}}{\bar A_{\ell}}\Big)=\hat{H}_{\infty,t,K}^\prime\Big(\frac{B_{t}}{\bar A_{t}}\Big)$.\\[0.8mm] 
\item $(\nu.2)$\;\textbf{One Interior and One Boundary Candidate from $\cup_{t=0}^{M}\Lambda_t^{(1)}$ for $t\neq \ell$:} Suppose there exists a minimizer of the form $\eta_{t,-k,j}
=
-\frac{b_{t,-k,j}}{a_{t,-k,j}},\;\;
a_{t,-k,j}\neq0$ such that $\eta_{t,-k,j}\neq -\frac{B_{\ell}}{\bar A_{\ell}}\;\implies \hat{H}_{\infty,\ell,K}^\prime\Big(-\frac{B_\ell}{\bar A_\ell}\Big)
=
\hat{H}_{\infty,t,K}^\prime\Big(-\frac{b_{t,-k,j}}{a_{t,-k,j}}\Big)$.\\[0.8mm]
\item $(\nu.3)$\;\textbf{One Interior and One Boundary Candidate from $\cup_{t=0}^{M}\Lambda_t^{(2)}$ for $t\neq \ell$:} Suppose there exists a minimizer of the form $\kappa_{t,-k,j}
=
\frac{(\bm W_{\infty,-k} - L \bm b_{t,-k})_j}{(L \bm a_{t,-k})_j \pm 1},$ with $(L \bm a_{t,-k})_j \pm 1\neq0$ such that $\kappa_{t,-k,j}\neq -\frac{B_{\ell}}{\bar A_{\ell}}\;\implies \hat{H}_{\infty,\ell,K}^\prime\Big(-\frac{B_\ell}{\bar A_\ell}\Big)
=
\hat{H}_{\infty,t,K}^\prime\big(\kappa_{t,-k,j}\big)$.\\[0.8mm]
\item $(\nu.4)$\;\textbf{Two Boundary Candidates  from $\cup_{t=0}^{M}\Lambda_t^{(r)}$ and $\cup_{\ell=0}^{M}\Lambda_\ell^{(s)}$} for all $s,r=1,2$ and any $\ell,t$ are not equal such that both gives equality in respective objective functions indexed by  $\hat{H}_{\infty,r,K}^\prime(\lambda)$ with $r\in\{\ell,t\}$.
\end{itemize}
\item In each of these above mentioned cases, we will establish that the equality of objective functions can hold on sets with probability measure $0$ utilizing Lemma \ref{lem:gausswn-} and \ref{lem:degeneracy-probability-zero}. This will complete the proof of the proposition. 
\end{itemize}

Now let's analyze along the lines of above mentioned objectives. We perform everything on a fixed interval $T_\ell$. On this interval, for each $k \in \{1,\dots,K\}$, the solution satisfies
\[
\hat{\bm{u}}_{\infty,\ell,-k}(\lambda)
=
\bm a_{\ell,-k} \lambda + \bm b_{\ell,-k},\;\; \lambda \in [\lambda_\ell, \lambda_{\ell+1}),
\]
for some vectors $\bm a_{\ell,-k}=-P_\ell\bar{\bm{c}}$ (independent of index $k$) and $\bm b_{\ell,-k}=P_\ell\bm{W}_{\infty,-k} \in \mathbb{R}^p$. Fix $\ell \in \{0,\dots,M\}$. Substituting this representation into (\ref{eqn:picw}), we get on that interval:
\[
\hat{H}'_{\infty,\ell,K}(\lambda)
=
\sum_{k=1}^{K}
\left[
\frac12 (\bm a_{\ell,-k}\lambda + \bm b_{\ell,-k})^\top
L
(\bm a_{\ell,-k}\lambda + \bm b_{\ell,-k})
-
(\bm a_{\ell,-k}\lambda + \bm b_{\ell,-k})^\top W_{\infty,k}
\right].
\]
Therefore, on $[\lambda_\ell,\lambda_{\ell+1})$, we obtain the full expansion:
\(
\hat{H}'_{\infty,\ell,K}(\lambda)
=
\frac12\bar{A}_\ell \lambda^2 + B_\ell \lambda + C_\ell,
\)
where the coefficients are given explicitly by
\[
\bar{A}_\ell
=
 \sum_{k=1}^K (\bm a_{\ell,-k})^\top L \bm a_{\ell,-k}=
K\, \bm a_{\ell}^\top L \bm a_{\ell},\;\;B_\ell
=
\sum_{k=1}^K
\left[
(\bm a_{\ell})^\top L \bm b_{\ell,-k}
-
(\bm a_{\ell})^\top W_{\infty,k}
\right],
\]
\[C_\ell
=
\sum_{k=1}^K
\left[
\frac12 (\bm b_{\ell,-k})^\top L \bm b_{\ell,-k}
-
(\bm b_{\ell,-k})^\top W_{\infty,k}
\right].
\]
 Since $L$ is positive definite, hence
\[
\bm a_\ell^\top L \bm a_\ell \ge 0,
\quad
\text{with equality iff } \bm a_\ell = \bm 0.
\]
Therefore,
\[
\bar{A}_\ell > 0
\quad \Longleftrightarrow \quad
\bm a_\ell \neq \bm 0
\quad \Longleftrightarrow \quad
P_\ell \bar{\bm{c}} \neq \bm 0.
\]
We now show that $P_\ell \bar{\bm{c}} \neq 0$. Note that,
\[
P_\ell \bar{\bm{c}} = 0
\quad \Longleftrightarrow \quad
\bar{\bm{c}} \in \mathrm{Ker}(P_{\ell}).
\]

Assume for contradiction that $P_\ell \bar{\bm{c}} = 0$. Recall that
\(P_\ell
=
L^{-1}
-
L^{-1}A_\ell^\top (A_\ell L^{-1}A_\ell^\top)^\dagger A_\ell L^{-1}.\) Then \[P_\ell \bar{\bm{c}} = 0
\;\Longleftrightarrow\;
\bar{\bm{c}}
=
A_\ell^\top (A_\ell L^{-1}A_\ell^\top)^\dagger A_\ell L^{-1}\bar{\bm{c}}\]
Define
\(\bm{z}
:=
(A_\ell L^{-1}A_\ell^\top)^\dagger A_\ell L^{-1}\bar{\bm{c}}.
\) Then the above condition is equivalent to
\[
\bar{\bm{c}} = A_\ell^\top \bm{z}
\quad \Longleftrightarrow \quad
\bar{\bm{c}} \in \operatorname{col}(A_\ell^\top).
\]
However, by construction of $A_\ell$, every row of $A_\ell$
corresponds to constraints involving only indices $j > p_0$.
Therefore, every vector in $\operatorname{col}(A_\ell^\top)$ has
zero entries in the first $p_0$ positions, i.e.
\[
(A_\ell^\top \bm{z})_j = 0 \quad \forall j \le p_0.
\]
On the other hand,
\(\bar{\bm{c}} =
(c_1,\dots,c_{p_0},0,\dots,0),
\quad c_j\in \{\pm 1\} \text{ for } j \le p_0\) and $p_0\ge 1$. This contradicts $\bar{\bm{c}} \in \operatorname{col}(A_\ell^\top)$.
Hence,
\[
P_\ell \bar{\bm{c}} \neq 0\implies \bar{A}_\ell>0.
\]
Therefore $\hat H'_{\infty,K}(\lambda)$ is strictly convex and piece wise quadratic on each interval $\{T_{\ell}\}_{\ell=0}^M$. Now before going into case-wise details, recall the idea of negating the contradiction. We assume that the two different minimizers from two different intervals $T_{\ell}\neq T_t$  always agree in two objective functions $\hat{H}^\prime_{\infty,\ell,K}(\cdot)=\hat{H}^\prime_{\infty,t,K}(\cdot)$ on a set with probability measure zero. Now the proof for case-wise negation of the possibilities mentioned in $(\nu.1)-(\nu.4)$ heavily relies on the difference of the matrix $P_\ell-P_t$. But it is possible that $P_\ell=P_t$ for any $\ell\neq t$. We will tackle this case separately and argue that, under this set-up, there always exists a global unique minimizer. Let us discuss this case first.\\

\emph{\bf{Regarding case $P_\ell=P_t$}:} Recall that on region $T_\ell$ the solution has the affine form:
\[
\hat{\bm{u}}_{\infty,\ell,-k}(\lambda)
=
P_\ell \bm{W}_{\infty,-k}-\lambda P_\ell \bar{\bm{c}},
\]
while on region $T_t$,
\[
\hat{\bm{u}}_{\infty,t,-k}(\lambda)
=
P_t \bm{W}_{\infty,-k}-\lambda P_t \bar{\bm{c}}.
\]
Since $P_\ell=P_t=P$, we obtain for every $\lambda$,
\[
\hat{\bm{u}}_{\infty,\ell,-k}(\lambda)
=
\hat{\bm{u}}_{\infty,t,-k}(\lambda)
=
P \bm{W}_{\infty,-k}-\lambda P \bar{\bm{c}}.
\]
Under the hypothesis, it's also true that, $\bar{A}_\ell=\bar{A}_t=A$,\;$B_\ell=B_t=B$ and $C_\ell=C_t=C$. Indeed this will imply that,
\[
\hat{H}_{\infty,K}(\lambda)=\hat{H}'_{\infty,\ell,K}(\lambda)
=\hat{H}'_{\infty,t,K}(\lambda)=
\frac12 A \lambda^2 + B \lambda + C,\;\; \text{for}\;\lambda\in T_\ell,T_t.
\]
Hence both regions generate exactly the same strictly convex quadratic polynomial. Therefore the unique unconstrained minimizer is:
\(\lambda^\star=-\frac{B}{A}.\) The minimizer on $[\lambda_\ell,\lambda_{\ell+1})$ is
\[
\lambda_\ell^{\mathrm{opt}}
=
\begin{cases}
\lambda^\star,
&
\lambda^\star\in[\lambda_\ell,\lambda_{\ell+1}),
\\[4pt]
\lambda_\ell,
&
\lambda^\star<\lambda_\ell,
\\[4pt]
\lambda_{\ell+1},
&
\lambda^\star\ge \lambda_{\ell+1},
\end{cases}
\]
and similarly
\[
\lambda_t^{\mathrm{opt}}
=
\begin{cases}
\lambda^\star,
&
\lambda^\star\in[\lambda_t,\lambda_{t+1}),
\\[4pt]
\lambda_t,
&
\lambda^\star<\lambda_t,
\\[4pt]
\lambda_{t+1},
&
\lambda^\star\ge \lambda_{t+1}.
\end{cases}
\]
Now consider set of all possible minimizers: $\{\lambda^\star,\lambda_{\ell},\lambda_{\ell+1},\lambda_{t},\lambda_{t+1}\}$.

\medskip

\noindent
{\bf Case 1:} Suppose
\(\lambda^\star \in [\lambda_\ell,\lambda_{\ell+1}).
\) Then the interval-wise minimizer on $T_\ell$ is
\(\lambda_\ell^{\mathrm{opt}} = \lambda^\star.
\) Now assume, towards contradiction, that there exists another index
$t \neq \ell$ such that the corresponding interval $T_t$ also produces a  minimizer, i.e.,
\[
\hat{H}_{\infty,K}(\lambda_t^{\mathrm{opt}})
=
\hat{H}_{\infty,K}(\lambda_\ell^{\mathrm{opt}})
=
\hat{H}_{\infty,K}(\lambda^\star).
\]

Since the intervals $\{[\lambda_j,\lambda_{j+1})\}_{j}$ form a partition of
$[0,\infty)$, we must have
\[
\lambda^\star \notin [\lambda_t,\lambda_{t+1}),
\qquad t \neq \ell.
\]
Therefore,
\(\lambda_t^{\mathrm{opt}} \neq \lambda^\star.\) However, $\hat{H}_{\infty,K}(\lambda)$ is strictly convex and admits the
unique global minimizer at $\lambda^\star$. Hence,
\[
\lambda_t^{\mathrm{opt}} \neq \lambda^\star
\quad\Longrightarrow\quad
\hat{H}_{\infty,K}(\lambda_t^{\mathrm{opt}})
>
\hat{H}_{\infty,K}(\lambda^\star).
\]
This contradicts the assumption that
\(
\hat{H}_{\infty,K}(\lambda_t^{\mathrm{opt}})
=
\hat{H}_{\infty,K}(\lambda^\star).
\) Therefore, no interval $T_t$ with $t \neq \ell$ can produce a competing
minimizer, and hence the minimizer is unique.
\medskip

\noindent
{\bf Case 2:}
$\lambda^\star$ belongs to neither interval. Then both interval minimizers occur at boundary points.

\medskip

\noindent
{\bf Case 2(a):}Suppose
\(\lambda^\star < \lambda_\ell,\;\text{and}\;\lambda^\star < \lambda_t.\) Then both interval-wise minimizers occur at the left boundary points:
\[
\lambda_\ell^{\mathrm{opt}} = \lambda_\ell,
\qquad
\lambda_t^{\mathrm{opt}} = \lambda_t.
\]

Assume, towards contradiction, that \(\hat{H}_{\infty,K}(\lambda_\ell)
=
\hat{H}_{\infty,K}(\lambda_t)\;\text{with}\;\;\lambda_\ell \neq \lambda_t.
\) Since $\hat{H}_{\infty,K}(\lambda)$ is strictly convex with unique minimizer at $\lambda^\star$, it is strictly increasing on $(\lambda^\star,\infty)$. Hence,
\[
\lambda_\ell \neq \lambda_t
\quad\Longrightarrow\quad
\hat{H}_{\infty,K}(\lambda_\ell)
\neq
\hat{H}_{\infty,K}(\lambda_t).
\]

This contradicts the assumption that both points yield the same minimal objective value. Therefore, at most one of the two candidates can be optimal, and no two distinct boundary points in this regime can simultaneously minimize $\hat{H}_{\infty,K}(\lambda)$.
\medskip

\noindent
{\bf Case 2(b):}
Suppose
\(\lambda^\star \ge \lambda_{\ell+1},\;\text{and}\;\lambda^\star \ge \lambda_{t+1}.\) Then both interval-wise minimizers occur at the right boundary points:
\[
\lambda_\ell^{\mathrm{opt}} = \lambda_{\ell+1},
\qquad
\lambda_t^{\mathrm{opt}} = \lambda_{t+1}.
\]

Assume, towards contradiction, that \(\hat{H}_{\infty,K}(\lambda_{\ell+1})
=
\hat{H}_{\infty,K}(\lambda_{t+1})
\;\text{with}\;\;\lambda_{\ell+1} \neq \lambda_{t+1}.
\) Since $\hat{H}_{\infty,K}(\lambda)$ is strictly convex with unique minimizer at $\lambda^\star$, it is strictly decreasing on $(-\infty,\lambda^\star)$. Hence,
\[
\lambda_{\ell+1} \neq \lambda_{t+1}
\quad\Longrightarrow\quad
\hat{H}_{\infty,K}(\lambda_{\ell+1})
\neq
\hat{H}_{\infty,K}(\lambda_{t+1}).
\]

This contradicts the assumption that both points yield the same minimal objective value. Therefore, at most one of the two candidates can be optimal, and no two distinct boundary points in this regime can simultaneously minimize $\hat{H}_{\infty,K}(\lambda)$.
\medskip

\noindent
{\bf Case 2(c):}
Without loss of generality, suppose
\(\lambda_{\ell+1}
\le
\lambda^\star
<
\lambda_t.\) Then the interval-wise minimizers are: \(
\lambda_\ell^{\mathrm{opt}} = \lambda_{\ell+1},\;\text{and}\;
\lambda_t^{\mathrm{opt}} = \lambda_t.\) Assume, towards contradiction, that \(\hat{H}_{\infty,K}(\lambda_{\ell+1})
=
\hat{H}_{\infty,K}(\lambda_t).
\) Now based on the discussions on the break points, for any $s\in\{\ell+1,t\}$, $\lambda_{s}$ can take one of the functional form: (i)\;$\lambda_s:=-\frac{b_{s,-k,j}}{a_{s,-k,j}}$ with $a_{s,-k,j}\neq 0$ and for all $j\notin Z_s$\;\;or (ii) $\lambda_s:=\frac{\Big(\bm{W}_{\infty,-k}-L\bm{b}_{s,-k}\Big)_j}{(L\bm{a}_{s,-k})_j\pm 1}$, for all $j\in Z_s$. Recall that, $\bm{a}_s=-P_s\bar{\bm{c}}$ and $\bm{b}_{s,-k}=P_s\bm{W}_{\infty,-k}$. It is important to mention that although we assumed $P_\ell=P_t$ but it is possible that $P_{\ell+1}\neq P_t$ since $\lambda_{{\ell+1}}\in T_{\ell+1}$ which is the next adjacent interval for $T_\ell$. Also recall that $\bar A_s>0$ and $B_s=\bm{\gamma}_s^\top\Big(\sum_{k=1}^K\bm{W}_{\infty,-k}\Big)$ with $\bm{\gamma}_s:=\Big[\frac{2K-3}{\sqrt{K-1}}-1\Big]P_s\bar{\bm{c}}$.\\

Based on the choice (i), define the vector
\(\bm{\rho}_{s,j}
:=
\frac{P_s e_j}{e_j^\top P_s \bar{\bm{c}}},
\) with $e_j$ being canonical basis vector in $\mathbb R^p$. Hence
\(-\frac{b_{s,-k,j}}{a_{s,-k,j}}
=
\frac{
e_j^\top P_sW_{\infty,-k}
}{
e_j^\top P_s \bar{\bm{c}}
}=
\bm{\rho}_{s,j}^\top \bm W_{\infty,-k}.
\) Similarly for case (ii), denote the vector, \(
\bm{\upsilon}_{s,j}
=\frac{1}{(L \bm{a}_s)_j \pm 1}\,(e_j - P_s^\top L e_j)\). Then \(\frac{\Big(\bm{W}_{\infty,-k}-L\bm{b}_{s,-k}\Big)_j}{(L\bm{a}_s)_j\pm 1}
=
\bm{\upsilon}_{s,j}^\top \bm W_{\infty,-k}\). Therefore, in either cases, $\lambda_s$ is a linear functional of $\bm{W}_{\infty,-k}$ and let us denote it as $\bm{\phi}_{s,j}^\top\bm{W}_{\infty,-k}$, where,
\[
\bm{\phi}_{s,j}=
\begin{cases}
 \bm{\rho}_{s,j} &\;\text{when we consider possibility (i)} \\
  \bm{\upsilon}_{s,j} &\;\text{when we consider possibility (ii)}
\end{cases}
\]
Next we argue that $\hat{H}_{\infty,K}(\lambda_{\ell+1})-\hat{H}_{\infty,K}(\lambda_t)$ is not an identically zero polynomial in deterministic arguments $(\bm\omega_{-1},...,\bm\omega_{-K})$. Then we lift the argument to random vectors $(\bm{W}_{\infty,-1},...,\bm{W}_{\infty,-K})$. Also note that, all these existing functional relationships will remain exactly same for $(\bm\omega_{-1},...,\bm\omega_{-K})$. Now let us precisely calculate,
\begin{align}\label{eqn:eff}
Q_{\ell+1,t}(\bm{\omega}_{-1},...,\bm\omega_{-K}):&=\hat{H}_{\infty,K}(\lambda_{\ell+1})-\hat{H}_{\infty,K}(\lambda_t)=\frac{A}{2}(\lambda_{\ell+1}^2-\lambda_t^2)+B(\lambda_{\ell+1}-\lambda_t)\nonumber\\
&=\Bigg\{\frac{A}{2}\Big[(\bm{\phi}_{\ell+1,j}^\top\bm{\omega}_{-k})+(\bm{\phi}_{t,j}^\top\bm{\omega}_{-k})\Big]+B\Bigg\}\Big[(\bm{\phi}_{\ell+1,j}^\top\bm{\omega}_{-k})-(\bm{\phi}_{t,j}^\top\bm{\omega}_{-k})\Big]
\end{align}
Now choose, $\bm\omega_{-1}=\bm{v},\;\bm\omega_{-2}=-\bm{v},\;\bm{\omega_{-k}}=\bm 0\;\text{for all}\;k\ge 3$. This will give us $B=0$. Now since the break points $\lambda_{s}$ are fixed at some $\omega_{-k}$ by construction, we can choose two different (or same) indices, for example at $k=1$ corresponding to $\lambda_{\ell+1}$ and $k=2$ corresponding to $\lambda_t$ without loss of generality such that, $$(\bm{\phi}_{\ell+1,j}^\top\bm{v})^2-(\bm{\phi}_{t,j}^\top(-\bm{v}))^2=(\lambda_{\ell+1}-\lambda_{t})(\lambda_{\ell+1}+\lambda_{t})\neq 0.$$
Existence of such $\bm{v}$ is possible. Otherwise the negation will imply that for all $\bm{v}\in\mathbb R^p$, 
\[
(\lambda_{\ell+1}-\lambda_{t})(\lambda_{\ell+1}+\lambda_{t})=0\implies \;\text{either}\;\lambda_{\ell+1}=\lambda_t\;\;\text{or}\;\;\lambda_{\ell+1}+\lambda_t=0
\]
Now since $\lambda_{\ell+1},\lambda_m>0$,\;$\lambda_{\ell+1}+\lambda_m=0$ will imply that $\lambda_{\ell+1}=0=\lambda_m$, which is infeasible. Therefore, all it boils down to $\lambda_{\ell+1}=\lambda_m$, in which case we have nothing to prove. Hence the minimizer is already unique. Thus for that choice $\bm v$ where minimizers are different, it is straight forward to see that $Q_{\ell+1,t}(\bm{\omega}_{-1},...,\bm\omega_{-K})\neq 0$ since $A>0$. Thus the polynomial is not identically zero. Rest of the argument follows from Lemma \ref{lem:gausswn-} and \ref{lem:degeneracy-probability-zero}.\\

Combining the above cases, whenever $P_\ell=P_t$, the two regions
correspond to the unique global minimizer. Now we will discuss the analysis when $P_\ell\neq P_t$.\\

\emph{\bf{Regarding Case $(\nu.1)$}:} Fix two distinct active-set regions $\ell\neq t$. Define $Q_{\ell, t}:\mathbb R^{kP}\mapsto\mathbb R$ such that
\[
Q_{\ell,t}(\omega_{-1},....,\omega_{-K})
:=
\bar A_\ell B_t^2
-
\bar A_t B_\ell^2
+
4\bar A_\ell \bar A_t(C_\ell-C_t).
\]
Then we will first show that $Q_{\ell,t}$ is not the zero polynomial in
\(\bigl\{\bm\omega_{-1},....,\bm\omega_{-K}\bigr\}\) identically. Note that $\bar A_\ell,\bar A_t>0$ are deterministic constants. It suffices to show that
there exists a deterministic choice of
\(
\bigl(\bm\omega_{-1},....,\bm\omega_{-K}\bigr)
\)
for which
\(Q_{\ell,t}\neq0.\) Recall that
\[
B_r
=
\bm\gamma_r^\top\Bigg(\sum_{k=1}^K
 \bm\omega_{-k}\Bigg),\;\;\text{with}\;\;
\bm\gamma_r
=\Big[\frac{2K-3}{\sqrt{K-1}}-1\Big]P_r\bar{\bm{c}}.
\]
Choose any vector $\bm v\in\mathbb R^p$ and set
\[
\bm\omega_{-1}=\bm v,
\quad
\bm\omega_{-2}=-\bm v,
\quad
\bm\omega_{-k}=\bm 0,
\quad k\ge3.
\]
Therefore
\(B_\ell=0,\;\;
B_t=0.\) Consequently,
\[
Q_{\ell,t}
=
4\bar A_\ell \bar A_t(C_\ell-C_t).
\]
Thus it remains to show that $C_\ell-C_t\neq0$ for a suitable choice of $\bm v$. Recall that
\[
C_r
=
\sum_{k=1}^K
\left[
\frac12(\bm b_{r,-k}^\top L \bm b_{r,-k}
-
\bm b_{r,-k}^\top \bm\omega_{k}
\right],
\]
with
\(\bm b_{r,-k}
=
P_r\bm\omega_{-k}.\) Using $P_r^\top LP_r=P_r$ (see Lemma \ref{lem:P-kernel}), we obtain
\[
C_r
=
\sum_{k=1}^K
\left[
\frac12
\bm\omega_{-k}^\top P_r \bm\omega_{-k}
-
\bm\omega_{-k}^\top P_r \bm\omega_{k}
\right].
\]
Furthermore,
\(\bm\omega_{k}
=
\frac1{\sqrt{K-1}}
\sum_{i=1}^K \bm\omega_{-i}
+
\frac{K-3}{\sqrt{K-1}}
\bm\omega_{-k}.\) This simplifies to
\(\bm\omega_{k}
=
\frac{K-3}{\sqrt{K-1}}
\bm\omega_{-k}.\) Hence
\[
C_r
=
\left(
\frac12
-
\frac{K-3}{\sqrt{K-1}}
\right)
\sum_{k=1}^K
\bm\omega_{-k}^\top P_r \bm\omega_{-k}.
\]

Under the chosen configuration, only the terms $k=1,2$ are nonzero.
Therefore
\[
C_r
=
\left(
\frac12
-
\frac{K-3}{\sqrt{K-1}}
\right)
\bigl(
\bm v^\top P_r \bm v
+
\bm v^\top P_r \bm v
\bigr)=
\Bigg[
1-\frac{2(K-3)}{\sqrt{K-1}}
\Bigg]
\bm v^\top P_r \bm v.
\]
It follows that
\[
C_\ell-C_t
=
\Bigg[
1-\frac{2(K-3)}{\sqrt{K-1}}
\Bigg]
\bm v^\top(P_\ell-P_t)\bm v.
\]
Now for every integer $K\ge2$,
\(
1-\frac{2(K-3)}{\sqrt{K-1}}
\neq0.
\) Since \(
P_\ell\neq P_t.
\) and the matrix $P_\ell-P_t$ is symmetric, hence, if
\[
\bm v^\top(P_\ell-P_t)\bm v=0
\quad
\text{for every }\bm v\in\mathbb R^p,
\]
then necessarily
\(P_\ell-P_t=0,\) contradicting $P_\ell\neq P_t$. Therefore there exists
\(\bm v\in\mathbb R^p\) such that
\(\bm v^\top(P_\ell-P_t)\bm v\neq0.
\) For this choice of $\bm v$,
\[
C_\ell-C_t\neq0.
\]
Hence
\(
Q_{\ell,t}
=
4\bar A_\ell \bar A_t(C_\ell-C_t)
\neq0.\) Thus $Q_{\ell,t}$ is not the zero polynomial identically. Rest of the argument follows from Lemma  \ref{lem:degeneracy-probability-zero} since $(\bm{W}_{\infty,-1},....,\bm{W}_{\infty,-K})$ jointly admits an absolutely continuous density (precisely non-singular Gaussian) (see Lemma \ref{lem:gausswn-}). Therefore we are done.\\

\emph{\bf{Regarding Case $(\nu.2)$}:} Fix an interval $T_\ell$ and an interior candidate: \(\lambda_{\infty,\ell}^{\rm int}
=
-\frac{B_\ell}{\bar A_\ell}.\) Fix another interval $T_t$ and a breakpoint candidate from
$\Lambda_t^{(1)}$:
\(\eta_{t,-k,j}
=
-\frac{b_{t,-k,j}}{a_{t,-k,j}},\;\;
a_{t,-k,j}\neq0.\) We show that for fixed $t\neq \ell$,
\[
\mathbf P\!\left[
\hat{H}_{\infty,K}^\prime\Big(-\frac{B_\ell}{\bar A_\ell}\Big)
=
\hat{H}_{\infty,K}^\prime\Big(-\frac{b_{t,-k,j}}{a_{t,-k,j}}\Big)
\right]
=
0.
\]

Since there are only finitely many intervals and coordinates,
a finite union argument then implies that scenario $(\nu.2)$
occurs with probability zero. Recall that
\(\bar A_\ell
>0.\) Moreover,
\[
\lambda_{\infty,\ell}^{\rm int}
=
-\frac{B_\ell}{\bar A_\ell}
=
-\frac1{\bar A_\ell}
\sum_{k=1}^K
\bm\gamma_\ell^\top \bm{W}_{\infty,-k}.
\]
Thus the interior candidate is a linear functional of the joint
Gaussian vector
\((\bm W_{\infty,-1},\ldots,\bm W_{\infty,-K}).\) Now based on previous calculation we can directly write for deterministic arguments $(\bm\omega_{-1},..,\bm\omega_{-K})$,
\begin{align}\label{eqn:mnjkn}
\hat{H}_{\infty,K}^\prime\left(\lambda_{\infty,\ell}^{int}=-\frac{B_\ell}{\bar A_\ell}\right)&=C_\ell-\frac{B_\ell^2}{4\bar A_\ell}\nonumber\\
&=\left(
\frac12
-
\frac{K-3}{\sqrt{K-1}}
\right)
\sum_{k=1}^K
\bm\omega_{-k}^\top P_\ell \bm\omega_{-k}-\frac{1}{4\bar A_\ell}\Bigg[\bm\gamma_\ell^\top\Bigg(\sum_{k=1}^K \bm\omega_{-k}\Bigg)\Bigg]^2
\end{align}
Based on previous discussions, \(\bm\rho_{t,j}
:=
\frac{P_t e_j}{e_j^\top P_t\bar{\bm{c}}}.
\) Hence
\(-\frac{b_{t,-k,j}}{a_{t,-k,j}}
=
\frac{
e_j^\top P_t\bm W_{\infty,-k}
}{
e_j^\top P_t\bar{\bm{c}}
}=
\bm\rho_{t,j}^\top \bm W_{\infty,-k},
\). It's simple to calculate that,
\begin{align}\label{eqn:vcfd}
 &\hat H'_{\infty,K}(\eta_{t,-k,j})
=
\frac{\bar A_t}{2} \,
(\bm\rho_{t,j}^\top \bm\omega_{-k})^2
+
B_t (\bm\rho_{t,j}^\top \bm\omega_{-k})
+
C_t\nonumber\\
&=\frac K2\, (\bar{\bm{c}}^\top P_t \bar{\bm{c}}) \,
(\bm\rho_{t,j}^\top\bm\omega_{-k})^2
+\Bigg[\bm\gamma_t^\top\Bigg(\sum_{k=1}^K \bm\omega_{-k}\Bigg)\Bigg](\bm\rho_{t,j}^\top \bm\omega_{-k})+\left(
\frac12
-
\frac{K-3}{\sqrt{K-1}}
\right)
\sum_{k=1}^K
\bm\omega_{-k}^\top P_t \bm\omega_{-k}
\end{align}
Define for $\ell\neq t$ and $j \notin Z_t$,
\[
Q_{\ell,t,j}(\bm\omega_{-1},..,\bm\omega_{-K})
:=\hat{H}_{\infty,K}^\prime\left(\lambda_{\infty,\ell}^{int}=-\frac{B_\ell}{\bar A_\ell}\right)-\hat H_{\infty,K}^\prime(\eta_{t,-k,j})
\]
Then the event of interest is $\{Q_{\ell,t,j}(\bm\omega_{-1},..,\bm\omega_{-K})=0\}$. Suppose for the contradiction, we assume that $Q_{\ell,t,j}(\bm\omega_{-1},..,\bm\omega_{-K})=0$ for all $(\bm\omega_{-1},..,\bm\omega_{-K})$. Choose any vector $\bm v\in\mathbb R^p$ and set
\[
\bm \omega_{-1}=\bm v,
\quad
\bm \omega_{-2}=-\bm v,
\quad
\bm\omega_{-k}=\bm 0,
\quad k\ge3.
\]  
Now since the threshold $\eta_{t,-k,j}$ is fixed at some $\bm\omega_{-k}$, choose $k=3$ without loss of generality such that $\bm\rho_{t,j}^\top\bm\omega_{-k}=0$. Then similar argument and calculation of case $(\nu.1)$ will give us that, for the above choice:
\[
Q_{\ell,t,j}(\bm\omega_{-1},..,\bm\omega_{-K})=\Bigg[
1-\frac{2(K-3)}{\sqrt{K-1}}
\Bigg]
\bm v^\top(P_\ell-P_t)\bm v.
\]
Rest of the steps are exactly similar to case $(\nu.1)$. Hence $Q_{\ell,t,j}$ is not the zero polynomial identically. Rest of the argument follows similarly from Lemma \ref{lem:gausswn-} and \ref{lem:degeneracy-probability-zero}.\\

\emph{\bf{Regarding Case $(\nu.3)$}:}
Fix an interval $T_\ell$ and an interior candidate: \(\lambda_{\infty,\ell}^{\rm int}
=
-\frac{B_\ell}{\bar A_\ell}.\) Fix another interval $T_t$, fix $j \in Z_t$ and a break point candidate from
$\Lambda_t^{(2)}$:
\(\kappa_{t,-k,j}
=
\frac{(\bm W_{\infty,-k} - L \bm b_{t,-k})_j}{(L \bm a_t)_j \pm 1},\;\;
(L \bm a_t)_j \pm 1\neq0.\) We show that for fixed $t\neq \ell$,
\[
\mathbf P\!\left[
\hat{H}_{\infty,K}^\prime\Big(-\frac{B_\ell}{\bar A_\ell}\Big)
=
\hat{H}_{\infty,K}^\prime\big(\kappa_{t,-k,j}\big)
\right]
=
0.
\]
For a deterministic argument $(\bm\omega_{-1},...,\bm\omega_{-K})$, following everything from equation (\ref{eqn:mnjkn}), we write as before:
\begin{align*}
\hat{H}_{\infty,K}^\prime\left(\lambda_{\infty,\ell}^{int}=-\frac{B_\ell}{\bar A_\ell}\right)&=C_\ell-\frac{B_\ell^2}{4\bar A_\ell}\\
&=\left(
\frac12
-
\frac{K-3}{\sqrt{K-1}}
\right)
\sum_{k=1}^K
\bm\omega_{-k}^\top P_\ell \bm\omega_{-k}-\frac{1}{4\bar A_\ell}\Bigg[\bm\gamma_\ell^\top\Bigg(\sum_{k=1}^K \bm\omega_{-k}\Bigg)\Bigg]^2    
\end{align*}
Recall, \(\frac{(\bm W_{\infty,-k} - L \bm b_{t,-k})_j}{(L \bm a_t)_j \pm 1}
=
\bm\upsilon_{t,j}^\top \bm W_{\infty,-k},\). It's simple to calculate that,
\begin{align}\label{eqn:vcfdx}
 \hat H'_{\infty,K}(\kappa_{t,-k,j})
&=
\frac{\bar A_t}{2} \,
(\bm\upsilon_{t,j}^\top \bm\omega_{-k})^2
+
B_t (\bm\upsilon_{t,j}^\top \bm\omega_{-k})
+
C_t\nonumber\\
&=\frac{\bar A_t}{2}\,
(\bm\upsilon_{t,j}^\top\bm\omega_{-k})^2
+\Bigg[\bm\gamma_t^\top\Bigg(\sum_{k=1}^K \bm\omega_{-k}\Bigg)\Bigg](\bm\upsilon_{t,j}^\top \bm\omega_{-k})+\left(
\frac12
-
\frac{K-3}{\sqrt{K-1}}
\right)
\sum_{k=1}^K
\bm\omega_{-k}^\top P_t \bm\omega_{-k}
\end{align}
Define for $\ell\neq t$,
\[
Q_{\ell,t,j}(\bm\omega_{-1},..,\bm\omega_{-K})
:=\hat{H}_{\infty,K}^\prime\left(\lambda_{\infty,\ell}^{int}=-\frac{B_\ell}{\bar A_\ell}\right)-\hat H_{\infty,K}^\prime(\kappa_{t,-k,j})
\]
Suppose for the contradiction, we assume that $Q_{\ell,t,j}(\bm\omega_{-1},..,\bm\omega_{-K})=0$ for all $(\bm\omega_{-1},..,\bm\omega_{-K})$. Choose any vector $bm v\in\mathbb R^p$ and set
\[
\bm\omega_{-1}=\bm v,
\quad
\bm\omega_{-2}=-\bm v,
\quad
\bm\omega_{-k}=\bm 0,
\quad k\ge3.
\]  
Rest of the proof is similar to case $(\nu.2)$.\\

\emph{\bf{Regarding Case $(\nu.4)$}:}
Fix an interval $T_\ell=[\lambda_\ell,\lambda_{\ell+1})$. On this interval, the objective admits the quadratic representation
\[
\hat H'_{\infty,K}(\lambda)
=
\frac{\bar A_\ell}{2} \lambda^2 + B_\ell \lambda + C_\ell,\;\;\text{with}\;\;\bar A_\ell = K\, (\bar{\bm{c}}^\top P_\ell \bar{\bm{c}})>0.
\]
Moreover, $B_\ell
=
\bm\gamma_\ell^\top
\Bigg(
\sum_{k=1}^K \bm W_{\infty,-k}
\Bigg)\;\;\text{with}\;\;
\bm\gamma_\ell
=\Bigg[\frac{2K-3}{\sqrt{K-1}}-1\Bigg]P_\ell \bar{\bm{c}}$\; and 
\[C_\ell
=
\left(
\frac12
-
\frac{K-3}{\sqrt{K-1}}
\right)
\sum_{k=1}^K
\bm W_{\infty,-k}^\top P_\ell \bm W_{\infty,-k}.
\]
\emph{\bf{Regarding Case $(\nu.4.1)$}:}
Fix an interval $T_\ell$ and a break point candidate from
$\Lambda_\ell^{(1)}$:
\(\eta_{\ell,-k,j}
=
-\frac{b_{\ell,-k,j}}{a_{\ell,j}},\)\(
a_{\ell,j}\neq0.\) Similarly for $t\neq \ell$, fix another $T_t$ with \(\eta_{t,-k,j}
=
-\frac{b_{t,-k,j}}{a_{t,j}},\;\;
a_{t,j}\neq0.\) We show that for fixed $t\neq \ell$,
\[
\mathbf P\!\left[
\hat{H}_{\infty,K}^\prime\big(\eta_{\ell,-k,j}\big)
=
\hat{H}_{\infty,K}^\prime\big(\eta_{t,-k,j}\big)
\right]
=
0.
\]
From calculations of case $(\nu.2)$, we just write for a deterministic arguments $(\bm\omega_{-1},..,\bm\omega_{-K})$,
\begin{align*}
\hat H'_{\infty,K}(\eta_{t,-k,j})
&=\frac{\bar A_t}{2} \,
(\bm\rho_{t,j}^\top\bm\omega_{-k})^2
+\Bigg[\bm\gamma_t^\top\Bigg(\sum_{k=1}^K \bm\omega_{-k}\Bigg)\Bigg](\bm\rho_{t,j}^\top \bm\omega_{-k})+\left(
\frac12
-
\frac{K-3}{\sqrt{K-1}}
\right)
\sum_{k=1}^K
\bm\omega_{-k}^\top P_t \bm\omega_{-k}    
\end{align*}
Similarly one can define $\hat H'_{\infty,K}(\eta_{\ell,-k,j})$. Define for $\ell\neq t$ and $j \notin Z_t, Z_\ell$,
\[
Q_{\ell,t,j}(\omega_{-1},..,\omega_{-K})
:=\hat H_{\infty,K}^\prime(\eta_{\ell,-k,j})-\hat H_{\infty,K}^\prime(\eta_{t,-k,j})
\]
Suppose for the contradiction, we assume that $Q_{\ell,t,j}(\bm\omega_{-1},..,\bm\omega_{-K})=0$ for all $(\bm\omega_{-1},..,\bm\omega_{-K})$. Choose any vector $\bm v\in\mathbb R^p$ and set
\[
\bm\omega_{-1}=\bm v,
\quad
\bm\omega_{-2}=-\bm v,
\quad
\bm\omega_{-k}=\bm 0,
\quad k\ge3.
\] 
Then rest of the argument is routine as before.

\begin{remark}
In fact we skip the remaining cases as the proof idea is exactly similar and repeated routine work as before. However for the sake of completeness, we just state those cases formally.  
\begin{itemize}
    \item Fix an interval $T_\ell$, $j \notin Z_\ell$ and a breakpoint candidate from
$\Lambda_\ell^{(1)}$:
\(\eta_{\ell,-k,j}
=
-\frac{b_{\ell,-k,j}}{a_{\ell,j}},\;\;
a_{\ell,j}\neq0.\) Similarly for $t\neq \ell$, fix $j \in Z_t$ and a break point candidate from
$\Lambda_t^{(2)}$:
\(\kappa_{t,-k,j}
=
\frac{(\bm W_{\infty,-k} - L \bm b_{t,-k})_j}{(L \bm a_t)_j \pm 1},\) with $(L \bm a_t)_j \pm 1\neq0.$ We can show that for fixed $t\neq \ell$,
\[
\mathbf P\!\left[
\hat{H}_{\infty,K}^\prime\big(\eta_{\ell,-k,j}\big)
=
\hat{H}_{\infty,K}^\prime\big(\kappa_{t,-k,j}\big)
\right]
=
0.
\]
\item Fix an interval $T_\ell$, $j \in Z_\ell$ and a break point candidate from
$\Lambda_\ell^{(2)}$:
\(\kappa_{\ell,-k,j}
=
\frac{(\bm W_{\infty,-k} - L \bm b_{\ell,-k})_j}{(L \bm a_\ell)_j \pm 1}\) with \(
(L \bm a_\ell)_j \pm 1\neq0.\) Similarly for $t\neq \ell$, fix $j \in Z_t$ and a break point candidate from
$\Lambda_t^{(2)}$:
\(\kappa_{t,-k,j}
=
\frac{(\bm W_{\infty,-k} - L \bm b_{t,-k})_j}{(L \bm a_t)_j \pm 1}.\) We can show that for fixed $t\neq \ell$,
\[
\mathbf P\!\left[
\hat{H}_{\infty,K}^\prime\big(\kappa_{\ell,-k,j}\big)
=
\hat{H}_{\infty,K}^\prime\big(\kappa_{t,-k,j}\big)
\right]
=
0.
\]
\end{itemize}
This concludes the claim of the proposition that, $\hat{\Lambda}_{\infty,K}$ is the global unique minimizer of $\hat{H}_{\infty,K}^\prime(\lambda)$. All it remains to establish that this global minimizer is well separated. It is important to mention that the set $\Omega^\star$ on which all these arguments should hold, one can consider that set as the all possible finite intersections of complement of these zero measure sets where these objective functions are equal.
\end{remark}
\emph{\bf{Well-separatedness of the minimizer}.} Let $\hat{\Lambda}_{\infty,K}(\omega)$ denote the unique global minimizer of the objective function 
$\hat{H}'_{\infty,K}(\lambda,\omega)$ established just before. Recall the finite candidate set
\[
\mathcal C
=
\{0\}
\;\cup\;
\Lambda
\;\cup\;
\left\{
-\frac{B_\ell}{\bar A_\ell}
:
\ell = 0,\dots,M,\;
-\frac{B_\ell}{\bar A_\ell} \in [\lambda_\ell,\lambda_{\ell+1})
\right\},
\]
and that $\hat{\Lambda}_{\infty,K}(\omega)\in \mathcal{C}$. We now establish separation. Fix $\varepsilon>0$ and let $\ell^\star$ be the unique index such that
$\hat{\Lambda}_{\infty,K}(\omega)\in T_{\ell^\star}$ (or equals a boundary point).
Recall that the partition $\{T_\ell\}_{\ell=0}^r$ satisfies
\[
[0,\infty) = \bigcup_{\ell=0}^M T_\ell=T_{\ell^\star} \;\cup\; \bigcup_{\ell \neq \ell^\star} T_\ell,
\qquad
T_\ell = [\lambda_\ell, \lambda_{\ell+1}),\quad \lambda_0=0,\;\; \lambda_{M+1}=\infty
\]
Hence, for any $\varepsilon > 0$, we can decompose the complement of the $\varepsilon$-neighborhood of the minimizer as
\begin{align*}
\{\lambda \in [0,\infty) : |\lambda - \hat{\Lambda}_{\infty,K}| > \varepsilon\}
&=
\underbrace{\Big\{\{\lambda : |\lambda - \hat{\Lambda}_{\infty,K}| > \varepsilon\} \cap T_{\ell^\star}\Big\}}_{:=C_{\ell^\star,\varepsilon}}\\
&\qquad\qquad\qquad\cup \underbrace{\bigcup_{\ell \neq \ell^\star}
\Big\{\{\lambda : |\lambda - \hat{\Lambda}_{\infty,K}| > \varepsilon\} \cap T_\ell\Big\}}_{:=C_{\ell^\star,\varepsilon}^c}.
\end{align*}
Define the finite gap
\(\delta(\omega)
:=
\min_{\lambda\in\mathcal C,\,
\lambda\neq
\hat{\Lambda}_{\infty,K}(\omega)}
\left[
\hat H'_{\infty,K}(\lambda,\omega)
-
\hat H'_{\infty,K}
(\hat{\Lambda}_{\infty,K}(\omega),\omega)
\right].\) Since $\mathcal{C}$ is finite and no two distinct candidates can attain the same value,
we have $\delta(\omega)>0.$ 
We split the argument into two cases.\\
\textbf{Case (c.1):} Suppose
\(
\lambda\in C_{\ell^\star,\varepsilon}.
\)
On the interval
\(T_{\ell^\star}\),
the global objective admits the quadratic representation
\[
\hat H'_{\infty,K}(\lambda,\omega)
=
\frac{\bar A_{\ell^\star}}2\lambda^2
+
B_{\ell^\star}\lambda
+
C_{\ell^\star},
\qquad
\bar A_{\ell^\star}>0.
\]
Hence
\(
\hat H'_{\infty,K}(\cdot,\omega)
\)
is
\(
\bar A_{\ell^\star}
\)-strongly convex on
\(T_{\ell^\star}\)  in the sense of definition \ref{def:strongconvexity}. Therefore, by Lemma~\ref{lem:strongconvexquadgrowth},
\[
\hat H'_{\infty,K}(\lambda,\omega)
\ge
\hat H'_{\infty,K}
(\hat{\Lambda}_{\infty,K}(\omega),\omega)
+
\frac{\bar A_{\ell^\star}}4
|\lambda-\hat{\Lambda}_{\infty,K}(\omega)|^2.
\]
Since
\(
|\lambda-\hat{\Lambda}_{\infty,K}(\omega)|>\varepsilon,
\)
we obtain
\[
\hat H'_{\infty,K}(\lambda,\omega)
\ge
\hat H'_{\infty,K}
(\hat{\Lambda}_{\infty,K}(\omega),\omega)
+
\frac{\bar A_{\ell^\star}}4
\varepsilon^2.
\]

\textbf{Case (c.2):} Suppose
\(
\lambda\in
C_{\ell^\star,\varepsilon}^{\,c}.
\)
Then there exists
\(
\tilde\ell\neq\ell^\star
\)
such that
\(
\lambda\in T_{\tilde\ell}.
\) Since
\(
\hat H'_{\infty,K}(\cdot,\omega)
\)
is strictly convex on
\(T_{\tilde\ell}\),
its minimum over
\(T_{\tilde\ell}\)
is attained at the unique candidate in
\(
\mathcal C\cap T_{\tilde\ell}.
\)
Therefore,
\[
\hat H'_{\infty,K}(\lambda,\omega)
\ge
\min_{\mu\in\mathcal C\cap T_{\tilde\ell}}
\hat H'_{\infty,K}(\mu,\omega).
\]
Since
\(
\tilde\ell\neq\ell^\star,
\)
the candidate attaining this minimum is distinct from the global minimizer. Hence,
by the definition of
\(
\delta(\omega),
\)
\[
\hat H'_{\infty,K}(\lambda,\omega)
\ge
\hat H'_{\infty,K}
(\hat{\Lambda}_{\infty,K}(\omega),\omega)
+
\delta(\omega).
\]

Combining the two cases, choose
\[
\eta(\varepsilon,\omega)
:=
\frac12
\min
\left\{
\frac{\bar A_{\ell^\star}}4
\varepsilon^2,
\;
\delta(\omega)
\right\}
>0.
\]
Then, for every
\(
\lambda
\)
satisfying
\(
|\lambda-\hat{\Lambda}_{\infty,K}(\omega)|>\varepsilon,
\)
we have
\[
\hat H'_{\infty,K}(\lambda,\omega)
\ge
\hat H'_{\infty,K}
(\hat{\Lambda}_{\infty,K}(\omega),\omega)
+
\eta(\varepsilon,\omega).
\]
Equivalently,
\[
\inf_{\{
|\lambda-\hat{\Lambda}_{\infty,K}(\omega)|
>\varepsilon
\}}
\hat H'_{\infty,K}(\lambda,\omega)
>
\hat H'_{\infty,K}
(\hat{\Lambda}_{\infty,K}(\omega),\omega)
+
\eta(\varepsilon,\omega),
\]
which proves the desired well-separatedness property. \hfill $\square$

\subsubsection{Proof of Theorem \ref{thm: lambdaCVconv}}\label{sec:thm4.1}
Recall from Table~\ref{tab:unified} that
\(n^{-1/2}\hat{\lambda}_{n,K}
\in
\operatorname*{Argmin}_{\lambda\ge0}
\hat{H}_{n,K}'(\lambda),
\) where
\begin{align}
\label{eqn:1}
\hat{H}_{n,K}'(\lambda)
&=
H_{n,K}(n^{1/2}\lambda)\nonumber\\
&=
\sum_{k=1}^{K}
\Bigg[
\frac{m}{2(n-m)}
\bigl(\hat{\bm{u}}_{n,-k}(\lambda)\bigr)^\top
\bm{L}_{n,k}
\bigl(\hat{\bm{u}}_{n,-k}(\lambda)\bigr)
-
\sqrt{\frac{m}{n-m}}
\bigl(\hat{\bm{u}}_{n,-k}(\lambda)\bigr)^\top
\bm{W}_{n,k}
\Bigg],
\end{align}
and, for each \(k\in\{1,\ldots,K\}\),
\begin{align}
\label{eqn:28}
\hat{\bm{u}}_{n,-k}(\lambda)
&=
(n-m)^{1/2}
\Bigl(
\hat{\bm{\beta}}_{n,-k}(n^{1/2}\lambda)
-
\bm{\beta}
\Bigr)
=
\operatorname*{Argmin}_{\bm{u}}
\hat{V}_{n,-k}(\bm{u},\lambda),
\nonumber\\
\hat{V}_{n,-k}(\bm{u},\lambda)
&=
\frac12
\bm{u}^\top
\bm{L}_{n,-k}
\bm{u}
-
\bm{u}^\top
\bm{W}_{n,-k}
+
n^{1/2}\lambda
\sum_{j=1}^{p}
\Bigl(
\bigl|
(n-m)^{-1/2}u_j+\beta_j
\bigr|
-
|\beta_j|
\Bigr).
\end{align}
The limiting objective functions
$\hat{V}_{\infty,-k}$ and
$\hat{H}_{\infty,K}'$
are defined in Table~\ref{tab:unified}. By Lemma \ref{lem:gausswn-}, we have \(\bm W_{n}^{(-)}:=(\bm{W}_{n,-1}^\top,...,\bm W_{n,-K}^\top)^\top
\xrightarrow{d}
(\bm{W}_{\infty,-1}^\top,...,\bm W_{\infty,-K}^\top)^\top:=\bm W_{\infty}^{(-)}.\)
Now by Dudley's almost sure representation theorem we can claim the following: 
\begin{enumerate}[label=(B.\roman*)]
\item there exist a new probability space $(\tilde{\Omega}, \tilde{\mathcal{F}}, \tilde{\mathbf{P}})$ and perfect maps $\pi_n, \pi_\infty : \tilde{\Omega} \rightarrow \Omega $.
\item there exist random vectors $\tilde{\bm{W}}^{(-)}_n := (\tilde{\bm{W}}_{n,-1}^\top,...,\tilde{ \bm{W}}_{n,-K}^\top)^\top$,\;and its limit 
$\tilde{\bm{W}}^{(-)}_{\infty} := (\tilde{\bm{W}}_{\infty,-1}^\top,...,\tilde{ \bm{W}}_{\infty,-K}^\top)^\top$ on 
$(\tilde{\Omega},\tilde{\mathcal{F}},\tilde{\mathbf P})$ such that $\tilde{\bm{W}}^{(-)}_n(\tilde\omega)=\bm{W}^{(-)}_n(\pi_n(\tilde\omega))$ and $\tilde{\bm{W}}^{(-)}_\infty(\tilde\omega)=\bm{W}^{(-)}_\infty(\pi_\infty(\tilde\omega))$.
    \item $\bm{W}^{(-)}_n
\stackrel{d}{=}
\tilde{\bm{W}}^{(-)}_n$ and $\bm{W}^{(-)}_\infty \stackrel{d}{=}
\tilde{\bm{W}}^{(-)}_\infty$.
\item $
\tilde{\bm{W}}^{(-)}_n
\xrightarrow{a.s}
\tilde{\bm{W}}^{(-)}_\infty.$
\end{enumerate}
Now for any $\lambda$, define the versions 
$\tilde{H}_{n,K}'(\lambda)\equiv \tilde{H}_{n,K}'(\tilde{\bm W}_n^{(-)}(\tilde{\omega}), \lambda)$ and $\tilde{H}_{\infty,K}'(\lambda)\equiv \tilde{H}_{\infty,K}'(\tilde{\bm W}_\infty^{(-)}(\tilde{\omega}), \lambda)$ on $(\tilde{\Omega}, \tilde{\mathcal{F}}, \tilde{\mathbf{P}})$ respectively by  \begin{align*}
&\tilde{H}_{n,K}'(\tilde{\bm W}_n^{(-)}(\tilde{\omega}), \lambda) := \hat{H}_{n,K}'(\bm W_n^{(-)}(\pi_n(\tilde{\omega})), \lambda)\;\;\\ &\qquad\qquad\qquad\text{and}\; \;\tilde{H}_{\infty,K}'(\tilde{\bm W}_\infty^{(-)}(\tilde{\omega}), \lambda) := \hat{H}_{\infty,K}'(\bm W_\infty^{(-)}(\pi_\infty(\tilde{\omega})), \lambda).
\end{align*}
Then define
\[
n^{-1/2}\tilde{\lambda}_{n,K}
\in
\operatorname*{Argmin}_{\lambda\ge0}
\tilde{H}_{n,K}'(\lambda) \;\; \text{and}\;\;
\tilde{\Lambda}_{\infty,K}
\in
\operatorname*{Argmin}_{\lambda\ge 0}
\tilde{H}_{\infty,K}'(\lambda).
\]
The above construction clearly gives $\tilde{\lambda}_{n,K} \equiv \tilde{\lambda}_{n,K}(\tilde{\bm{W}}_n^{(-)}(\tilde{\omega})) = \hat{\lambda}_{n,K}(\bm{W}_n^{(-)}(\pi_n(\tilde{\omega})))$, i.e., $\tilde{\lambda}_{n,K}\circ \tilde{\bm{W}}_n^{(-)} = \hat{\lambda}_{n,K} \circ \hat{\bm{W}}_n^{(-)} \circ \pi_n$. Similarly, $\tilde{\Lambda}_{\infty, K} \circ \tilde{\bm{W}}_\infty^{(-)} = \hat{\Lambda}_{\infty, K} \circ \hat{\bm{W}}_n^{(-)} \circ \pi_{\infty}$. 

Due to the boundedness established in Theorem \ref{thm:cvthm}, it is enough to prove that, for every bounded continuous function
\(g:\mathbb{R}\to\mathbb{R}\),
$$\mathbf{E}
\bigl[
g(n^{-1/2}\hat{\lambda}_{n,K})
\bigr]
\longrightarrow
\mathbf{E}
\bigl[
g(\hat{\Lambda}_{\infty,K})
\bigr],\;n\to\infty,$$ to establish
\(n^{-1/2}\hat{\lambda}_{n,K}\xrightarrow{d}
\hat{\Lambda}_{\infty,K}\). Since $\pi_n, \pi_\infty$ are perfect maps, the boundedness of $g$ implies that
\begin{align}
\label{eqn:dudley}
&\left|
\mathbf{E}
\bigl[
g(n^{-1/2}\hat{\lambda}_{n,K})
\bigr]
-
\mathbf{E}
\bigl[
g(\hat{\Lambda}_{\infty,K})
\bigr]
\right|\\
&= \left|
\mathbf{E}
\bigl[
g\circ(n^{-1/2}\hat{\lambda}_{n,K}) \circ \bm{W}_n^{(-)}
\bigr]
-
\mathbf{E}
\bigl[
g\circ(\hat{\Lambda}_{\infty,K})\circ \bm{W}_\infty^{(-)}
\bigr]
\right|
\nonumber\\
=&\;\left|
\tilde{\mathbf{E}}
\bigl[
g\circ(n^{-1/2}\hat{\lambda}_{n,K}) \circ \bm{W}_n^{(-)}\circ \pi_n
\bigr]
-
\mathbf{E}
\bigl[
g\circ(\hat{\Lambda}_{\infty,K})\circ \bm{W}_\infty^{(-)}\circ \pi_\infty
\bigr]
\right|\nonumber\\
 \leq\; & \tilde{\mathbf{E}}\left|
\bigl[
g(n^{-1/2}\tilde{\lambda}_{n,K})
\bigr]
-
\bigl[
g(\tilde{\Lambda}_{\infty,K})
\bigr]
\right|.
\end{align}
To prove the RHS in (\ref{eqn:dudley}) to converge to $0$, it is enough to prove that
\begin{align}\label{eqn:10001}
n^{-1/2}\tilde{\lambda}_{n,K}
\xrightarrow{\tilde{\mathbf P}}
\tilde{\Lambda}_{\infty,K},
\end{align}
due to continuity of $g$ and the bounded convergence theorem. 
In that regard note that for any fixed $\varepsilon > 0$,
\begin{align}\label{eqn:100002}
  \tilde{\mathbf{P}}\Big(|n^{-1/2}\tilde{\lambda}_{n, K} - \tilde{\Lambda}_{\infty, K}| > \varepsilon\Big)&\leq \; \tilde{\mathbf{P}}\Big(|n^{-1/2}\tilde{\lambda}_{n,K} - \tilde{\Lambda}_{\infty,K}| > \varepsilon,n^{-1/2}\tilde{\lambda}_{n,K}\leq K_{\varepsilon}\Big)+\varepsilon\nonumber\\
  &\leq \;\tilde{\mathbf{P}}\Big(\sup_{0\leq \lambda \leq K_{\varepsilon}}\Big|\tilde{H}_{n, K}^\prime(\lambda) - \tilde{H}_{\infty, K}^\prime(\lambda)\Big|  >  \tilde{\eta}(\varepsilon)/2 \Big) + \varepsilon,
\end{align}
where $\tilde{\eta}(\varepsilon) = \inf_{|\lambda - \tilde{\Lambda}_{\infty,K}| > \varepsilon} \tilde{H}_{\infty, K}^\prime(\lambda) - \tilde{H}_{\infty, K}^\prime(\tilde{\Lambda}_{\infty, K})$ and $\tilde{\mathbf{P}}\big[0\leq n^{-1/2}\tilde{\lambda}_{n,K}, \tilde{\Lambda}_{\infty, K}\leq K_{\varepsilon}\big]\geq 1-\varepsilon$. Such an $K_{\varepsilon}$ exists due to Theorem \ref{thm:cvthm} and the almost sure uniqueness of $\tilde{\Lambda}_{\infty, K}$ (cf. Proposition \ref{prop:C.4}). Moreover, the last line in (\ref{eqn:100002}) is properly defined since the involved quantities are actually $\tilde{\mathcal{F}}$ measurable, as is established in Remark \ref{remark:measurableeta} and \ref{remark:measurablesup}. Now the almost sure well-seperatedness of $\tilde{\Lambda}_{\infty, K}$ (cf. Proposition \ref{prop:C.4}) implies the following:
$$\text{for any}\; \delta>0,\;\text{there exists}\; c_{\delta}> 0\; \text{such that}\; \tilde{\mathbf P}(\tilde{\eta}(\varepsilon) > 2c_\delta)>1-\delta.$$
Thus taking $\delta = \varepsilon$, from (\ref{eqn:100002}) we have
\begin{align}\label{eqn:100003}
  \tilde{\mathbf{P}}\Big(|n^{-1/2}\tilde{\lambda}_{n, K} - \tilde{\Lambda}_{\infty, K}| > \varepsilon\Big) 
  \leq \tilde{\mathbf{P}}\Big(\sup_{0\leq \lambda \leq K_{\varepsilon}}\Big|\tilde{H}_{n, K}^\prime(\lambda) - \tilde{H}_{\infty, K}^\prime(\lambda)\Big|  > c_{\varepsilon} \Big) + 2\varepsilon.
\end{align}
Since $\varepsilon > 0$ is arbitrary, for LHS in (\ref{eqn:100003}) to converge to $0$, as $n \rightarrow \infty$, it is enough to claim that

\begin{align}\label{eqn:100004}
\sup_{0\leq \lambda \leq K_{\varepsilon}}\Big|\tilde{H}_{n, K}^\prime(\lambda) - \tilde{H}_{\infty, K}^\prime(\lambda)\Big|\xrightarrow{\tilde{P}} 0,
\end{align}
i.e., $\tilde{H}_{n, K}^\prime(\lambda)$ converges to $ \tilde{H}_{\infty, K}^\prime(\lambda)$ uniformly on $[0,K_{\varepsilon}]$, in probability.
To that end, observe that
\begin{enumerate}[label=(\roman*)]
\item
The sequence of objective functions $\{\tilde{H}^\prime_{n,K}(\lambda)\}_{n\geq 1}$ is stochastically equicontinuous in $\lambda$, due to Proposition \ref{prop:eqicon} and equation (\ref{eqn:mkop}). 
\item
The objective function $\tilde{H}^\prime_{\infty,K}(\lambda)$ is stochastically continuous in $\lambda$, due to stochastic continuity of $\tilde{\bm{u}}_{\infty,-k}(\lambda)$ in $\lambda$ for all $k$. See Remark \ref{remark:uinftyeqicon} for details.
\item For every fixed $\lambda\in[0,K_\varepsilon]$,
\(\big[\tilde H'_{n,K}(\lambda)-\tilde H'_{\infty,K}(\lambda)\big] \xrightarrow{a.s.} 0.\)

\item $\{\tilde{H}^\prime_{n,K}(\lambda)-\tilde{H}^\prime_{\infty,K}(\lambda)\}_{n\geq 1}$ is stochastically equicontinuous in $\lambda$, due to (i) and (ii).
\end{enumerate}
Therefore, (\ref{eqn:100004}) is true due to Theorem~2.1 of
\citet{newey1991uniform}, since the space 
$[0,K_\varepsilon]$ is compact. Therefore, the proof is complete. \hfill $\square$

\subsubsection{Proof of Theorem \ref{thm:distconvcv}}\label{sec:thm4.2}
We are going to show that 
\begin{align}\label{T.1}
n^{1/2}\big(\hat{\bm{\beta}}_n(\hat{\lambda}_{n,K})-\bm{\beta}\big)= \mbox{Argmin}_{\bm{u}}V_n(\bm{u},\bm W_n,\hat{\lambda}_{n,K})\xrightarrow{d}\mbox{Argmin}_{\bm{u}}V_\infty(\bm{u},\bm W_\infty,\hat{\Lambda}_{\infty,K}),
\end{align}
where 
\begin{align*}
V_{\infty}(\bm{u},\bm W_{\infty},\hat{\Lambda}_{\infty,K})&\;=\bigg[(1/2)\bm{u}^\top\bm{L}\bm{u} -\bm{u}^\top \bm{W}_{\infty}\bigg] + \bigg[\hat{\Lambda}_{\infty,K}\Big\{\sum_{j=1}^{p_0}u_jsgn({{\beta}_{j}})+\sum_{j=p_0+1}^{p}|u_j|\Big\}\bigg],
\end{align*}
and
\begin{align*}
&V_n(\bm{u},\bm W_n,\hat{\lambda}_{n,K})\;=\bigg[(1/2)\bm{u}^\top\bm{L}_n\bm{u} - \bm{u}^\top\bm{W}_n\bigg] + \bigg[\hat{\lambda}_{n,K}\sum_{j=1}^{p}\Big(|\beta_{j}+\dfrac{u_{j}}{n^{1/2}}|-|\beta_{j}|\Big)\bigg]\\
&=\bigg[(1/2)\bm{u}^\top\bm{L}\bm{u} -\bm{u}^\top \bm{W}_{n}\bigg]+\bigg[n^{-1/2}\hat{\lambda}_{n,K}\Big\{\sum_{j=1}^{p_0}u_jsgn({{\beta}_{j}})+\sum_{j=p_0+1}^{p}|u_j|\Big\}\bigg] + R_n(\bm{u})\\
&= V_{\infty}(\bm{u}, \bm{W}_n, n^{-1/2}\hat{\lambda}_{n,K}) + R_n(\bm{u}),
\end{align*}
In the above, $\bm{L}_n=n^{-1}\sum_{i=1}^{n}\bm{x}_i{x}_i^\top$ and $\bm{W}_n=n^{-1/2}\sum_{i=1}^{n}\varepsilon_i\bm{x}_i$ where $\bm{L}$ is the limit of $\bm{L}_n$ and $\bm{W}_\infty$ is the weak limit of $\bm{W}_n$. Now to prove (\ref{T.1}), we are going to employ Lemma \ref{lem:argmin}. Hence we require the finite dimensional distribution convergence of $V_n(\cdot,\bm W_n,\hat{\lambda}_{n,K})$ to that of $V_{\infty}(\cdot,\bm W_{\infty},\hat{\Lambda}_{\infty,K})$ and the stochastic equi-continuity of $\{V_n(\cdot,\bm W_n,\hat{\lambda}_{n,K})\}_{n\geq 1}$.\\
First we establish the finite dimensional distribution convergence of $\{V_n(\cdot,\bm W_n,\hat{\lambda}_{n,K})\}_{n\geq 1}$. Note that $R_n(\bm{u}) = o_p(1)$ for any fixed $\bm{u}$ due to Theorem \ref{thm: lambdaCVconv} and condition (C.1). Hence it is enough to show that for any fixed $\bm{u} \in \mathbb{R}^p$, $V_{\infty}(\bm{u}, \bm{W}_n, \hat{\lambda}_{n,K})$ converges in distribution to $V_{\infty}(\bm{u}, \bm{W}_\infty, \hat{\Lambda}_{\infty,K})$. Now fix an $\bm{u}\in \mathbb{R}^p$ and hence without loss of generality we can drop $\bm{u}$ from the argument of all the aforementioned functions. Note that $\bm{W}_n =[K(K-1)]^{-1/2}\sum_{k=1}^{K}\bm{W}_{n, -k}$ and $n^{-1/2}\hat \lambda_{n, K}$ is a measurable function of $\bm{W}_n^{(-)} = (\bm{W}_{n, -1}^\top, \dots, \bm{W}_{n, -K}^\top)^\top$ (cf. Remark \ref{remark:measurabilitywrtcv}). Hence 
if we define $$\psi_n = (\bm{W}_n^\top, n^{-1/2}\hat{\lambda}_{n, K})^\top\; \text{and}\; \psi_\infty = (\bm{W}_\infty^\top, \hat{\Lambda}_{\infty, K})^\top
$$ as $\sigma(\bm{W}_n^{(-)})$ measurable functions, 
then we can write
$$V_{\infty}(\bm{W}_n, \hat{\lambda}_{n,K}) 
\equiv V_{\infty}\circ\psi_n\;\;\text{and}\;\;V_{\infty}(\bm{W}_\infty, \hat{\Lambda}_{\infty,K}) 
\equiv V_{\infty}\circ\psi_\infty.$$ 
Now based on (B.i) - (B.iv) and the constructions of the versions of $\tilde{\lambda}_{n, K}$ and $\tilde{\Lambda}_{\infty, K}$ afterwards, done in the proof of Theorem \ref{thm: lambdaCVconv}, we can define $$\tilde{\psi}_n = (\tilde{\bm{W}}_n^\top, n^{-1/2}\tilde{\lambda}_{n, K})^\top\; \text{and}\; \tilde{\psi}_\infty = (\tilde{\bm{W}}_\infty^\top,\tilde{\Lambda}_{\infty, K})^\top$$ respectively as versions of $\psi_n$ and $\psi_{\infty}$ on the new probability space $(\tilde{\Omega}, \tilde{\mathcal{F}}, \tilde{\mathbf{P}})$. Subsequently, we have $V_{\infty}\circ \tilde{\psi}_j$ as the version of $V_{\infty}\circ \psi_j$ for $j = n$ and $\infty$, on the new probability space $(\tilde{\Omega}, \tilde{\mathcal{F}}, \tilde{\mathbf{P}})$. 
Therefore, for any bounded continuous function $g:\mathbb{R}\times\mathbb{R}\to\mathbb{R}$ we have
\begin{align}\label{eqn:000007}
 &\Big{|}\mathbf{E}\big[g(V_{\infty}(\bm{W}_n, \hat{\lambda}_{n,K}))\big]-\mathbf{E}\big[g(V_{\infty}(\bm{W}_\infty, \hat{\Lambda}_{\infty,K}))\big]\Big{|}=\;\Big{|}\mathbf{E}\big[g\circ V_{\infty}\circ\psi_n\big]-\mathbf{E}\big[g\circ V_{\infty}\circ\psi_\infty\big]\Big{|} \nonumber\\
 =\;&\Big{|}\tilde{\mathbf{E}}\big[g\circ V_{\infty}\circ \tilde{\psi}_n \big]-\tilde{\mathbf{E}}\big[g\circ V_{\infty}\circ\tilde{\psi}_\infty\big]\Big{|} \leq\; \tilde{\mathbf{E}}\Big{|}\big[g\circ V_{\infty}\circ \tilde{\psi}_n \big]-\big[g\circ V_{\infty}\circ\tilde{\psi}_\infty\big]\Big{|}
\end{align}
The second equality is due to the boundedness of $g$ and the perfectness of the maps $\pi_n, \pi_\infty$. Since $(g\circ V_{\infty})(\cdot)$ is a continuous function and $g$ is bounded, it is enough to establish that $\tilde\psi_n  \xrightarrow{\tilde{\mathbf{P}}}\tilde{\psi}_\infty,$ to claim that the last line in (\ref{eqn:000007}) converges to $0$. Note that $\tilde\psi_n  \xrightarrow{\tilde{\mathbf{P}}}\tilde{\psi}_\infty$ is an immediate consequence of the following:
\begin{enumerate}[label=(\roman*)]
    \item $\tilde{\bm{W}}_n^{(-)}\xrightarrow{a.s.} \tilde{\bm{W}}_\infty^{(-)}$
    \item  $\tilde{\bm{W}}_n = [K(K-1)]^{-1/2}\sum_{k=1}^{K}W_{n, -k}\circ\pi_n$ and $\tilde{\bm{W}}_\infty = [K(K-1)]^{-1/2}\sum_{k=1}^{K}W_{\infty, -k}\circ\pi_\infty$ .
    \item $n^{-1/2}\tilde{\lambda}_{n, K}\xrightarrow{\tilde{\mathbf{P}}} \tilde{\Lambda}_{\infty, K}$, done in the proof of Theorem \ref{thm: lambdaCVconv}.
\end{enumerate}
Thus, we have established the finite dimensional distribution convergence of $V_n(\cdot,\hat{\lambda}_{n,K})$. Again, $\{V_n(\cdot,\hat{\lambda}_{n,K})\}_{n\geq 1}$ is stochastically equi-continuous in $\bm{u}$ on compact sets, due to Lemma \ref{lem:stcontwrtu}. Moreover, under the regularity condition (C.1), $V_\infty(\cdot,\hat{\Lambda}_{\infty,K})$ has almost sure unique minimum. Therefore combining all these we can claim that \eqref{T.1} is true based on Lemma \ref{lem:argmin}. Therefore the proof is complete. \hfill $\square$

\subsubsection{Proof of Theorem \ref{thm:bootconsistency}}\label{sec:thm5.1}
Define the set A as:
\[
A:=\Big\{\|(\hat{\bm{\beta}}_n(\hat{\lambda}_{n,K})-\bm{\beta})\|=o\big(n^{-1/2}\log n\big)\Big\}\cap\Big\{\mathcal{L}\big(\bm{W}_n^{*}\mid \mathcal{E}\big)
\xrightarrow{d_*} N\big(0,\bm{S}\big)\Big\}
\]
where $\bm{W}_n^{*}=n^{-1/2}\sum_{i=1}^{n}\Big\{y_i-\bm{x}_i^\top\tilde{\bm{\beta}}_n(\hat{\lambda}_{n,K})\Big\}\bm{x}_i\Big(\frac{G_i^*}{\mu_{G^*}}-1\Big)$, $\mathcal{E}$ is the sigma-field generated by $\{\varepsilon_1,..,\varepsilon_n\}$ and $\mathcal{L}(\cdot|\mathcal{E})$ denotes the conditional law or distribution of the underlined random vector given the data. Now note that $\mathbf P(A)=1$ due to Lemma \ref{lem:asconcentration} and \ref{lem:bnormal}. It is enough to prove that on this set $ A$, $$n^{1/2}\big[\hat{\bm{\beta}}_{n}^*\big(\hat{\lambda}_{n,K}^*\big)-\tilde{\bm{\beta}}_n(\hat{\lambda}_{n,K})\big]=\mbox{argmin}_{\bm{u}}V_n^*(\bm{u},\hat{\lambda}_{n,K}^*)\xrightarrow{d_*}\mbox{Argmin}_{\bm{u}}V_\infty(\bm{u},\hat{\Lambda}_{\infty,K}),$$
where
\begin{align}\label{eqn:1396}
&V_n^*(\bm{u},\hat{\lambda}_{n,K}^*)=\underbrace{(1/2)\bm{u}^\top\bm{L}_n\bm{u}-\bm{u}^\top\bm{W}_n^*}_{\tilde{\ell}_{1n}^*(\bm{u}, \bm{W}_n^*)}+\underbrace{\hat{\lambda}_{n,K}^*\sum_{j=1}^{p}\Big\{ \big|\tilde{\beta}_{nj}(\hat{\lambda}_{n,K})+\frac{u_{j}}{n^{1/2}}\big|-\big|\tilde{\beta}_{nj}(\hat{\lambda}_{n,K})\big|\Big\}}_{\tilde{\ell}_{2n}^*(\bm{u},\hat{\lambda}_{n,K}^*)}   , 
\end{align}
and the form of $V_\infty(\bm{u},\hat{\Lambda}_{\infty,K})$ is available in the proof of Theorem \ref{thm:distconvcv}. Moreover, note that for any $\omega \in A$, $sgn(\tilde{\bm{\beta}}_{nj}(\hat{\lambda}_{n,K})) =sgn(\beta_j)$, for any $j$ such that $\beta_j \neq 0$, for large enough $n$. Thus, for any fixed $\omega \in A$, we can one by one establish Theorem \ref{prop:cv}, Theorem \ref{thm:cvthm}, Theorem \ref{thm: lambdaCVconv} in the Bootstrap regime following the exact proof steps. Then one can establish (\ref{eqn:1396}) by replacing $\bm{W}_n^{(-)}$ with $\bm{W}_n^{(-)*} = (\bm{W}_{n, -1}^{*\top}, \dots, \bm{W}_{n, -K}^{*\top})^\top$ in the proof of Theorem \ref{thm:distconvcv} and following the proof steps. Therefore, the proof is complete. \hfill $\square$

\bibliographystyle{amsplain}

\end{document}